\documentclass[prb,onecolumn,nofootinbib,citeautoscript,10pt,notitlepage]{revtex4-2}
\usepackage{dsfont}
\usepackage{float}
\usepackage{array}
\usepackage{slashed,bbold}

\usepackage{amsmath,amssymb,bm} 
\usepackage{graphicx}

\usepackage[tight]{subfigure}

\usepackage[papersize={8.5in,11in}]{geometry}
\usepackage{overpic}

\usepackage[pdftex,colorlinks=true,linkcolor=blue,citecolor=blue,urlcolor=blue]{hyperref}

\usepackage[usenames, dvipsnames]{color}

\def \bea{\begin{align}}
\def \eea{\end{align}}
\def \nn{\nonumber \\}
\newcommand{\beq}{\begin{equation}}
\newcommand{\eeq}{\end{equation}}

\newcommand{\bq}{\boldsymbol{Q}}

\begin{document}

\title{Induction of non-Fermi liquids by critical cavity photons at the onset of superradiance}

\author{Ipsita Mandal}
\email{ipsita.mandal@snu.edu.in}

\affiliation{Department of Physics, Shiv Nadar Institution of Eminence (SNIoE), Gautam Buddha Nagar, Uttar Pradesh 201314, India}

\begin{abstract}
We investigate the emergence of a non-Fermi liquid (NFL) at a putative quantum critical point signalling the onset of superradiance, in a set-up involving cavity quantum electrodynamics. Although the finiteness of the cavity, being bounded by reflecting mirrors, endows the cavity photons with an effective mass, they become massless at the continuous phase-transition point. We consider the matter part coming from the fermions hopping on a honeycomb lattice near half-filling, featuring doped Dirac cones at two sets of inequivalent valleys. This choice is dictated by the presence of a fermion-boson interaction vertex, which can give rise to Landau damping of the critical bosons, eventually leading to an NFL phase for the fermions. To set up the quantum effective action, we identify the hot-spots of the generically anisotropic (trigonally-warped) Fermi surfaces, which give the sets of points having parallel/antiparallel tangent vectors. The cavity photons act as charge density wave (CDW) order parameters, connecting pairs of hot-spots belonging to the Fermi surface of a single valley. With these ingredients, we set upon identifying NFL phases, using the tools of dimensional regularization and renormalization-group-flow equations. Our final results indicate that stable NFL phases exist in the low-energy limit, for the projections of the flows along the CDW coupling constant.
\end{abstract}

\maketitle

\tableofcontents

\section{Introduction}

The enormous development in theoretical tools, available to explain the behaviour of solid state matter, has the Landau's Fermi liquid theory at our disposal, which has been extremely successful in describing normal metals. However, there exists an extensive number of metallic states where the formalism fails because the crucial ingredients, namely well-defined long-lived fermionic quasiparticles, get destroyed by strong interactions mediated between the quasiparticles.
These systems are commonly known as non-Fermi liquids (NFLs), due to the breakdown of the perturbative framework of the Landau's Fermi liquid theory. One important example involves the mediation of strong electron-electron correlations via gapless bosonic quantum fields, arising in widely-varied contexts.
The first representative NFL to be studied in this category is a finite density of nonrelativistic fermions interacting with transverse $U(1)$ gauge field bosons \cite{holstein}, where it was shown that, electromagnetic fields, interacting with a metal, give rise to an NFL phase. 
Eventualy, similar behaviour has been found to be ubiquitous in systems comprising finite-density fermions coupled with (a) artificial gauge field(s) emerging in various settings \cite{baskaran,larkin, PhysRevLett.63.680,leenag, blok, ubbens, nayak1, sudip}; (b) massless order-parameter bosons emerging at quantum critical points \cite{max-isn, ogankivfr, metzner, delanna, kee, lawler1, rech, wolfle, maslov, quintanilla, yamase1, yamase2, halboth, jakub, zacharias, eaKim, huh, denis, ips-lee, ips-uv-ir2, ips-subir, ips-sc, max-sdw, chubukov1, Chubukov, shouvik2, ips-c2, andres1, andres2}.

Due to the inherent strongly-interacting nature of the NFLs, it is always a challenging task to come up with a controlled approximation to theoretically extract the physical characteristics. To address this issue satisfactorily, there have been considerable efforts to devise quantum field theoretic (QFT) techniques amenable to analytical treatments \cite{holstein, reizer,leenag, HALPERIN, polchinski, ALTSHULER, sudip, eaKim, nayak, nayak1, lawler1, SSLee,
max-sdw, max-isn, chubukov1, Chubukov, mross, jiang, ips2, ips3, Shouvik1, denis, shouvik2, ips-lee, ips-uv-ir2, ips-subir, ips-sc, ips-c2, andres1, andres2, Lee_2018, ips-fflo, ips-u1, ips-rafael}. Using various regularization techniques, it has been shown that while such theories living in
two spatial dimensions give rise to a strong NFL behaviour, their counterparts residing in three dimensions showcase a marginal Fermi-liquid behaviour \cite{ips-lee, ips-u1}.
To illustrate the ubiquitousness of the NFL phases, we would like to point out that NFL fixed points are also known to exist in two-dimensional (2d) and three-dimensional (3d) semimetals (with twofold or multifold bands-crossings), when the chemical potential cuts a nodal point in the presence of a long-ranged (i.e., unscreened) Coulomb potential \cite{abrikosov, moon-xu, rahul-sid, ips-rahul, ips-qbt-sc, ips-hermann, ips-hermann2, ips-hermann3, juricic, ips-birefringent, ips-hermann-review}.

When the NFL phases result from the strong interactions between the itinerant fermions and the order-parameter bosons becoming massless at a quantum critical point, there are two possible scenarios: (I) the critical bosons' momenta are centred about zero momentum, causing the quasiparticles to lose coherence across the entire Fermi surface \cite{max-isn, ogankivfr, metzner, delanna, kee, lawler1, rech, wolfle, maslov, quintanilla,yamase1, yamase2, halboth,
jakub, zacharias, eaKim, huh, denis, ips-lee, ips-uv-ir2, ips-subir, ips-sc, ips-sc_err}; (II) the bosonic momentum is centred around a finite wavevector denoted by $ \mathbf Q $. In the latter case, $ \mathbf Q $ is equivalent to a nesting vector, which connects two points on the Fermi surface related by $\xi (\mathbf  Q  + \mathbf G + \mathbf k_0) = \xi (\mathbf k_0) $, where (a) $\mathbf G $ is either a null vector or a nonzero vector belonging to the space of the reciprocal lattice vectors; and (b) $\xi (\mathbf k_0)$ is the dispersion at a point $\mathbf k_0 $ on the Fermi surface. The two points coupled by the boson with momentum $\mathbf Q$ are referred to as the hot-spots, as they source ordering instabilitites like
spin density wave (SDW) and charge density wave (CDW). For massless bosons, depending on the nature of the fermion-boson coupling, an NFL behaviour emerges locally in the vicinity of the hot-spots \cite{max-sdw, chubukov1,Chubukov, shouvik2,ips-c2, andres1, andres2, ips-2kf}.

In this paper, we consider a system where a CDW boson carries a momentum $ \mathbf  Q $ equal to a nesting vector, creating pairs of hot-spots on the Fermi surfaces, with each pair having antiparallel Fermi velocities (or, equivalently, tangent vectors). For such scenarios, the CDW and SDW orderings are caused by a well-understood singularity resulting from an enhanced phase space for the low-energy effective excitations. For the hot-spots related by an inversion symmetry, the nesting-vector nature of $ \mathbf  Q$ forces its magnitude to be equal to $2k_F$ \cite{max-sdw, metzner1, metzner2, ips-2kf}, where $k_F $ is the the magnitude of the local Fermi momentum vector. However, the case that we will describe here has two distinct Fermi surfaces, which are related by an inversion symmetry, implying that $ |\mathbf  Q|  $ is equal to the sum of the magnitudes of the Fermi momenta at the paired hot-spots within each Fermi surface, whose curvature values are generically unequal. The nature of the fermion-boson coupling is such that it cannot couple the hot-spots of the two Fermi surfaces related by the inversion symmetry. The setting also involves $\mathbf  Q $ not being commensurate with any of the underlying reciprocal lattice vectors.

Observing the growing prowess and versatility of cavity-confined photons to induce strong electron-electron couplings via light-matter interactions \cite{jaksch, kollath, eckstein, farokh, demler, diehl, ahana}, we will investigate the emergence of putative NFLs in the context of cavity quantum electrodynamics (QED) involving 2d crystal lattices \cite{roux, piazza_qed, zhang, piazza_superrad, bhaseen, zhai, basko, polini, pascal, peng}. The continuously developing field of cavity QED shows the promise of achieving a strong coupling of the cavity photons with fermionic lattices, which encompass 2d layered heterostructures \cite{sentef} and synthetic ultracold atomic arrays \cite{piazza_qed, roux, zhang}. Due to the finite size of the cavity, bounded by the cavity mirrors, the photonic spectrum has an effective mass. Nevertheless, we can tune the system towards a phase showing superradiance, when the cavity photons become massless right at the point of this continuous phase transition \cite{piazza_superrad, bhaseen, zhai, basko, polini, pascal, peng}. The superradiant phase is one which is characterized by a ground state containing a macroscopically large number of coherent photons, which can be thought of as a photon condensate. Effectively, the lattice atoms are then coupled to a single-mode spatially-uniform electromagnetic field. The important point to note is that the transition requires a very strong coupling between the atoms and the cavity field. The superradiant phase transition has been observed experimentally in various set-ups like optically
pumped gas \cite{ref22cav}, photo-excited semiconducting quantum dots \cite{ref23cav, ref24cav}, and pumped ultracold gases trapped in an ultrahigh-finesse optical cavity \cite{ref25cav}. However, these experiments involve an external drive, and an equilibrium version is yet to be designed. Despite this fact, we will adopt an optimistic attitude, with the conviction that the desired set-up will be successfully engineered in the near future.

The nature of the crystal lattice affects the phases that can appear in the cavity QED platforms described above. In particular, Peng and Piazza \cite{peng} have studied the cases of the square and honeycomb lattices at the onset of superradiance, and inferred that the nature of the fermion-boson coupling shows the possibility of having NFL phases for the latter. They have considered the Fermi surfaces of the slightly-doped Dirac cones that emerge near half-filling of a honeycomb lattice (such as graphene). They have assumed perfectly circular Fermi surfaces, and have concluded that the entire Fermi surfaces become ``hot'' (like the critical Fermi surface of the Ising-nematic quantum critical point \cite{max-isn, denis, ips-lee, ips-uv-ir2}), coupling in pairs of antipodal patches to the critical cavity-photon mode. Using a random-phase approximation (RPA), they have demonstrated that the fermion self-energy
scales with the frequency $\omega $ as $\omega^{2/3}$, which is a typical signature of an NFL phase.
Since the RPA is an uncontrolled approximation for NFLs, our aim is to describe this system using a controlled approximation. There is, in addition, a second crucial reason for revisiting the QFT set-up of this problem. The assumption of a circular Fermi surface for a doped Dirac cone holds only for vanishingly small doping. Indeed, in the generic scenarios, the leading correction to the isotropic dispersion is given by the so-called trigonal warping of the band around each of the $K_+ $ and $ K_-$ points \cite{ando1998, dresselhaus, neto}. There is no reason to assume a circular Fermi surface, because the crystal symmetries (of the underlying triangular lattice) only ensure that the dispersion always retains a threefold rotational symmetry. The upshot of this observation is that the entire Fermi surfaces around the $K_\pm $ valleys \textit{do not} become hot, as all points on the trigonally-warped Fermi surfaces do not have antipodal partners with antiparallel tangent vectors. The cavity-photon mode acts like a CDW order parameter, connecting only three pairs of hot-spots for each Fermi surface, as illustrated in Fig.~\ref{fig_fs}. Consequently, we have an effective CDW order parameter boson whose momentum $\mathbf Q $ is incommensurate, with its magnitude being equal to that of the vector connecting the two hot-spots with antiparallel Fermi velocities, but unequal curvature in general.

In order to formulate a controlled QFT description of the hot-spot theory, we will implement the analytical approach of dimensional regularization, in the same spirit as done in earlier works \cite{denis, ips-lee, ips-uv-ir2, ips-sc, ips-subir, ips-fflo, ips-u1, ips-2kf, Lee_2018}. The steps involve increasing the co-dimension of the Fermi surface to a generic value, and identifying the value of the upper critical dimension $d=d_c$ at which the interactions become marginal. This allows us to compute the physical observables in a systematic perturbative approximation (about $d_c$) using the perturbative parameter $\epsilon = d_c -2 $, despite the quasiparticle-weight of the original fermionic excitations being driven to zero by the strong fermion-boson interactions.

The paper is organized as follows. In Sec.~\ref{secmodel}, we set up the effective low-energy Euclidean action in the Matsubara frequency space, describing the fermionic excitations at the conjugate hot-spots of the Fermi surfaces around the $K_{\pm}$ valleys, all having parallel/antiparallel tangent vectors. Sec.~\ref{seconeloop} is devoted to the derivations of the one-loop bosonic and fermionic self-energies, as well as the one-loop vertex-corrections. Using the one-loop results, the renormalization-group (RG) flows are determined in Secs.~\ref{secrg} and \ref{secrg2}, by curing the ultraviolet divergences through defining counterterm actions. The RG equations enable us to calculate the fixed points of the theory in the infrared (IR) limit. In particular, our final results show that the system exhibits stable NFL fixed points. The two separate RG-related sections deal with the cases of non-circular and circular patches, respectively. Finally, we conclude with a summary and some discussions in Sec.~\ref{secsum}. The appendices show some identities involving useful integrals, as well the details of some intermediate steps.

\section{Model}
\label{secmodel}

The noninteracting tight-binding Hamiltonian in the honeycomb lattice, which we consider for the cavity QED treatment, can be taken from that of graphene, which has been widely studied in the literature \cite{neto}. Exactly at half-filling, the Dirac cones emerge at each of the valleys denoted by $K_\varsigma $, where $ \varsigma =\pm $. We note that the two inequivalent Dirac points, located at $K_+$ and $K_-$, are exchanged on applying the time-reversal symmetry.
We are interested in the low-energy dynamics of the fermions with momenta near $K_\varsigma $, which is obtained by taking the continuum limit of tigh-binding model. On doping away from the Dirac points, the system represents a Fermi liquid with a nonparabolic dispersion and singly-connected convex Fermi surfaces. For very low doping, the dispersion is linear and isotropic, thus falling into the category of the so-called of Dirac Fermi liquid (DFL) \cite{maslov-dfl}.

\begin{figure}[t]
\begin{center}
\includegraphics[width = 0.65 \textwidth]{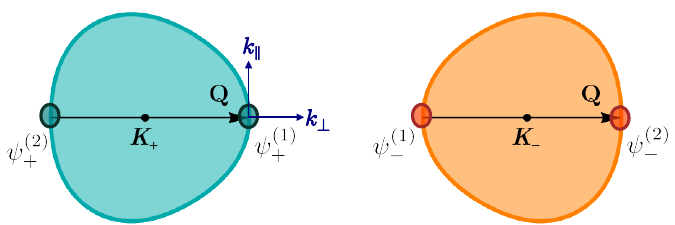} 
\end{center}
\caption{Schematics of the two Fermi surfaces with trigonal warping, located around the two adjacent valleys $K_+$ and $K_-$, respectively. We show the two hot-spots, on each Fermi surface, connected by the wavevector $ \boldsymbol{\mathcal Q}  $ (indicated by the black arrow), which is incommensurate with the underlying reciprocal lattice vectors. For the valley $K_+$ ($K_-$), the fermionic fields in the vicinity of the right and left hot-spots are designated as $\psi_+^{(1)}$ ($\psi_-^{(2)}$) and $\psi_+^{(2)}$ ($\psi_-^{(1)}$), respectively. They interact with the critical cavity-photon modes whose momenta are centred at $ \pm \boldsymbol{\mathcal Q}$.
\label{fig_fs}}
\end{figure}

The dispersion is obtained by expanding the momentum vector $\mathbf P $ around a $ K_\varsigma $-point as $\mathbf P = \mathbf  K_\varsigma + \mathbf p $, where $\mathbf K_\varsigma $ is the momentum at $K_\varsigma$ and $| \mathbf  p| /|\mathbf  K_\varsigma | \ll 1$. At leading order in $| \mathbf  p| /
|\mathbf  K_\varsigma |$, it gives us the linear-in-$|\mathbf p|$ dispersion, characteristic of a DFL. Retaining terms up to order $ \left( | \mathbf  p| /|\mathbf  K_\varsigma | \right)^2 $, we get the corrections leading to the trigonal warping of the bands \cite{ando1998, dresselhaus, neto} around the $K_\varsigma$-point. Choosing a diagonal electron-hole basis, the final form of the Hamiltonian (suppressing the spin degrees of freedom) can be written as \cite{maslov-dfl}
\begin{align}
\label{eqheh}
H_{0} =\sum_{ \varsigma,{ \mathbf{p}} } 
\left[
\left (\epsilon_{\varsigma ,\mathbf{p}, + }-\mu_c \right ) 
\alpha^{\dagger}_{\varsigma, {\mathbf{p}}} \,
\alpha^{\phantom{\dagger}}_{\varsigma, {\mathbf{p}} } 
+ \left (\epsilon_{\varsigma ,\mathbf{p}, -}-\mu_c \right ) 
\beta^{\dagger}_{\varsigma, {\mathbf{p}} }\, 
\beta^{\phantom{\dagger}}_{\varsigma, {\mathbf{p}} }
\right ] ,
\end{align}
with $\mu_c $ being the chemical potential.
Here, the dispersion, taking into account trigonal warping, is captured by
\begin{align}
\label{twd}
\epsilon_{\varsigma, {\mathbf{p}} ,\lambda } (\theta_p) &=
\epsilon^\mathrm{D}_{ {\mathbf{p}}, \lambda}
+\epsilon^\mathrm{TW}_{\varsigma, {\mathbf{p}}, \lambda } (\theta_p)\,, \quad
\epsilon^\mathrm{D}_{ {\mathbf{p}} ,\lambda } = 
 \lambda \, v_\mathrm{D} \, |\mathbf{p}|\,,
\quad
\epsilon^\mathrm{TW}_{\varsigma, {\mathbf{p}} ,\lambda } (\theta_p) = 
 \lambda \, \varsigma\, 
\frac{v_\mathrm{D} \,  |\mathbf p|^2}  { 2 \, \tilde \rho } 
\cos \big(3 \theta_p \big)\,,
\quad \theta_{p}= \arctan\bigg (\frac{p_y} {p_x} \bigg)\,,
\end{align}
and $ \alpha^{\dagger}_{\varsigma, {\mathbf{p}} }  $ ($ \beta^{\dagger}_{\varsigma, {\mathbf{p}} } $) represents the creation operator for an electron (hole) in
the conduction (valence) band located in the vicinity of the $K_\varsigma$ point. Furthermore,
$\lambda=\pm $, $v_\mathrm{D}$ represents the Fermi velocity, and $\tilde \rho $ is the warping parameter which is inversely proportional to the nearest-neighbour distance. Note that we have chosen a convention/coordinate system such that there is no relative sign between the $\epsilon^\mathrm{D}_{ {\mathbf{p}}, \lambda}$ and $\epsilon^\mathrm{TW}_{\varsigma, {\mathbf{p}}, \lambda }$ terms for the valley $K_+$. For not-too-large warping, the two Fermi surfaces at the two valleys remain globally convex, as shown in Fig.~\ref{fig_fs}.

\subsection{Identifying hot-spots paired by CDW ordering}

Since the coupling between the lattice electrons and the cavity photons is determined via the Peierls substitution, the fermion-boson vertex is basically given by the current operator \cite{peng}.
The gradient part of the current operator for the Hamiltonian in Eq.~\eqref{eqheh} is readily obtained from the relation $ \mathbf{j} = -\frac{\delta}{\delta \mathbf{A}}H_0(\mathbf{p}-e\mathbf{A}) \,,$ where $\mathbf A $ is the vector potential. Since the occupied states in the valence band do not contribute to the current, we set $\lambda = +$, and the intraband part of the current operator reduces to
\begin{align}
\label{eqcur}
 \mathbf{j} =\sum_{\varsigma,{\mathbf{p}} } {\bf v}_{\varsigma,{\mathbf{p}}} \, 
 \alpha_{\varsigma,{\mathbf{p}}}^{\dagger} \, \alpha^{\phantom{\dagger}}_{\varsigma,{\mathbf{p}}}
 \,, \quad {\bf v}_{\varsigma,{\mathbf{p}}} = 
 {\boldsymbol{\nabla}}_{\mathbf k} \epsilon_{\varsigma ,\mathbf{p}, +}\,.
\end{align}
Since the polarization function, used to investigate a putative CDW gap-opening, arises from the current-current correlator, we find that there is no possibility of intervalley scattering. In other words, a CDW boson can only connect hot-spots on the same valley, when its interaction vertex with the fermions involves the current operator $\mathbf{j}$.

We do not repeat here the computation of the polarization function, as it is a pretty standard exercise (that can be found in the literature) which illustrates the opening of a CDW gap. The instability is caused by an enhanced phase space available for the low-energy effective excitations, when we consider scattering of electrons between the hot-spots having (anti)parallel tangent vectors. In particular, our case is very much similar to the computations of the current-current correlators in Ref.~\cite{maslov-dfl}, where the authors have studied doped monolayer graphene in the presence of an effective four-fermion interaction, caused by a bare Coulomb potential. They have also taken into account the trigonal-warping effects. The only difference from their analysis is that we do not have any four-fermion terms and, hence, no possibility of intervalley scatterings. If such a term were present, we would have the scope of CDW instabilitites with a $\boldsymbol Q$-vector connecting time-reversed patches of the two inequivalent Fermi surfaces (centred around $K_+$ and $K_-$), similar to the cases studied in Ref.~\cite{levitov}. We would like to point out that the derivations shown in the supplemental material of Ref.~\cite{peng} are not applicable here, because they have assumed a circular Fermi surface at each valley, with the pairs of hot-spots having the same curvature. As explained earlier, for a noncircular Fermi surface, the hot-spots connected by $\boldsymbol Q $ do not have the same curvature.

\subsection{Patch theory using time-reversed partners}

In order to construct a patch theory, we need to construct two-component
spinors comprising time-reversed partners~\cite{denis}, which here are
effectively the fermionic operators on the two conjugate valleys around
$K_+ $ and $K_-$. Hence, at each valley, we have three pairs of hot-spots, each pair
connected by a CDW boson. These three pairs are related by a rotation
of $2\pi/3$, as a consequence of the underlying threefold rotational
symmetry of the lattice. Consequently, the set of three different $\mathbf Q $-vectors, at which the three CDW bosons are centred, are related by the same rotation of $2\pi/3$. Since the cavity photons have no self-interactions, we have three decoupled systems, each comprising two pairs of hot-spots, all having parallel or antiparallel tangent vectors, and interacting with one type of CDW bosons. Hence, it is sufficient to analyze the characteristics of one of them, which we now pursue in the remainder of the paper.
The corresponding action involving the patches at $\theta_p =0 $ and $\theta_p = \pi $, written in terms of the so-called patch coordinates \cite{denis,ips-lee,ips-u1, ips-rafael, ips-fflo, ips-2kf}, is given by
\begin{align}
 S &= \sum_{\substack{ \varsigma = \pm \\ 
n = 1, 2 }  }
 \int_{k} \left \lbrace \psi^{(n)}_{\varsigma } (k) \right  \rbrace^{\dagger}
  \left [ -i\, k_0 -(-1)^n
  \, \varsigma \,v_F^{(n)} \,k_1 
  + \kappa^{(n)} \, k_2^2 \right ] \psi_{\varsigma }^{(n)} (k) 
	+ 
\frac{1} {2}	\int_k \phi(k) \left( k_0^2 + k_1^2 + k_2^2 \right) \phi(-k) \nn
& \qquad + \frac{e} {2}  \int_k  \int_q \,
\Big[ \,\phi(q) \left \lbrace 
\psi^{(1)}_{+ } (k+q) \right \rbrace^{\dagger} 
\psi^{(2)}_{+ }(k) 
+ \phi (-q) \left \lbrace 
\psi^{(2)}_{+ } (k-q) \right \rbrace^{\dagger} 
\psi^{(1)}_{+ }(k) 
\,\Big] \nn
& \qquad + \frac{ e} {2}  \int_k  \int_q \,
\Big[ \,\phi (q) \left \lbrace 
\psi^{(2)}_{- } (k+q) \right \rbrace^{\dagger} 
\psi^{(1)}_{- }(k) 
+ \phi (-q) \left \lbrace 
\psi^{(1)}_{- } (k-q) \right \rbrace^{\dagger} 
\psi^{(2)}_{- }(k) 
\,\Big]\,.
\label{eqs0}
\end{align}
Here, $k=(k_0,\boldsymbol k)$ denotes the three-vector comprising the Matsubara space frequency $k_0 $ and the spatial momentum vector $\boldsymbol k =(k_1, k_2) \equiv (k_\perp, k_\parallel)$, $\int_k \equiv \int dk_0\, d^d{\mathbf k} /(2\,\pi)^{d+1} $, and $d =2 $ is the number of spatial dimensions. In the vicinity of the point $K_\varsigma$, the fermionic degrees of freedom about the two hot-spots are represented by $\psi_\varsigma^{(1)}(k)$ and $\psi_\varsigma^{(2)}(k)$, respectively, as shown in Fig.~\ref{fig_fs}. The field $\phi (k)$ refers to the CDW bosonic field, emerging from the cavity photons, carrying frequency $k_0$ and momenta $  \mathbf Q  + \boldsymbol k$. At the superradiance quantum critical point, the cavity bosons become massless, which is reflected in the purely bosonic part of the action. We rescale the fermionic momenta such that, for the fields $\psi_+^{(1)}(k)$ and $\psi_-^{(1)}(k)$, we set $v_F^{(1)} = \kappa^{(1)}= 1$. To simplify notations, we then use the symbols $v_F^{(2)} = \upsilon $ and $\kappa^{(2)}= \kappa $. We would like to point out that $\kappa $ can be negative when the curvature at the corresponding hot-spots is negative. Although the bosonic velocity, in general, has a distinct value, we have set the bare velocity of the bosons equal to unity, because the actual value of the bosonic velocity does not matter in the low-energy effective theory. This results from the fact that the dynamics of the bosons at the quantum critical point is dominated by the particle-hole excitations of the Fermi surface at low energies.

\begin{figure}[t]
\begin{center}
\includegraphics[width = 0.25 \textwidth]{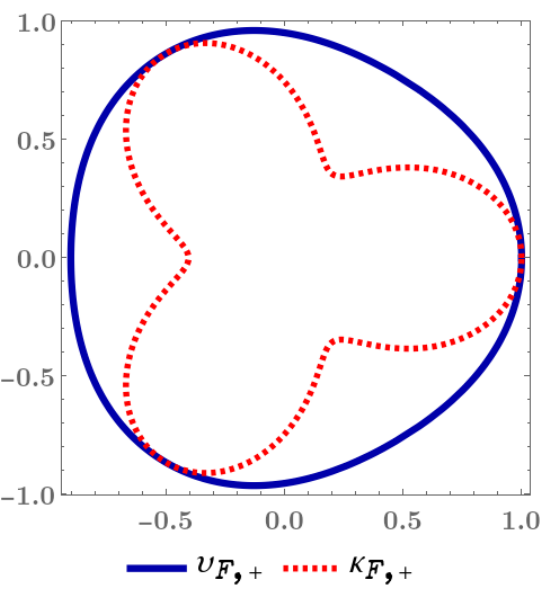} 
\end{center}
\caption{Typical parameters of the Fermi surface at valley $K_+$, when all the hot-spots have a positive curvature.
\label{figfs2}}
\end{figure}

In our notations, before rescaling, the Fermi velocity at a point with angular coordinate $ \theta_p$ is given by $ \upsilon_{F, \varsigma} (\theta_p)  = v_D \left [ 1 +  {\varsigma} \cos\big (3 \theta_p \big) /  \rho \right ]$ (with $\rho \simeq \tilde \rho \, v_D / \mu_c $), and we denote half the curvature as $\kappa_\varsigma (\theta_p)$.
These two quantities appear as $v_F^{(n)}$ and $\kappa^{(n)}$, respectively, when we consider the hot-spots at $\theta_p =0$ and $\theta_p = \pi $ [cf. Fig.~\ref{fig_fs}].
After using the freedom of rescaling of the momenta, we are now effectively using the rescaled versions, given by $ \upsilon_{F, \varsigma} (\theta_p) \rightarrow \upsilon_{F, \varsigma} (\theta_p)/ \upsilon_{F, +} (0)$ and $\kappa_\varsigma (\theta_p) \rightarrow \kappa_\varsigma (\theta_p) /\kappa_+ (0)$. This is demonstrated in Fig.~\ref{figfs2} for the valley $K_+$. Clearly, it shows that both $ v_F^{(1)}\equiv \upsilon_{F,+} (0)$ and $\kappa^{(1)} \equiv \kappa_+ (0) $ (at the right-hand and left-hand hot-spots of $K_+$ and $K_-$, respectively) are set to unity, while the two remaining conjugate hot-spots feature $ v_F^{(2)}\equiv \upsilon =  \upsilon_{F,+} (\pi )$ and $\kappa^{(2)} \equiv \kappa =\kappa_+ (\pi) $, such that $1 -\kappa/\upsilon \geq 0 $. Here, we consider the warping parameter regime such that the Fermi surfaces remain convex globally.

Using the patch coordinates involves the crucial ingredient that $\mathbf K $ has dimension one and $ k_d$ has dimension $1/2$. From the kinetic parts of the action, we infer that the engineering dimensions of the fermions and the bosons evaluate to $ \Big [ \psi^{(n)}_{\varsigma } \Big ] =[\phi_\pm] = -7/4 $. Plugging these values into the interacting part of the action, we get $[e] = 1/4$, which immediately tells us that $e$ is a relevant coupling. Hence, a non-Fermi liquid is expected to emerge, analogous to the systems studied in Refs.~\cite{denis, ips-lee, ips-u1, ips-2kf}. In order to obtain a controlled approximation,
we employ the technique of dimensional regularization, which involves extending the co-dimension of the Fermi surface as an intermediate mathematical step \cite{denis,ips-lee,ips-u1, ips-2kf}. This enables us to determine the upper critical dimension $d=d_c$, at which the one-loop fermion self-energy shows a logarithmic divergence in terms of a Wilsonian cut-off $\Lambda $, and the system behaves as a marginal Fermi liquid. 

In order to preserve the analyticity of the theory in momentum space (which translates to locality in the position space) with general co-dimensions, we define the following two spinors \cite{denis,ips-lee,ips-u1}:
\begin{align}
 \Psi_1^T(k) = \left [
\psi_{+}^{(1)}(k)  \quad  
\left \lbrace \psi_{-}^{(1)} \right \rbrace^ \dagger(-k) \right ]
\text{ and }
\Psi_2^T(k) = \left [
\psi_{-}^{(2)}(k)  \quad  
\left \lbrace \psi_{+}^{(2)} \right \rbrace^ \dagger(-k) \right ], 
\end{align}
and their conjugates
\begin{align}
 \bar \Psi_n \equiv \Psi_n^\dagger \,\gamma_0 \text{ for } n \in \lbrace 1,2 \rbrace\,.
\end{align} 
In terms of the above spinors, we are now equipped to
write down an action that describes the patches of the one-dimensional Fermi surface, in the vicinity of the hot-spots,
embedded in a $d$-dimensional momentum space \cite{denis,ips-lee,ips-u1, ips-2kf, ips-fflo}. The explicit form of the low-energy effective action is captured by
\begin{align}
\label{eqs1}
S  &=   \sum_n  \int_k \bar \Psi_n  (k) \,i
\left [  \mathbf  \Gamma \cdot \mathbf  K  +  \gamma_{d-1} \, \delta_k^{(n)} \right  ] \Psi_n (k)  
 +
 \frac{1}{2}\int_k k_d^2 \,  \phi (k) \, \phi (-k) 
 \nn & \hspace{ 0.5 cm } 
- \left[ 
\frac{ i\, e \, \mu^{x_e/2} } {2} 
\int_{k} \int_q \phi (q) \,
 \bar{\Psi}_1 (k+q) \, {\Psi}_2(-k) 
+ \text{h.c.} \right ] ,
\nn x_e &=  \frac{ 5 } {2} - d \,, \quad 
\delta_k^{(1)} = k_{d-1} +  k_d^2 \,, \quad
\delta_k^{(2)} = \upsilon \,  k_{d-1} + \kappa\, k_d^2 \,.
\end{align}
The $(d-1)$-dimensional vector $\mathbf  K ~\equiv ~(k_0, k_1,\ldots, k_{d-2})$ describes
the frequency and the $(d-2)$-components 
of the momentum vector, with the latter arising from the added co-dimensions. The original momentum components along the $k_1$- and $k_2$-directions have been relabelled as $k_{d-1}$ and $k_d$, respectively.
Overall, we have now an auxuliary system in a $d$-dimensional momentum space, with the set
$ \lbrace k_1, \cdots ,k_{d-1} \rbrace $ representing the
$(d-1)$-components perpendicular to the Fermi surface, while $k_d$ is oriented along the local tangent. The symbol $\mathbf  \Gamma \equiv (\gamma_0, \gamma_1,\ldots, \gamma_{d-2})$ represents a $(d-1)$-dimensional vector of matrices, as we take a scalar product of $\mathbf  \Gamma $ with $\mathbf  K$ in the kinetic term for the fermions.
Regarding the matrix dimensions of the components of $\mathbf  \Gamma $, we note that, in practice, 
it is sufficient to consider only the $2 \times 2$ Pauli matrices, such that
$\gamma_0 = \sigma_y $ and $ \gamma_{d-1} = \sigma_x$. This is because, in the end, we are interested in continuing to $d=2$, which is the actual physical spatial dimension of the system under consideration.
We have introduced a floating mass scale $\mu \sim \Lambda $, raised to the  power $x_e/2$, in order to render the coupling constant $ e$ dimensionless.

The non-interacting parts of the patch-action in Eq.~\eqref{eqs1}
are invariant under the scale transformations
\begin{align}
\mathbf  K & =  \frac{\mathbf  K'}{b} \,, 
\quad k_{ d-1 } =\frac{k_{ d-1 }'}{b} \,,
\quad k_d = \frac{k_d'}{\sqrt{b}} \,,
\quad
\Psi_n (k)  =  b^{ \frac{2 d + 3} {4}}  \, \Psi'_n (k') \,, \quad 
\phi  (k) = b^{\frac{2 d + 3}{4}}  \, \phi' (k')\,.
 \end{align}
This follows from the fact that $ [\mathbf K] = 1 $ and $ [k_d] =1/2$, which are the characteristics of the patch coordinates. We note that the fermions around the two hot-spots, which interact strongly with the bosons, have
$ |k_d| \gg k_{d-1}$, because a scattering event away from the Fermi surface costs a lot of energy.
In the kinetic part for the bosons, only the term proportional to $k_d^2 $ 
is retained, because the part involving $\left( {\mathbf  K}^2 + k_{d-1}^2 \right)$ 
is irrelevant (for an RG flow torwads the IR energy scales) under the scaling relations shown above.

From Eq.~\eqref{eqs1}, we get the bare propagator for the fermions as
\begin{align}
\label{propf}
G_n (k)  \equiv  \left\langle \Psi_n(k)\,  \bar{\Psi}_n(k) \right\rangle_0
= -i\, \frac{\mathbf  \Gamma \cdot \mathbf  K +
\gamma_{d-1} \,\delta_k^{(n)}} 
{\mathbf  K^2  + \delta_k^2} \,.
\end{align}
The bare bosonic propagator is given by
\begin{align}
D_{(0)}(k) 
= \frac{1}{k_d^2  }  \,.
\end{align}

The value of $x_e$ tells us that the coupling constant $e$ becomes marginal at the upper critical dimension $d_c=5/2$. In other words, $e$ is relevant for $d<5/2$ and irrelevant for $d >5/2$. Our aim is to access the interacting phase perturbatively in $d=5/2-\epsilon$, using $\epsilon $ as the perturbative parameter. In particular, this implies that at the end of our systematic $\epsilon$-expansion, we have to set $\epsilon=1/2$ for our original two-dimensional theory.

\section{One-loop Feynman diagrams}
\label{seconeloop}

In this section, we show the results for all the Feynman diagrams at the one-loop order.

\subsection{One-loop boson self-energy}
\label{oneloopbos}

Here, we compute the one-loop self-energy of the boson, which is given by
\begin{align}
\label{bosloop}
\Pi_1 (q) & = - \frac{
\left ( i \, e  \,\mu^{\frac {x_e} {2} } \right )^2} {2}
 \int_k \mbox{Tr}
\left[  G_1 (k+q)\, G_2 (k) \right] 
\nn & = e^2 \, \mu^{x_e}  \int_k 
\frac{\mathbf  K \cdot (\mathbf  K +{\boldsymbol Q}) + \delta_k^{(1)} \, \delta_{k+q}^{(2)}
}
{ \left [\mathbf  K^2 + \delta_k^2 \right ] 
\left [(\mathbf  K +{\boldsymbol Q})^2 + \delta_{k+q}^2 \right ]}
\quad [ \text{where } \delta_k^{(1)} = k_{d-1} + k_d^2  \text{ and }
\delta_{k+q}^{(2)} = \upsilon\left( k_{d-1} + q_{d-1} \right) 
+  \kappa \left( k_d + q_d \right)^2]
\nn & = e^2 \, \mu^{x_e}  \int_k 
\frac{\mathbf  K \cdot (\mathbf  K +{\boldsymbol Q}) 
+ \upsilon \,\delta_k^{(1)} \, 
 \left \lbrace  k_{d-1} + q_{d-1} 
+  \frac{\kappa} {\upsilon} \left( k_d + q_d \right)^2 \right \rbrace
}
{ \upsilon^2
\left [ \mathbf  K^2 + \delta_k^2 \right ] 
\left [ \frac{(\mathbf  K +{\boldsymbol Q}) ^2 }
{\upsilon^2} + 
 \left \lbrace k_{d-1} + q_{d-1} 
+  \frac{\kappa} {\upsilon} 
\left( k_d + q_d \right)^2  \right \rbrace^2 \right ]} \,.
\end{align}

We now perform the integration over $k_{d-1}$ using
Eq.~(\ref{equseint}), shown in Appendix~\ref{appint},
to obtain 
\begin{align}
\label{eqpi0}
\Pi_1 (q)  &  = \frac{e^2 \, \mu^{x_e}} {2} 
\int \frac{ d k_{d} \, d\mathbf  K}{(2 \, \pi)^d} 
\frac{ \left(  {\mathbf  K}
+ \frac{|\mathbf  K + {\boldsymbol Q}|} {\upsilon}  \right) 
\; \left[ \,\mathbf  K \cdot (\mathbf  K +{\boldsymbol Q}) 
+ {\mathbf  K}\;  \frac{ |\mathbf  K +{\boldsymbol Q}|} {\upsilon}   \right]  
}
{ 
\upsilon^2\, {\mathbf  K}\; \frac{ |\mathbf  K +{\boldsymbol Q}|} {\upsilon}
\, 
\left[ \left \lbrace 
 q_{d-1} 
+  \frac{\kappa} {\upsilon} \left( k_d + q_d \right)^2 
-k_d^2 \right \rbrace^2
 + \left ( {\mathbf  K}
+\frac{ |\mathbf  K +{\boldsymbol Q}|} {\upsilon} \right)^2
\right] }    \nn
&  = \frac{e^2 \, \mu^{x_e}} {2} 
\int \frac{ d k_{d} \, d\mathbf  K}{(2 \, \pi)^d} 
\frac{ \left(  {\mathbf  K}
+ \frac{|\mathbf  K + {\boldsymbol Q}|} {\upsilon}  \right) 
\; \left[ \,\mathbf  K \cdot (\mathbf  K +{\boldsymbol Q}) 
+ {\mathbf  K}\;\frac{ |\mathbf  K +{\boldsymbol Q}|} {\upsilon} \,  \right]  
}
{ \upsilon \, {\mathbf  K}\; |\mathbf  K +{\boldsymbol Q}|\,
\left[ \Upsilon^2 (k,q) +  \left ( {\mathbf  K}
+\frac{ |\mathbf  K +{\boldsymbol Q}|} {\upsilon} \right)^2
\right ] }  \,,
\end{align}
where
\begin{align}
\Upsilon(q,k) =  \begin{cases}
\left (1 - \frac {\kappa} {\upsilon} \right)
\left [ k_d^2 - \frac {\upsilon \, e_q}
{  \upsilon - \kappa  } \right ]
& \text{ for } \upsilon \neq \kappa \\
 \delta_q^{(1)} + 2\, k_d \, q_d 
& \text{ for } \upsilon = \kappa 
\end{cases} \,,
\end{align}
and
\begin{align}
e_q = \frac {\kappa \,  q_d^2} 
{\upsilon  - \kappa  }   \,
+ \,q_ {d-1} \,.
\end{align}

For the case of $\upsilon = \kappa $, we get
\begin{align}
\label{eqPi1}
\Pi_1 (q)  & = \frac{e^2 \, \mu^{x_e}} {4} 
\int  \frac{  \, d\mathbf  K}{(2 \, \pi)^d} 
\int_{-\infty }^\infty dk_d \,
\frac{ \left(  {\mathbf  K}
+ \frac{|\mathbf  K + {\boldsymbol Q}|} {\upsilon}  \right) 
\; \left[ \,\mathbf  K \cdot (\mathbf  K +{\boldsymbol Q}) 
+ {\mathbf  K}\; |\mathbf  K +{\boldsymbol Q}| \,  \right]  
}
{\sqrt {|q_d |}\; \upsilon \, {\mathbf  K}\; |\mathbf  K +{\boldsymbol Q}|\,
\left[   k_d^2 +  \left ( {\mathbf  K}
+\frac{ |\mathbf  K +{\boldsymbol Q}|} {\upsilon} \right)^2
\right ] }   
 = \frac{e^2 \, \mu^{x_e}} { 8 \, |q_d| \, \upsilon } 
\; I_1 (d, {\boldsymbol Q}) \, ,
\end{align}
where
\begin{align}
\label{eqi1}
I_1 (d, {\boldsymbol Q}) = \int  \frac{  \, d^{d-1}\mathbf  K}{(2 \, \pi)^{d-1} } 
\left[ \frac{ \upsilon\,
 \mathbf  K \cdot (\mathbf  K +{\boldsymbol Q}) }
{ {\mathbf  K}\; |\mathbf  K +{\boldsymbol Q}|}
+ 1 \right ] . 
 \end{align}

For the case of $ \upsilon \neq \kappa $, we change variables to $u = k_d^2 $, which gives the Jacobian factor as $ 1/\left( 2 \,\sqrt {u}  \right) = 1/\left( 2 \,|k_d| \right) $.
From the denominator of the second factor in the integrand, we find that it forces the dominant contribution to the integral to come from $ u \sim  e_q $, in the regime $|\boldsymbol Q| \ll \kappa \,q_d^2 $. We have assumed $e_q$ to be a positive quantity, remembering that the typical energy scales impose the constraint of $q_d \gg q_{d-1}$. Using the above, we approximate $ |k_d| $ by $\sqrt {\upsilon\,e_q /\left(\upsilon -\kappa \right) }$ in the Jacobian.

The next steps for evaluating the remaining integrals can be found in Appendix~\ref{apponeloopbos}.
The final answer takes the form of
 \begin{align}
\label{api}
\Pi_1 (q) = - \,\beta_d \, e^2 \, \mu^{x_e}  \,
 \frac{  |{\boldsymbol Q}|^{ d - 1} }
{ f(q)} \,, \quad
f(q) = \begin{cases}
\sqrt {\upsilon\,( \upsilon-\kappa) } 
 \,\sqrt e_q\, \Theta(e_q)
& \text{ for } \upsilon \neq \kappa \\
 2\,|q_d| & \text{ for } \upsilon = \kappa \\
\end{cases} \,,
 \end{align}
where
\begin{align}
\label{betad}
\beta_d = \frac{  \Gamma^2 \big (\frac{d} {2} \big )}
{ 2^{d} \, \pi^{ \frac{d-1} {2} }\;
| \cos \big (  \frac{\pi \,d} {2} \big ) |  
\; \Gamma(\frac{d-1}{2}) \,\Gamma (d)} \,.
\end{align}

We note that since the bare boson propagator, $D_{(0)} (k) $, is independent of $\mathbf K$, the loop-integrals involving it are ill-defined, unless one resums a series of diagrams that provides a nontrivial dispersion along these frequency and momentum components. Hence, in all loop calculations involving bosonic propagators, we will include the lowest-order finite correction $ \Pi_1 (k) $ from the one-loop bosonic self-energy, which is proportional to $ |\mathbf K|^{d-1} /  f(k) $. Thereby we use the dressed bosonic propagator
\begin{align}
\label{eqbos1}
D_{(1)} (k) = \frac{1}
{ \left[ D_{(0)}(k) \right]^{-1}- \, \Pi_1 (k) } \,,
\end{align}
which is equivalent to rearranging the perturbative loop-expansions such that the one-loop finite part of the boson self-energy, dependent on $ \mathbf K $, is included at the zeroth order. Here,
$ \Pi_1 (k) $ represents the so-called {\textit{Landau-damping term}}, which leads to the signature $\text{sgn} (k_0) |k_0|^{2/3} $-dependence of the fermionic self-energies, characterizing the non-Fermi liquid behaviour in various quantum critical systems \cite{max-isn, max-sdw, denis, ips-lee, ips-uv-ir2, ips-fflo, ips-u1}. The Landau-damped part also plays the most significant role in inducing unconventional superconductivity in this kind of non-Fermi liquid systems \cite{ips2, ips3, ips-sc}.

\subsection{One-loop fermion self-energies}
\label{secferm}

There are two one-loop fermion self-energies to be computed, which are captured by the following:
\begin{align}
\label{eqsigma1}
\Sigma_1 (q) &= \frac{  \left (i\,e \, \mu^{\frac{x_e} {2} } \right )^2 }  {2} 
\int_k  G_2 (q-k)\, D_{(1)} (k) 
=   \frac{i \,e^2 \,  \mu^{x_e}} {2}  \int_k
\frac{1 } {k_d^2
+   \beta_d \, e^2 \, \mu^{x_e} \,
 \frac{  |{\mathbf K}|^{d- 1} } { f(k)} }
  \,\frac{\gamma_{d-1} \, \delta_{q-k}^{(2)} 
+ \mathbf  \Gamma \cdot (\mathbf  Q -\mathbf  K)}
{(\mathbf  Q -\mathbf  K)^2 + \left[\delta_{q-k}^{(2)} \right]^2
}  \, ,
\end{align}
and
\begin{align}
\label{eqsigma2}
\Sigma_2 (q) &= \frac{ \left (i\,e \, \mu^{\frac{x_e} {2} } \right )^2  }  {2} 
\int_k  G_1 (q-k)\, D_{(1)} (k) 
 = 
 \frac{i \,e^2 \,  \mu^{x_e}} {2}  \int_k
\frac{1 } {k_d^2
+   \beta_d \, e^2 \, \mu^{x_e}  \,
 \frac{  |{\mathbf K}|^{ d - 1} } { f(k)} } 
\,\frac{\gamma_{d-1} \, \delta_{q-k}^{(1)} 
+ \mathbf  \Gamma \cdot (\mathbf  Q -\mathbf  K)}
{(\mathbf  Q -\mathbf  K)^2 + \left[\delta_{q-k}^{(1)} \right]^2 }  \, ,
\end{align}
where 
\begin{align}
\delta_{q-k}^{(2)} & = -\, \upsilon \, k_{d-1}
+ \delta_q^{(2)} + \kappa \left( k_d^2 - 2  \, q_d \, k_d \right)
=  \upsilon
\left [ q_{d-1} +
\frac{  \kappa \, q_d^2   } 
{\upsilon} + \frac{ \kappa} {\upsilon} 
\left( k_d^2 - 2  \, q_d \, k_d \right) - \, k_{d-1}
\right ].
\end{align}
While the details are relegated to Appendix~\ref{appferm}, we show here the final results.
Setting $d = d_c-\epsilon$, we get the following expressions for the singular parts:
\begin{enumerate}

\item $ \upsilon \neq \kappa $:
\begin{align}
\Sigma_1 (q) & = -\, 
 \frac{  {\mathcal U}_1 \,e^{\frac{4} {3}} 
 }   { \epsilon}  \,
\frac{ \left [ 
\kappa  \, (2  \,\upsilon - \kappa)
\right]^{\frac{1} {6}} }
{\upsilon}
\, i\left( \mathbf{\Gamma} \cdot \boldsymbol Q \right)
+\mathcal{O}\big(\epsilon^0\big) \,,\quad
\Sigma_2 (q)  = -\, 
 \frac{  {\mathcal U}_1 \,e^{\frac{4} {3}} 
 }   { \upsilon^{2/3} \, \epsilon}
\, i\left( \mathbf{\Gamma} \cdot \boldsymbol Q \right)
+\mathcal{O}\big(\epsilon^0\big) \,,\quad
{\mathcal U}_1 = \frac {\left[
\Gamma\big (\frac {1} {4} \big) \,
\Gamma\big (\frac {5} {4} \big) \right ]^{1/3}
} 
{ 6 \times  {3}^{1/6} \, \pi^{4/3}} \,.
\end{align}

\item $ \upsilon = \kappa $:
\begin{align}
\Sigma_2 (q) =\upsilon \,\Sigma_1 (q)\,,\quad
\Sigma_1 (q)  = -\, 
 \frac{  {\mathcal U}_2 \,e^{\frac{4} {3}} 
 }   { \upsilon \, \epsilon}
\, i\left( \mathbf{\Gamma} \cdot \boldsymbol Q \right)
+\mathcal{O}\big(\epsilon^0\big) 
\,,\quad
{\mathcal U}_2 = \frac{2^{1/3}}
{  3^{7/6} } \,.
\end{align}

\end{enumerate}
Here, the logarithmic divergences (in the language of the Wilsonian language) are parametrized by poles at $  \epsilon =0 $.

\subsection{One-loop vertex-corrections}

In general, the one-loop fermion-boson vertex functions 
[$\Gamma_{12} (q, p)$ and $\Gamma_{21} (q,p)$] depend on two external frequency-momenta $q$ and $p$. 
In order to extract the leading $1/\epsilon$ 
divergence, however, it is sufficient to look at the $p \rightarrow 0$ limit, where we need to evaluate the simpler loop-integral, viz.
\begin{align}
\Gamma_{n_1 n_2} (q,0)& = \frac{ e^2 \,\mu^{x_e}} {2} 
\int_k   G_{n_1} (k)\, G_{n_2} (k) \,   D_{(1)} (k-q) 
\nn &= \frac{ e^2 \,\mu^{x_e}} {2} 
\int_k D_{(1)} (k-q)\, 
\frac{\delta_k^{ (n_1) } \,\delta_k^{(n_2)} +  \mathbf  K^2 
- \gamma_{d-1} \left( \mathbf  \Gamma \cdot \mathbf  K \right)
\left[ \delta_k^{ (n_1) } + \delta_k^{(n_2)} \right ]
}
{\left[ {\mathbf  K}^2 + \left \lbrace \delta_k^{(n_1)} \right \rbrace ^2 \right ]
\left[ {\mathbf  K}^2 + \left \lbrace \delta_k^{(n_2)} \right \rbrace ^2 \right ]
}\,.
\end{align}
Since $\Gamma_{12} (q,0) = \Gamma_{21} (q,0)$, we need to evaluate just the expression
\begin{align}
\label{eqgamint0}
\Gamma_{12} (q,0) &= \frac{ e^2 \,\mu^{x_e}} {2} 
\int_k   
\frac{\delta_k^{ (1) } \,\delta_k^{(2)} + \mathbf  K^2 
- \gamma_{d-1} \left( \mathbf  \Gamma \cdot \mathbf  K \right)
\left[ \delta_k^{ (1) } + \delta_k^{(2)} \right ]
}
{\left[ {\mathbf  K}^2 + \left \lbrace \delta_k^{(1)} \right \rbrace ^2 \right ]
\left[ {\mathbf  K}^2 + \left \lbrace \delta_k^{(2)} \right \rbrace ^2 \right ]
} \,.
\end{align}
Clearly, it vanishes trivially for $\upsilon = \kappa $. Hence, we focus on the case of $\upsilon \neq \kappa $. The details of the intermediate steps are shown in Eq.~\eqref{eqgamint}
of Appendix~\ref{appvertex}. The final answer turns out to be zero for any divergent contribution.

\section{RG flows for $\upsilon \neq  \kappa $}
\label{secrg}

In our QFT language, the action in Eq.~\eqref{eqs1} is referred to as the \textit{physical action}, defined at an energy scale $ \mu \sim \Lambda $, because it is supposed to consist of the fundamental Lagrangian with nondivergent quantities. However, the loop integrals lead to terms diverging lograithmically in $ \Lambda$, or with a positive power of $\Lambda$. In order to cure these ultraviolet divergences, we employ the renormalization procedure. Here, we have set up the grounds to apply dimensional regularization as the regularization method. In this formalism, the divergent terms are the singular terms arising in the $\epsilon \rightarrow 0$ limit, which we have evaluated at the one-loop order. In this section, we will use the minimal subtraction ($ {\rm MS}$) renormalization scheme to control the ultraviolet divergences \cite{thooft, weinberg}, which involves cancelling the divergent parts of the loop-contributions via the addition of the appropriate counterterms. 

The \textit{counterterm} action, designed to absorb the singular terms, follows the structure of the original action in Eq.~\eqref{eqs1}, such that
\begin{align}
\label{actcount}
S_{CT}  = &     \int_k \bar \Psi_1  (k) \,i
\left [ A_1\, \mathbf  \Gamma \cdot \mathbf  K  + 
\gamma_{d-1} \left( A_2 \, k_{d-1} + A_3\,k_d^2 \right)
 \right  ] \Psi_1 (k)  
 +  \int_k \bar \Psi_2  (k) \,i
\left [ A_4 \, \mathbf  \Gamma \cdot \mathbf  K  +  \gamma_{d-1} \left(
A_5 \, \upsilon \,k_{d-1} 
+ A_6 \, \kappa \, k_d^2  \right) \right  ] \Psi_2 (k)  
 \nn & + \frac{1}{2} \int_k A_7 \, k_d^2 \,  \phi (k) \, \phi(-k) 
- \left[ \frac{ i\, e \, \mu^{x_e/2} } {2} 
\int_{k} \int_q A_8 \, \phi (q) \,
 \bar{\Psi}_1 (k+q) \, {\Psi}_2(-k) 
+ \text{h.c.} \right ].
\end{align}
The coefficients appearing in the counterterms are given by the power series
\begin{align}
A_{\zeta} = 
\sum_{ n=1} ^\infty \frac{Z^{(n)}_{ \zeta}}
{\epsilon^n}  \text{  with }  
\zeta \in [1, 8 ]\,,
\end{align}
which must cancel the divergent pieces $\propto 1/\epsilon^n$, arising from the loop-level Feynman diagrams. Because of the existence of a $(d-1)$-dimensional rotational invariance in the space perpendicular to the Fermi surface, each term in $ {\mathbf  \Gamma} \cdot {\mathbf  K}$ is renormalized in the same way.

With the above ingredients, we can formally subtract off $S_{CT}$ from the so-called \textit{bare} action 
\begin{align}
\label{actren}
S_{\text{bare}}  = &  \sum_n \int_{k^B} {\bar{\Psi}}^B_n (k^B)
\, i \left[ 
\,{\mathbf  \Gamma} \cdot { \mathbf  K^B} 
+   \gamma_{d-1} \,\delta^{(n)}_{k^{B}} \right ] \Psi^B_n (k^B)
+ \frac{1} {2}\int_{ k^B} \left (k^B_d \right)^2
\phi^B (k^B) \,\, \phi^B (-k^B) 
 \nn & \,
- \left [\frac{ i\, e^B } {2} 
\int_{k^B} \int_{q^B} \phi^B (q^B) \,
 \bar \Psi^B_1  (k^B+q^B) \, \gamma_{d-1} \,  \Psi^B_2 (-k^B)  
+ \text{h.c.} \right ] ,
\end{align}
so that the \textit{physical} effective action ($ S $), by definition, consists of only well-behaved nondivergent quantum parameters. On the other hand, $S_{\text{bare}}$ consists of the \textit{bare quantities}, which can be divergent.
The superscript ``$B$'' has been used to denote the bare fields, couplings, frequency, and momenta. 
This prescription allows us to obtain the physical observables from the renormalized coupling constant(s), which are determined by the RG flow equations. The latter describe the evolution of the coupling constants as derivatives of the floating energy scale $\mu \,e^{- l} $ (alternatively, with respect to an increasing logarithmic length scale $l$).

The form of the RG flow equations are obtained from relating the bare quantities to the so-called renormalized quantities
(i.e., the ones without the superscript ``$B$''). We do so by defining first the multiplicative $Z_\zeta $-factors as follows:
\begin{align}
S_{\text{bare}}  = & S + S_{CT}\,, 
\quad Z_{\zeta}  =  1 + A_{\zeta}\,,
\end{align}
\begin{align}
&  {\mathbf  K}^B =    {\mathbf  K} \, , \quad
k_{d-1}^B =\frac{Z_2} {Z_1} \, k_{d-1}  \, , 
\quad  k^B_d  =  \sqrt{\frac{Z_3} {Z_1}} \,k_d \,, \quad
\Psi_n^B(k^B)  =   Z_{\Psi_n}^{1/2}\, \Psi_n(k)\,, 
\quad \phi_{\pm}^B(k^B) =  Z_{\phi}^{1/2}\, \phi_{\pm}\,,
\end{align}
and
\begin{align}
& Z_{\Psi_1}  = Z_ 1 \left( \frac  {Z_1} {Z_ 2} \right)
 \sqrt {\frac {Z_ 1} {Z_3}} \, ,\quad
Z_{\Psi_2}  =  Z_4 \left( \frac  {Z_1} {Z_ 2} \right)
 \sqrt {\frac {Z_ 1} {Z_3}} \, ,\quad
Z_{\phi}  = Z_7 \left( \frac  {Z_1} {Z_ 2} \right)
\left( \frac  {Z_1} {Z_ 3} \right)^{3/2} \,, \nn
& \upsilon^B =  \frac { Z_ 5} 
{  Z_ 4}   \left( \frac  {Z_1} {Z_ 2} \right) \upsilon \,,\quad
\kappa^B=  \frac { Z_ 6} {  Z_ 4 } 
\left( \frac  {Z_1} {Z_ 3} \right) \kappa\,,\quad
e^B=  Z_{e}\,e\,\mu^{\frac{\epsilon} {2} }\,, \quad
Z_{e}= 
\frac {  Z_ 8 \; \sqrt {\frac {Z_1} {Z_2}} }
 { \left ( \frac {Z_1} {Z_3}  \right)^{1/4}\, 
 \sqrt {Z_1 \, Z_ 4 \, Z_7}
 } \, .
\end{align}
Observing that there exists a freedom to change the renormalization of the fields and the renormalization of momenta
without affecting the action, we have exploited it by requiring $ \mathbf  K^B  = \mathbf  K $, which is equivalent to measuring the scaling dimensions of all the other quantities relative to the scaling dimension of $ \mathbf K $. In the end, we have succeesed to obtain $S$ as the renormalized action (also known as the Wilsonian effective action), which comprises renormalized nondivergent quantities.

\subsection{RG-flow equations from one-loop results}

At one-loop order, the divergent contributions lead to
\begin{align}
\label{eqZvals}
Z_1 & = 1- 
 \frac{  {\mathcal U}_1 \,e^{\frac{4} {3}} 
 }   { \epsilon}  \,
\frac{ \left [ 
\kappa  \, (2  \,\upsilon - \kappa)
\right]^{\frac{1} {6}} }
{\upsilon}\,, \quad
Z_4 = 1 -\, \frac{  {\mathcal U}_1 \,e^{\frac{4} {3}} 
 }   { \upsilon^{2/3} \, \epsilon}
\,,\nn
Z_2 &  = Z_3 =  Z_5=  Z_6 =Z_7 =Z_8 = 1 \,, \quad
{\mathcal U}_1 = \frac {\left[
\Gamma\big (\frac {1} {4} \big) \,
\Gamma\big (\frac {5} {4} \big) \right ]^{1/3}
} 
{ 6 \times   {3}^{1/6} \, \pi^{4/3}} \,. 
\end{align}

In general, we need to define two dynamical critical exponents for the fermions, defined by
\begin{align}
& z = 1 +
\frac{\partial \ln \big ( \frac{Z_1} {Z_2} \big ) }  
{\partial \ln \mu}\,, \quad 
\tilde z = 1 +
\frac{\partial \ln \big ( \frac{Z_1} {Z_3} \big ) } 
{\partial \ln \mu}\,.
\end{align}
However, at the one-loop level, the result $ Z_2 = Z_3 =1$ forces them to be equal. Hence, we just set $\tilde z  = z$
for the leading-order loop-level correction.
By definition, the anomalous dimensions for the fermions and the bosons are given by
\begin{align}
\eta_{\psi_n}  = \frac{1} {2} 
\frac{\partial \ln Z_{\psi_n} }  
{\partial \ln \mu}  \text{ and }
\eta_\phi  = \frac{1} {2} \frac{\partial \ln Z_\phi }  
{\partial \ln \mu} \,,
\end{align}
respectively.
Lastly, we have the beta functions for the three coupling constants as follows:
\begin{align}
\beta_e  =  \frac{ d  e }  
{ d\ln \mu}\,,  \quad 
\beta_\upsilon  =  \frac{ d  \upsilon }  
{ d \ln \mu} \,, \quad
\beta_\kappa  =  \frac{ d  \kappa }  
{ d \ln \mu} \,.
\end{align}

The prime reason for the introduction of the \textit{ad hoc} mass scale $\mu $ is to regularize the theory, thus eliminating the ultraviolet divergences emerging from the integrals of the loop-level Feynman diagrams. However, since physical observables, in the end, must be independent of $\mu $, as $\mu$ is no way a parameter of the fundamental theory, the bare parameters in $S_{\rm bare}$ must not dependent on it either. Invoking this constraint, in conjunction with the requirement that the regular (i.e., nondivergent) parts of the final expressions should have the expansions as
\begin{align}
\label{eqexp}
& z = z^{(0)} \,, \quad
\eta_{\psi_n} = \eta_{\psi_n}^{ (0)}  + \eta_{\psi_n}^{ (1)} \,\epsilon\,, \quad
\eta_\phi = \eta_\phi^{ (0)}  + \eta_\phi^{ (1)} \,\epsilon\,, \nn
& \beta_e = \beta _e^{ (0)}  + \beta _e^{ (1)} \,\epsilon\,, \quad
\beta_\upsilon = \beta_\upsilon^{ (0)}  + \beta_\upsilon^{ (1)} \,\epsilon\,, \quad
\beta_\kappa = \beta_\upsilon^{ (0)}  + \beta_\kappa ^{ (1)} \,\epsilon \,,
\end{align}
in the limit $\epsilon  \rightarrow 0 $, the differential equations governing the RG-flow equations are obtained.
We need to carry out the following steps sequentially:
(1) impose the condition that $\frac{ d }{d \ln \mu} \,(\mbox{bare quantity}) =0$;
(2) plug in the expressions shown in Eqs.~\eqref{eqZvals} and \eqref{eqexp}; 
(3) expand each equation in powers of $\epsilon$;
and (4) match the coefficients of the regular powers of $\epsilon $ on both sides of the resulting equations
to determine the expressions for all the quantities shown in Eq.~\eqref{eqexp}.
This elaborate exercise leads to
\begin{align}
& \beta_\upsilon^{(1)} = \beta_\kappa^{(1)} =\eta_{\psi_n}^{ (1)}
= \eta_\phi^{ (1)} = 0\,,\quad
 \beta_e^{(1)} = -\frac{e} {2} \,,\quad
z  = 1 +
 \beta_e^{(1)} \, \frac{  \partial Z_1^{(1)} }{ \partial e} \, , \nn
& \beta_e^{(0)} = - \frac{e} {4} \left[ 
3 \left( z-1\right) + e \left(  
\frac{\partial Z_1^{(1)} } { \partial e}
+ \frac{ \partial Z_4^{(1)} }  { \partial e}
 \right) \right ],\quad
 \beta_\upsilon^{(0)} =\upsilon \left( 1-z
 + \beta_e^{(1)} \, \frac{ \partial Z_4^{(1)} }  { \partial e}
 \right),\nn
& \beta_\kappa^{(0)} = \kappa \left( 1-z
 + \beta_e^{(1)} \, \frac{ \partial Z_4^{(1)} }  { \partial e}
 \right) ,\quad
\eta_{\psi_1}^{ (0)} =\frac{ 3 \left( z-1\right)
+ 2\,\beta_e^{(1)} \, \frac{  \partial Z_1^{(1)} }{ \partial e}
}  {4} \,,\nn
& \eta_{\psi_2}^{ (0)} =\frac{ 3 \left( z-1\right)
+ 2\,\beta_e^{(1)} \, \frac{  \partial Z_4^{(1)} }{ \partial e}
}  {4} \,,\quad 
\eta_{\phi}^{ (0)} =\frac{ 5\left(z-1\right)
} {4}\,.
\end{align}
The solutions are readily found to be
\begin{align}
\label{eqbeta}
& z = 1 +  \frac {2 \; \mathcal{U}_1 \,
\tilde {e} \, \kappa^{1/6}  \, (2 \, \upsilon - \kappa) ^{1/6}
} 
{3 \, {\upsilon} }\,, \quad
\eta_{\psi_1} = \frac {5 \; \mathcal{U}_1 \,
\tilde {e} \, \kappa^{1/6}  \, (2 \, \upsilon - \kappa) ^{1/6}
} 
{ 6 \, {\upsilon} } \,, \quad
\eta_{\psi_2} =  \frac {
  \mathcal{U}_1 \,\tilde {e}  
\left[ 2\, {\upsilon}^{1/3}  + 
3\, \kappa^{1/6}  \, (2 \, \upsilon - \kappa) ^{1/6} 
 \right ]
} {6 \, {\upsilon} } \,, \nn
\eta_{\phi} & = \frac {5 \; \mathcal{U}_1 \,
\tilde {e} \, \kappa^{1/6}  \, (2 \, \upsilon - \kappa) ^{1/6}
} 
{ 6 \, {\upsilon} } \,, 
 \quad \frac{ \beta_e} {e}
 =  
 \frac { \mathcal{U}_1 \,\tilde {e}\,
 \left[ 2\, {\upsilon}^{1/3}
- \kappa^{1/6}  \, (2 \, \upsilon - \kappa) ^{1/6} 
 \right ]  }
  {6 \, \upsilon } -\frac{\epsilon} {2} \,,\nn
 \beta_\upsilon &= \frac {2 \;
\mathcal{U}_1 \,\tilde {e} 
\left[ {\upsilon}^{1/3}  - \kappa^{1/6}  \, (2 \, \upsilon - \kappa) ^{1/6} 
 \right ]
} {3}  \,,\quad
  \beta_\kappa = \frac {2 \;
  \mathcal{U}_1 \,\tilde {e} \,\kappa 
\left[ {\upsilon}^{1/3}  - \kappa^{1/6}  \, (2 \, \upsilon - \kappa) ^{1/6} 
 \right ]
} {3 \, {\upsilon} }   \, ,
 \end{align}
 where
\begin{align}
\tilde e = e^{4/3} \,.
\end{align}
Since we are interested in the behaviour at the IR energy scales, we determine the RG flows with respect to the logarithmic length scale $l$, which are given by the derivatives
\begin{align}
\frac{ d e}{ d l} \equiv - \,\beta_e\,,\quad
\frac{ d \upsilon} {d l} \equiv - \, \beta_\upsilon \,,\quad
\frac{ d \kappa} {d l} \equiv - \,\beta_\kappa \,.
\end{align}

\begin{figure}[t]
\centering
\includegraphics[width=0.3 \textwidth]{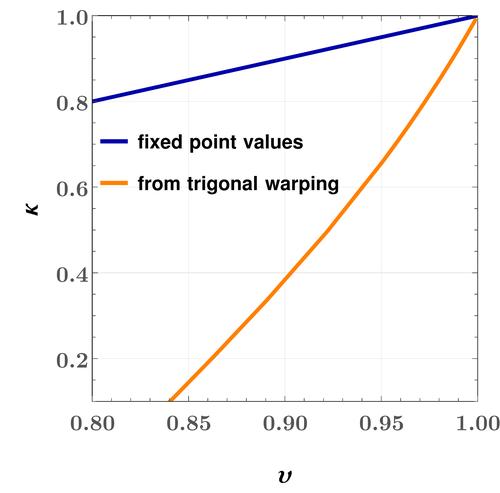}
\caption{\label{figsolns}
The relation between $\upsilon$ and $\kappa$, obtained from the trigonal-warping relation (orange curve) and fixed-point equation $ \kappa = \upsilon$ (blue curve). The orange curve has been obtained as a parameteric plot obeying the relations $\upsilon = \frac {\rho - 1} {\rho + 1}$ and $\kappa = \frac {(\rho - 10) \, (\rho + 1)^2} {(\rho - 1)^2  (\rho + 10)}$, by varying the warping parameter $\rho $ in the range $[10, 1500] $.
}
\end{figure}

\subsection{Fixed points}
 
 Let us denote the coupling constants at the fixed point by the superscript ``$*$''.
To obtain the non-Gaussian fixed point (i.e., where $\tilde e \neq 0 $), the last two expressions of Eq.~\eqref{eqbeta} tell us that we must have 
\begin{align}
{\upsilon^*}^{1/3}  = (\kappa^*)^{1/6}  \, (2 \, \upsilon^* - \kappa^*) ^{1/6} 
\Rightarrow \kappa^* = \upsilon^* \,.
\end{align} 
Plugging this value into $\beta_e  = 0,$ we get
\begin{align}
 \tilde e^*  = \frac {3  \, (\upsilon^*)^{2/3} \, \epsilon} 
 { \mathcal U_1  }  \,.
\end{align}
Basically, it is a fixed line in the space spanned by $ \lbrace e, \,\upsilon, \,  \kappa \rbrace$, rather than a fixed point, because of the relation $ \kappa^* = \upsilon^*$.

For a trigonally-warped Fermi surface, for a given value of $\upsilon$, the possible value of $\kappa $ can be obtained from $\epsilon_{\varsigma ,{\mathbf{p}} , + } $ [cf. Eq.~\eqref{twd}] by scanning the space of the warping parameter. Since the Fermi-surface curves are given by $ \left [ 1 + \frac {\varsigma} { \rho} \cos\big (3\theta _p\big) \right ] \times \text{constant} $, we have the relations $\upsilon = \frac {\rho - 1} {\rho + 1}$ and $\kappa = \frac {(\rho - 10) \, (\rho + 1)^2} {(\rho - 1)^2  (\rho + 10)}$. It is apparent that a flat patch (with $\kappa = 0 $) arises for  $\rho = 10 $. This indicates that, for $\upsilon <9/11 =0.818182  $, each of the two Fermi surfaces becomes concave at the hot-spots labelled with the superscript ``$(2)$'', thus featuring a negative value of $\kappa $. Since our analysis does not hold for concave patches of Fermi surfaces, we are restricted to focus on $\upsilon \gtrsim 0.82  $. Let us now compare the starting value of $\kappa $ connected with a given value of $\upsilon $, coming from the noninteracting Fermi-surface equation, with that obtained from the solutions at the fixed points. We find that the fixed-point value of $\kappa $ is always greater than the $\kappa$-value given by the trigonal-warping relation (see Fig.~\ref{figsolns} for a pictorial representation). Noting that $\kappa = 0 $ indicates a flat patch of the Fermi surface, increasing the value of $\kappa $ (compared to the value obtained from the trigonal-warping relation) gives rise to a higher degree of curving of the Fermi-surface patch around the hot-spots labelled with the subscripts ``$(2)$''.

In a generic situation, for a given $\upsilon$, we consider a small deviation from the fixed-point values parameterized by $\lbrace \delta \tilde e, \, \delta \kappa \rbrace$. We plug this in into the expressions for $\frac{d \tilde e}{dl}$ (which is the negative of the beta function for $\tilde e$, readily obtained from $\frac{d e}{dl}$) and $ \frac{d \kappa} {dl}$, and expand up to linear order in the deviation parameters. From these equations, we construct the stability matrix $\mathcal M$, from the coefficients of $\lbrace \delta \tilde e, \, \delta \kappa \rbrace $ of the two linearized equations. The eigenvalues of $\mathcal M $ contain the information about the stability of the concerned fixed point (as indicated in its nomenclature). For the non-Gaussian fixed points, the eigenvalue of $\mathcal M $ along the $\tilde e$-direction is always negative, while it is zero along the $\kappa $-direction. This implies that, while we have a stable fixed point for the flows projected along the $\tilde e$-axis, the stability is of a neutral nature for the flows projected along the $\kappa $-axis. The same behaviour is observed for the stability matrix of $\lbrace \delta \tilde e, \, \delta \upsilon\rbrace$, when we hold $\kappa $ fixed.

\section{RG flows for $\kappa = \upsilon $}
\label{secrg2}

For the special case of $\kappa = \upsilon $,
the counterterm action should be written as
\begin{align}
\label{actcount2}
S_{CT}  = &     \int_k \bar \Psi_1  (k) \,i
\left [ A_1\, \mathbf  \Gamma \cdot \mathbf  K  + 
\gamma_{d-1} \left( A_2 \, k_{d-1} + A_3\,k_d^2 \right)
 \right  ] \Psi_1 (k)  
 +  \int_k \bar \Psi_2  (k) \,i
\left [ A_4 \, \mathbf  \Gamma \cdot \mathbf  K  +  
\gamma_{d-1}\, A_5 \, \upsilon \left( k_{d-1} 
 + k_d^2  \right) \right  ] \Psi_2 (k)  
 \nn & + \frac{1}{2} \int_k A_7 \, k_d^2 \,  \phi (k) \, \phi(-k) 
- \left[ \frac{ i\, e \, \mu^{x_e/2} } {2} 
\int_{k} \int_q A_8 \, \phi (q) \,
 \bar{\Psi}_1 (k+q) \, {\Psi}_2(-k) 
+ \text{h.c.} \right ].
\end{align}

\begin{align}
\label{eqZvals}
Z_1 & = 1- 
 \frac{  {\mathcal U}_2 \,e^{\frac{4} {3}} 
 }   { \upsilon \, \epsilon}  \,, \quad
Z_4 =  1- 
 \frac{  {\mathcal U}_2 \,e^{\frac{4} {3}} 
 }   { \epsilon} \,,\quad
Z_2  = Z_3 = Z_5 =  Z_7 =Z_8 = 1 \,, \quad
{\mathcal U}_2 = \frac{2^{1/3}}
{  3^{7/6} } \,. 
\end{align}
The solutions are readily found to be
\begin{align}
& z = 1 +  \frac {2 \; \mathcal{U}_2\,
\tilde {e} } 
{3 \, {\upsilon} }\,, \quad
\eta_{\psi_1} = \frac {5 \; \mathcal{U}_2 \,
\tilde {e} } 
{ 6 \, {\upsilon} } \,, \quad
\eta_{\psi_2} =  \frac {
  \mathcal{U}_2 \,\tilde {e} \,(3+2\,\upsilon)  } 
  {6 \, {\upsilon} } \,, \nn
\eta_{\phi} & = \frac {5 \; \mathcal{U}_2 \,\tilde {e} } 
{ 6 \, {\upsilon} } \,, 
 \quad \frac{ \beta_e} {e}
 =  
 \frac { \mathcal{U}_2 \,\tilde {e}\,
 \left ( 2\, {\upsilon} -1 \right )  }
  {6 \, \upsilon } -\frac{\epsilon} {2} \,,\quad
 \beta_\upsilon = \frac {2 \;
\mathcal{U}_2 \,\tilde {e} \,(\upsilon-1)
} {3}    \, .
 \end{align}
The fixed points are given by 
\begin{align}
{\upsilon^*}  = 1 \,,\quad
 \tilde e^*  = \frac {3  \, \epsilon} 
 { \mathcal U_2 }  \,.
\end{align}
In this case, the stability matrix is given by
$\text{diag}\left( -\epsilon/2 , \, -2\, \epsilon \right) + \mathcal{O}(\epsilon^2)$. Since it has both the eigenvalues negative, it is a stable fixed point in the $\tilde e \,\upsilon$-plane. Since $\kappa = \upsilon =1 $ holds for a circular patch on either Fermi surface, the coupling with the CDW boson pushes the hot-spots with label ``$(2)$'' to have the same nature as hot-spots with label ``$(1)$''.

From the generic case of $\kappa \neq \upsilon$, we have seen that the fixed points are characterized by $\kappa^* = \upsilon^* $. Hence, we are then forced to look at the $\kappa = \upsilon$ case under RG flow, even when we start with $\kappa \neq \upsilon$. In the end, all the hot-spots tend to have the circular-patch nature, which exists in the absence of trigonal warping.

\section{Summary and concluding remarks}
\label{secsum}

In this paper, we have examined a putative quantum critical point at the onset of superradiance, in a set-up designed to engineer cavity QED. Considering a honeycomb lattice near half-filling, featuring doped Dirac cones, we have identified CDW wavevectors connecting hot-spots located on the emerging Fermi surfaces. The results have been dictated by taking into account the trigonal warping of the Fermi surfaces. With these ingredients, we have set upon identifying NFL phases, predicted via RPA calculations \cite{peng}. A controlled approximation, using the tools of dimensional regularization and RG-flow equations, have helped us conclude that these must be stable NFL phases in the IR-energy limit, when we project the flow lines along the $e$-direction.

The engineering of cavity QED allows for enhanced tunability of electronic properties and control over unwanted effects, thus promising to be a versatile platform for observing strongly-correlated many-body effects via light-matter interactions. As a consequence, our investigations are of considerable relevance, on the merit of establishing the possibilities of realizing NFL phases in such cavity QED systems. In the future, it will be worthwhile to study the pairing instabilities in these settings, in line with the studies carried out in Refs.~\cite{ips-sc, ips-qbt-sc}.

\section*{Acknowledgments}
We are grateful to Sung-Sik Lee, Carsten Timm, and Rafael M. Fernandes for providing key insights. We thank Francesco Piazza and Peng Rao for useful discussions.

\appendix

\section{Useful integrals}
\label{appint}

In this appendix, we list some integration formulas which are useful for performing the loop calculations shown in the main text.

\begin{align}
 \int_{-\infty}^{\infty} 
 \frac{dx} {2\,\pi} \,
 \frac{1}
{\left [\left( x+a_1 \right)^2 + A^2 \right ] 
 \left  [\left( x + a_2 \right)^2 + B^2 \right ]} 
&= \frac{|A|+|B|}
{2 \,|A| \,|B| \left [\left( a_1 - a_2 \right)^2 +(|A|+|B|)^2 \right ]
}\,,  \\
\int_{-\infty}^{\infty} 
 \frac{dx} {2\,\pi} \,
\frac{\left( x+a_1 \right) \,\left( x + a_2 \right) }
{ \left [\left( x+a_1 \right)^2 + A^2 \right ]
\left  [\left( x + a_2 \right)^2 + B^2 \right ]} 
&=
\frac{|A|+|B|}
{2 \left [\left( a_1 - a_2 \right)^2 +(|A|+|B|)^2 \right ] } \,,   \\
 \int_{-\infty}^{\infty}  \frac{dx} {2\,\pi} \,
\frac{\left( x+a_1 \right) }
{\left [\left( x+a_1 \right)^2 + A^2 \right ] 
\left  [\left( x + a_2 \right)^2 + B^2 \right ]} &=
\frac{ a_1- a_2  }
{2 \,|B| \left [(a_1- a_2)^2 +(|A|+|B|)^2 \right ]}\,, \\
 \int_{-\infty}^{\infty} 
 \frac{dx} {2\,\pi} \,\frac{\left( x + a_2 \right) }
{ \left [\left( x+a_1 \right)^2 + A^2 \right ] 
\left  [\left( x + a_2 \right)^2 + B^2 \right ]} &=
\frac{ a_2 -a_1 }
{2 \,|A| \left  [\left( a_1 - a_2 \right)^2 +(|A|+|B|)^2 \right ]} \, ,\\
\int_0^{\infty} \frac{dx} {2\,\pi} \,
 \frac{x^m} {x^3 + A}  = 
 \frac{1}  { 6 \, \left | \sin \big( (m+1) \,\pi/3 \big) \right | 
 A^{\frac{2-m}{3}} } & \text{ for } 
0< m+1 < 3 \,.
\label{equseint}
\end{align}

The Feynman parametrization is given by
 \begin{align}
 \label{feynm}
\frac{1}{A^{\alpha} \, B^{\beta}}= 
\frac{\Gamma (\alpha +\beta)} {\Gamma (\alpha)  \, \Gamma (\beta)}
\int_0^1 \frac{x^{\alpha-1}\,(1-x)^{\beta-1}\,dx}
{\left[ x \,A +(1-x) \,B\right]^{\alpha+\beta}} \,.
 \end{align}

\vspace{10 mm}
The surface area or, more accurately, the $(N-1)$-dimensional volume of an $(N-1)$-sphere (i.e., $S^{N-1}$), forming the boundary of the $ N $-ball of radius unity, is given by
\begin{align}
\label{angint}
S^{N-1}=\int  d \Omega_N = \frac{ 2 \,\pi^{N/2} } { \Gamma{(N/2)}} \, .
\end{align}

\section{One-loop boson self-energy}
\label{apponeloopbos}

In this appendix, we explain how we arrive at the final expressions for the one-loop bosonic self-energy, starting with Eq.~\eqref{eqpi0}.

For the case of $\upsilon = \kappa $, we first evaluate the integral $I_1 (d, {\boldsymbol Q})$ [cf. Eq.~\eqref{eqi1}]. The ($d-1$)-dimensional integral in $I_1 (d,{\boldsymbol Q})$ can be computed 
using the Feynman parametrization shown in Eq.~\eqref{feynm}.
Plugging in $\alpha=\beta=1/2$, 
$A= | \mathbf  K + {\boldsymbol Q}|^2 $, and $B={\mathbf  K}^2 $, and using $\int_0^1 \frac{dx}
{\sqrt{x \, (1-x)}} = \frac{1}{\pi}$, we get
 \begin{align}
\label{eqi1app} 
I_1 (d, {\boldsymbol Q}) = \frac{1}
{ \pi \, (2 \,\pi)^{d-1}  } 
\int_0^1 \frac{dx\ }{\sqrt{x \, (1-x)}}\,
\int {d \mathbf  K} \,
\left\{
\frac{
\upsilon\, \mathbf  K \cdot (\mathbf  K + {\boldsymbol Q})}
{   x \, |\mathbf  K  + {\boldsymbol Q}|^2 +  (1-x)\, \mathbf  K^2 } 
+ 1 \right\}. 
 \end{align}
Introducing the new variable $\mathbf  U = \mathbf  K + x \, {\boldsymbol Q} \,,$ we get $\, x \, |\mathbf  K  + {\boldsymbol Q}|^2 +  
(1-x) \mathbf  K^2 \ = \mathbf  U ^2 + x \, (1-x) \,{\boldsymbol Q}^2 \,,$ thus leading to
\begin{align}
I_1 (d, {\boldsymbol Q}) &= \frac{1}
{\pi \, (2 \,\pi)^{d-1}  } 
\int_0^1 \frac{dx }{\sqrt{x \, (1-x)}}\,
\int {d^{d-1} \mathbf  U} \,
\left \{ \upsilon\,
\frac{   \mathbf  U^2 + 
 (1-2\,x) \, \mathbf  U \cdot {\boldsymbol Q} -x\, (1-x){\boldsymbol Q}^2}
{   \mathbf  U ^2 + x  \,(1-x)  \, {\boldsymbol Q}^2}
+ 1 \right\}  \nn
&= \frac{1}
{\pi \, (2 \,\pi)^{d-1}  } 
\int_0^1 \frac{dx }{\sqrt{x \, (1-x)}}\,
\int {d^{d-1} \mathbf  U} \,
\frac{   (1+\upsilon) \,\mathbf  U^2 + (1-\upsilon) \,x\, (1-x){\boldsymbol Q}^2}
{   \mathbf  U ^2 + x  \,(1-x)  \, {\boldsymbol Q}^2}
\nn & = \int_0^1 dx\,
\frac { 2^{2-d} \, \pi^{\frac {d-1} {2} }
\,\upsilon   \,
|\bq|^{d -1}
\, \left [ x\,(1 - x) \right]^{ \frac{d-2} {2} }
|\sec \big (\frac {\pi \,  d} {2} \big)| }
 { \Gamma \big (\frac {d-1} {2} \big )}
\nn & = 
\frac { 2^{2-d} \, \pi^{\frac {d-1} {2} }
\,\upsilon   \,
|\bq|^{d -1}
\,|\sec \big (\frac {\pi \,  d} {2} \big)|
\;\Gamma^2(d/2) }
 { \Gamma \big (\frac {d-1} {2} \big )
 \,\Gamma[d]}\,.
\end{align}

For the case of $ \upsilon \neq \kappa $, we change variables to $u = k_d^2 $ [in Eq.~\eqref{eqpi0}], which gives the Jacobian factor as $ 1/\left( 2 \,\sqrt {u}  \right) = 1/\left( 2 \,|k_d| \right) $.
From the denominator of the second factor in the integrand, we find that it forces the dominant contribution to the integral to come from $ u \sim  e_q $, in the regime $|\boldsymbol Q| \ll \kappa \,q_d^2 $, where
\begin{align}
e_q = \frac {\kappa \,  q_d^2}
 {\upsilon  -  {\kappa}} + \,q_ {d-1} \, .
\end{align}
We have assumed $e_q$ to be a positive quantity remembering that the typical energy scales impose the constraint of $q_d \gg q_{d-1}$. Using the above, we approximate $ |k_d| $ by $\sqrt {\upsilon\,e_q /\left(\upsilon -\kappa \right) }$ in the Jacobian. This leads to
\begin{align}
\Pi_1 (q)  &  \simeq \frac{e^2 \, \mu^{x_e} 
}  {2} 
\int  \frac{  \, d\mathbf  K} {(2 \, \pi)^d} 
\int_0^\infty du \,
\frac{
\sqrt{\upsilon-\kappa} \,
 \left(  {\mathbf  K}
+ \frac{|\mathbf  K + {\boldsymbol Q}|} {\upsilon}  \right) 
\; \left[ \,\mathbf  K \cdot (\mathbf  K +{\boldsymbol Q}) 
+ {\mathbf  K}\; \frac{ |\mathbf  K +{\boldsymbol Q}|} {\upsilon} \,  \right]  
}
{\sqrt {\upsilon\,e_q} \,
  {\mathbf  K}\;  |\mathbf  K +{\boldsymbol Q}|\,
\left[ 
\left ( \frac {\upsilon- \kappa} {\upsilon} \right)^2
\left(u - \frac {\upsilon \, e_q}
{  \upsilon - \kappa  } \right)^2 +  \left ( {\mathbf  K}
+\frac{ |\mathbf  K +{\boldsymbol Q}|} {\upsilon} \right)^2
\right ] }    \nn
&  \simeq \frac{e^2 \, \mu^{x_e} 
}  {2} 
\int  \frac{  \, d\mathbf  K} {(2 \, \pi)^d} 
\int_{-\infty}^\infty du \,
\frac{
\sqrt{\upsilon-\kappa} \,
 \left(  {\mathbf  K}
+ \frac{|\mathbf  K + {\boldsymbol Q}|} {\upsilon}  \right) 
\; \left[ \,\mathbf  K \cdot (\mathbf  K +{\boldsymbol Q}) 
+ {\mathbf  K}\; \frac{ |\mathbf  K +{\boldsymbol Q}|} {\upsilon} \,  \right]  
}
{\sqrt {\upsilon\,e_q} \;
{\mathbf  K}\; |\mathbf  K +{\boldsymbol Q}| \,
\left[ 
\left ( \frac {\upsilon- \kappa} {\upsilon} \right)^2
u^2 +  \left ( {\mathbf  K}
+\frac{ |\mathbf  K +{\boldsymbol Q}|} {\upsilon} \right)^2
\right ] }    \nn
& = \frac{e^2 \, \mu^{x_e} } 
{ 4 \, \sqrt e_q  \,\upsilon^{3/2} \, \sqrt{\upsilon-\kappa }} 
 \; I_1 (d, {\boldsymbol Q}) \, .
\end{align}

Finally, gathering the above expressions gives us
\begin{align}
\label{api_app}
\Pi_1 (q) & = - \,\beta_d \, e^2 \, \mu^{x_e}  \,
 \frac{  |{\boldsymbol Q}|^{ d - 1} }
{ f(q)} \,, \quad
f(q) = \begin{cases}
\sqrt {\upsilon\,( \upsilon-\kappa) } 
 \,\sqrt e_q\, \Theta(e_q)
& \text{ for } \upsilon \neq \kappa \\
 2\,|q_d| & \text{ for } \upsilon = \kappa \\
\end{cases} \,,
\nn & \text{where }
\beta_d   = \frac{  \Gamma^2 \big (\frac{d} {2} \big )}
{ 2^{d} \, \pi^{ \frac{d-1} {2} }\;
| \cos \big (  \frac{\pi \,d} {2} \big ) |  
\; \Gamma(\frac{d-1}{2}) \,\Gamma (d)} \,.
\end{align}

\section{One-loop fermion self-energies}
\label{appferm}

In this appendix, we compute the one-loop fermion self energies, which are given by Eqs.~\eqref{eqsigma1} and \eqref{eqsigma2}, shown in Sec.~\ref{secferm}.

\subsection{Computing $\Sigma_1 $ for $\upsilon \neq \kappa $}

In this subsection, we outline the evaluation of Eq.~\eqref{eqsigma1} for $\upsilon \neq \kappa $. We divide the full expression into two parts, viz. $\Sigma_{1,1} (q)$  and $\Sigma_{1,2} (q)$, which are proportional to $\mathbf \Gamma $ and $\gamma_{d-1} $, respectively.   


For extracting the divergent term proportional to $ \boldsymbol{\Gamma} \cdot \boldsymbol Q $, in order to tackle the integrals, we set the external momentum components $q_d $ and $q_{d-1}$ to zero. Hence, remembering that $e_k =  k_{d-1} 
+ \frac { \kappa \,  k_d^2} {\upsilon  - \kappa }$, we need to deal with
\begin{align}
\delta_{-k}^{(2)} =  { \kappa} \,
 k_d^2  - \, {\upsilon}\, k_{d-1} 
=    { \kappa}  \, k_d^2  
+ \frac {  \upsilon \,\kappa \,  k_d^2}
 { \upsilon -  {\kappa}  }
- \,{\upsilon}\, e_k 
=  \upsilon \, B
 \left(    k_d^2 -\frac{ e_k} {B} \right)  ,
\text{ where }
B = \frac {\kappa \left (2 - \frac {\kappa} {\upsilon} \right)} 
{ \upsilon -  {\kappa} }\,.
\end{align}
Changing variables from $ \lbrace k_{d-1}, k_d^2 \rbrace $ to $ \lbrace e_k, k_d^2 \rbrace $, we obtain the form
\begin{align}
\Sigma_{1,1} (\boldsymbol Q, 0,0 ) & =  
\frac{i \,e^2 \,  \mu^{x_e}} {2}  \int
\frac{{ d^{d-1}\mathbf K}} {(2\, \pi)^{d+1}}
\int_{-\infty}^\infty de_k \int_0^\infty dk_d \,
\frac{2} {k_d^2
+  \frac{\beta_d \, e^2 \, \mu^{x_e} \,\Theta(e_k)}
{\sqrt {\upsilon\,( \upsilon-\kappa) }}
 \frac{  |{\mathbf K}+ \boldsymbol Q |^{ d - 1} } 
 { \sqrt{ e_k}  }
}
  \,\frac{
 \mathbf  \Gamma \cdot \mathbf  K }
{ \mathbf  K^2 + \upsilon^2 \,B^2
\left ( k_d^2 - \frac{  e_k} {B} \right )^2
}  \, .
\end{align}

Defining $ u = k_d^2  $, the above simplifies to
\begin{align}
\Sigma_{1,1} (\boldsymbol Q, 0,0 )  & \simeq  \frac{i \,e^2 \,  \mu^{x_e}}
 { 2}   \int
\frac{{ d^{d-1}\mathbf K}} {(2\, \pi)^{d+1}}
\int_{e_k>0} \frac{ de_k} {\sqrt {k_{d}}}  \int_0^\infty d u \,
\frac{1 } {
u +    \frac{\beta_d \, e^2 \, \mu^{x_e} }
{\sqrt {\upsilon\,( \upsilon-\kappa) }} \,
 \frac{  |{\mathbf K +\boldsymbol Q}|^{ d - 1} } 
 { \sqrt{ e_k}  }
}
  \,\frac{
 \mathbf  \Gamma \cdot \mathbf  K }
{ {\mathbf  K}^2 + \upsilon^2 \, B^2 \left ( u- \frac{  e_k} {B}  \right)^2 }  \nn
& \simeq  \frac{i \,e^2 \,  \mu^{x_e} }
 { 2 }  
\int
\frac{{ d^{d-1}\mathbf K}} {(2\, \pi)^{d+1}}
\int_{e_k>0} \frac{ de_k}   {\sqrt {e_k/B} }  
\int_{-\infty}^\infty d u \,
\frac{1 } 
{\frac{ e_k} {B^2}  +   \frac{\beta_d \, e^2 \, \mu^{x_e} }
{\sqrt {\upsilon\,( \upsilon-\kappa) }}  
\, \frac{  |{\mathbf K +\boldsymbol Q}|^{ d - 1} } 
 { \sqrt{ e_k}  }
}
  \,\frac{ \mathbf  \Gamma \cdot \mathbf  K }
{ {\mathbf  K}^2 + \upsilon^2 \, B^2 \, u^2}  \nn
& =  \frac{i \,e^2 \,  \mu^{x_e}\, \sqrt B}
 {  2 \,\upsilon  }  
\int \frac{{ d^{d-1}\mathbf K}} {(2\, \pi)^{d}}
\int_{e_k>0}  de_k  \, \frac{\mathbf  \Gamma \cdot \mathbf  K}
 { 2 \, |\mathbf K |
 \left[  e_k^{3/2} +   \frac{\beta_d \, e^2 \, \mu^{x_e} }
{\sqrt {\upsilon\,( \upsilon-\kappa) }} 
 \, B \, |{\mathbf K +\boldsymbol Q}|^{ d - 1}  \right ]} \nn
&  =  \frac{i \,e^2 \,  \mu^{x_e}\, B^{\frac{1}{6}}
}
 {  2 \times 3\,\sqrt  3\,\upsilon 
 \left [   \frac{\beta_d \, e^2 \, \mu^{x_e} }
{\sqrt {\upsilon\,( \upsilon-\kappa) }}  \right ]^{\frac{1}{3}}
 }  
\int
\frac{{ d^{d-1}\mathbf K}} {(2\, \pi)^{d-1}}
 \, \frac{ \mathbf  \Gamma \cdot \mathbf  K}
 { |\mathbf K |
  \; |{\mathbf K +\boldsymbol Q}|^{ \frac{d - 1} {3}}  } \,.
\end{align}
In the second line, observing that the denominator of the second term restricts the dominant contribution of $u$ to $e_k/B $, we have approximated $u$ in the denominator of the first term by $e_k/B $ and $d k_d $ as $ \frac{du}   {2 \, \sqrt { e_k/B}} $. All these steps render the integrals amenable to a final analytical expression. Using Eq.~\eqref{feynm} by setting $\alpha = (d-1)/{6}$ and $\beta = 1/2$,
we get
\begin{align}
\Sigma_{1,1} (\boldsymbol Q, 0,0 ) 
&  =  \frac{i \,e^{\frac{4}{3}} \,  \mu^{ \frac{2\,x_e} {3} }\, B^{\frac{1}{6}}
}
 { 2 \times 3\,\sqrt  3 \;\upsilon 
\,\beta_d^{\frac{1}{3}}
 }  
\int_0^1 dx\,
\frac{\Gamma \big( \frac{d+2}{6}\big) \;
x^{ - \frac{1}{2}} \left( 1-x \right)^{\frac{d-7}{6}}
} 
 { \sqrt \pi \; \Gamma \big( \frac{d-1}{6}\big)}
\int \frac{{ d^{d-1}\mathbf K}} {(2\, \pi)^{d-1}}
\frac{\mathbf  \Gamma \cdot \mathbf  K }
{   
\left[ x \, |\mathbf  K  + {\boldsymbol Q}|^2 +  (1-x) \,\mathbf  K^2
\right ]^{\frac{d+2}{6}}
 } \nn
&  = \frac{i \,e^{\frac{4}{3}} \,  \mu^{ \frac{2\,x_e} {3} }\, B^{\frac{1}{6}}
}
 { 2 \times 3\,\sqrt  3 \;\upsilon 
\,\beta_d^{\frac{1}{3}}
 }  
\int_0^1 dx\,
\frac{\Gamma \big( \frac{d+2}{6}\big) \;
x^{\frac{d-7}{6}} \left( 1-x \right)^{ - \frac{1}{2}} 
} 
 { \sqrt \pi \; \Gamma \big( \frac{d-1}{6}\big)}
\int \frac{{ d^{d-1}\mathbf U }
} {(2\, \pi)^{d-1}}
\frac{\mathbf  \Gamma \cdot \left( \mathbf  U -x \,\boldsymbol Q \right ) }
{   
\left[  \mathbf  U^2 + x \,  (1-x) \, | {\boldsymbol Q}|^2
\right ]^{\frac{d+2}{6}}
 } \nn
\nn & =
- \,i\left( \mathbf  \Gamma \cdot \boldsymbol Q \right)
\frac{ e^{\frac{4} {3}}  
\left [ 
\kappa  \, (2  \,\upsilon - \kappa)
\right]^{\frac{1} {6}} }
{\upsilon}
\frac { 
\Gamma\big (\frac {x_e} {3}  \big) \;
\Gamma\big (\frac {d} {2} \big) \;
\Gamma\big (\frac {d + 2} {6} \big)}
 { 3 \,\sqrt {3}  \times
 2^{\frac {2 \left( d + 2 \right) } {3}  } \, \pi^{\frac {d + 1} {2}}
 \; \Gamma\big (\frac {5 \, d -  2} {6} \big)\,
\beta_d^{\frac{1}{3}} } 
  \left(  \frac{\mu} { |\boldsymbol Q| } \right)^{\frac{2\, x_e} {3} } \,.
\end{align}

For extracting any divergent term proportional to $ \gamma_{d-1} $, we proceed by setting $\boldsymbol Q = \mathbf 0$, to evaluate
\begin{align}
\Sigma_{1,2} (\mathbf  0, q_{d-1}, q_d ) & =
\frac{i \,e^2 \,  \mu^{x_e}} {2}  \int_k
\frac{1 }    {k_d^2
+  2\,\beta_d \, e^2 \, \mu^{x_e} \,\Theta(e_k)
 \frac{  |{\mathbf K}|^{ d - 1} }  {\sqrt e_k}       }
  \, \frac{\gamma_{d-1} \, \delta_{q-k}^{(2)} 
}
{ {\mathbf  K}^2 + \left[\delta_{q-k}^{(2)} \right]^2
} \,.
\end{align}
In order to make progress, we need some more simplifications, which we implement by setting $q_d$ to zero. Hence,
\begin{align}
\Sigma_{1,2} (\mathbf  0, q_{d-1}, 0) 
& =  \frac{i \,e^2 \,  \mu^{x_e}} {2}  \int
\frac{{ d^{d-1}\mathbf K}} {(2\, \pi)^{d+1}}
\int_{e_k>0} de_k \int_0^\infty dk_d \,
\frac{2} {k_d^2
+  \frac{\beta_d \, e^2 \, \mu^{x_e} }
{\sqrt{\upsilon\,(\upsilon + \kappa )}}\,
 \frac{  |{\mathbf K}|^{ d - 1} } 
 { \sqrt{ e_k}  }
} \,\frac{
 \gamma_{d-1} \,\upsilon \,B
\left ( k_d^2 - \frac{  e_k-q_{d-1}} {B} \right )  }
{ \mathbf  K^2 + \upsilon^2 \,B^2
\left ( k_d^2 - \frac{  e_k-q_{d-1}} {B} \right )^2
} \nn
& \simeq  \frac{i \,e^2 \,  \mu^{x_e}} {2}  \int
\frac{{ d^{d-1}\mathbf K}} {(2\, \pi)^{d+1}}
\int_{e_k>0} \frac{ de_k} {\frac{  e_k-q_{d-1}} {B} }
 \int_{-\infty}^\infty du \,
\frac{1} { \frac{  e_k-q_{d-1}} {B} 
+  \frac{\beta_d \, e^2 \, \mu^{x_e} }
{\sqrt{\upsilon\,(\upsilon + \kappa )}}\,
 \frac{  |{\mathbf K}|^{ d - 1} } 
 { \sqrt{ e_k}  }
} \,\frac{
 \gamma_{d-1} \,\upsilon \,B
\left ( u - \frac{  e_k-q_{d-1}} {B} \right )  }
{ \mathbf  K^2 + \upsilon^2 \,B^2
\left ( u - \frac{  e_k-q_{d-1}} {B} \right )^2
} \nn
& = 0\,.
\end{align}

\subsection{Computing $\Sigma_2 $ for $\upsilon \neq \kappa $}

In this subsection, we outline the evaluation of Eq.~\eqref{eqsigma2} for $\upsilon \neq \kappa $. For extracting the divergent term proportional to $ \mathbf \Gamma \cdot \boldsymbol Q $, in order to tackle the integrals, we set the external momentum components $q_d $ and $q_{d-1}$ to zero, such that we need to deal with
\begin{align}
\delta_{-k}^{(1)} = k_d^2  - \, k_{d-1} 
=      k_d^2  
+ \frac { \kappa \,  k_d^2}
 {\upsilon- {\kappa}    }
- \, e_k 
=   \tilde B  \, k_d^2 -\,e_k  \,, \text{ where }
\tilde B = \frac {\upsilon} {\upsilon - \kappa} \,.
\end{align}
Proceeding in the same way as for the evaluation of Eq.~\eqref{eqsigma1}, we get the final expression to be
\begin{align}
\Sigma_{2} (q)
 & =
- \,i\left( \mathbf \Gamma \cdot \boldsymbol Q \right)
e^{\frac{4} {3}} \, \upsilon^{\frac{1}{3}}
\left [ 
\frac {\upsilon} {\upsilon - \kappa}\right]^{\frac{1} {6}}
\frac { 
\Gamma\big (\frac {x_e} {3}  \big) \;
\Gamma\big (\frac {d} {2} \big) \;
\Gamma\big (\frac {d + 2} {6} \big)}
 {3 \,\sqrt {3}  \times
 2^{\frac {2\, (d + 4)} {3}  } \, \pi^{\frac {d + 1} {2}}
 \; \Gamma\big (\frac {5 \, d -  2} {6} \big)\,
\beta_d^{\frac{1}{3}} } 
  \left(  \frac{\mu} { |\boldsymbol Q| } \right)^{\frac{2\, x_e} {3} } ,
\end{align}
with the $\gamma_{d-1}$-part being nonsingular.

Setting $d = d_c-\epsilon$, we get the singular part to be
\begin{align}
\Sigma_2 (q) & = -\, 
 \frac{  {\mathcal U}_1 \,e^{\frac{4} {3}} 
 }   { \upsilon^{2/3} \, \epsilon}
\, i\left( \mathbf{\Gamma} \cdot \boldsymbol Q \right)
+\mathcal{O}\big(\epsilon^0\big) \,,
\end{align}
where the logarithmic divergence (in the language of the Wilsonian language) is parametrized by a pole at $  \epsilon =0 $.
 
\subsection{Computing $\Sigma_1 $ and $\Sigma_2 $ for $\upsilon =\kappa $}

In this subsection, we outline the evaluation of Eqs.~\eqref{eqsigma1} and \eqref{eqsigma2} for the special case of $ \kappa =\upsilon $. Here, we have to deal with
\begin{align}
\delta_{q-k}^{(2)} =  {\upsilon} \left[ 
 \left(k_d+q_d \right)^2  - \, k_{d-1}+q_{d-1} \right ]
 =
-\, {\upsilon}  \left[  k_{d-1} -q_{d-1}
- \left(k_d+q_d \right)^2  \right ].
\end{align}
Plugging it in, we get
\begin{align}
\label{eqsigma11}
\Sigma_1 (q) &= \frac{  \left (i\,e \, \mu^{\frac{x_e} {2} } \right )^2 }  {2} 
\int_k  G_2 (q-k)\, D_{(1)} (k) 
= \frac{i \,e^2 \,  \mu^{x_e}} {2}  \int_k
\frac{1 } 
{ k_d^2
+   \beta_d \, e^2 \, \mu^{x_e} \,
 \frac{  |{\mathbf K}|^{d- 1} } { 2\,|k_d|} 
 }
  \,\frac{
  \mathbf  \Gamma \cdot (\mathbf  Q + \mathbf  K)
  - \gamma_{d-1} \,\upsilon  \left[  k_{d-1} -q_{d-1}
- \left(k_d+q_d \right)^2  \right ] 
}
{(\mathbf  Q + \mathbf  K)^2 + 
\upsilon^2  \left[  k_{d-1} -q_{d-1}
- \left(k_d+q_d \right)^2  \right ]^2
} \nn
& = \frac{i \,e^2 \,  \mu^{x_e}} {2}  \int_k
\frac{1 } 
{ k_d^2
+   \beta_d \, e^2 \, \mu^{x_e} \,
 \frac{  |{\mathbf K}|^{d- 1} } { 2\,|k_d|} 
 }
  \,\frac{
  \mathbf  \Gamma \cdot (\mathbf  Q + \mathbf  K)
  - \gamma_{d-1} \,\upsilon \,  k_{d-1}
}
{ (\mathbf  Q + \mathbf  K)^2 + \upsilon^2  \, k_{d-1}^2
} \nn
& =  \frac{i \,e^2 \,  \mu^{x_e}} {2 \,\upsilon}  \int
\frac{{ d^{d-1}\mathbf K}} {(2\, \pi)^{d}}
\int_{0}^\infty dk_d \,
\frac{ k_d } 
{ k_d^3
+   \beta_d \, e^2 \, \mu^{x_e} \,\frac{  |{\mathbf K}|^{d- 1} } { 2} 
 }
  \,\frac{
  \mathbf  \Gamma \cdot (\mathbf  Q + \mathbf  K) }
{ | \mathbf  Q + \mathbf  K| }\nn
& =  \frac{i \,e^{4/3} \,  \mu^{\frac{ 2\, x_e} {3}}}
 {\upsilon \,\beta_d^{1/3} }  \int
\frac{{ d^{d-1}\mathbf K}} {(2\, \pi)^{d-1}}
  \,\frac{
  \mathbf  \Gamma \cdot (\mathbf  Q + \mathbf  K) }
{ |\mathbf  Q + \mathbf  K| \,|\mathbf  K|^{\frac{d-1} {3}} }  \nn
& = - \,i\left( \Gamma \cdot \boldsymbol Q \right)
\frac{ e^{\frac{4} {3}}  }
{\upsilon}
\frac { 
\Gamma\big (\frac {x_e} {3}  \big) \;
\Gamma\big (\frac {d} {2} \big) \;
\Gamma\big (\frac {d + 2} {6} \big)}
 { 3 \,\sqrt {3}  \times
 2^{\frac {2 \, d + 3 } {3}  } \, \pi^{\frac {d + 1} {2}}
 \; \Gamma\big (\frac {5 \, d -  2} {6} \big)\,
\beta_d^{\frac{1}{3}} } 
  \left(  \frac{\mu} { |\boldsymbol Q| } \right)^{\frac{2\, x_e} {3} }  \,,
\end{align}
and
\begin{align}
\Sigma_2 (q) =\upsilon \,\Sigma_1 (q)\,.
\end{align}
Setting $d = d_c-\epsilon$, we get the singular parts to be captured by
\begin{align}
\Sigma_1 (q) & = -\, 
 \frac{  {\mathcal U}_2 \,e^{\frac{4} {3}} 
 }   { \upsilon \, \epsilon}
\, i\left( \mathbf{\Gamma} \cdot \boldsymbol Q \right)
+\mathcal{O}\big(\epsilon^0\big) 
\,,\quad
{\mathcal U}_2 = \frac{2^{1/3}}
{  3^{7/6} } \,,
\end{align}
where the logarithmic divergence (in the language of the Wilsonian language) is parametrized by a pole at $  \epsilon =0 $.

\section{One-loop vertex-corrections}
\label{appvertex}

In this appendix, we elaborate on the details of performing the integral, shown in  Eq.~\eqref{eqgamint0}, for the nontrivial case of $\upsilon \neq \kappa $.
In order to be able to extract the divergent part, we take the $q_d =0 $ limit, which gives rise to an integral of the form
\begin{align}
\label{eqgamint}
\Gamma_{12} (q,0) \vert_{q_d =0 }
&= \frac{ e^2 \,\mu^{x_e}} {2}  \int_k 
\frac{1 } 
{ k_d^2
+  \frac{\beta_d \, e^2 \, \mu^{x_e} \,\Theta(e_k)}
{\sqrt{\upsilon \,(\upsilon-\kappa)}}
 \frac{  |{\mathbf K}|^{ d - 1} } {\sqrt e_k}
 }
\frac{\delta_{k+q}^{ (1) } \,\delta_{k+q}^{(2)}+ {\mathbf  K}^2 
- \gamma_{d-1} \left( \mathbf  \Gamma \cdot \mathbf  K \right)
\left[ \delta_{k+q}^{ (1) } + \delta_{k+q}^{(2)} \right ]
}
{\left[ {\mathbf  K}^2 + \left \lbrace \delta_{k+q}^{(1)} \right \rbrace ^2 \right ]
\left[ {\mathbf  K}^2 + \left \lbrace \delta_{k+q}^{(2)} \right \rbrace ^2 \right ]
} 
\nn &= \frac{ e^2 \,\mu^{x_e} } {2} 
 \int_k D_{(1)} (k)
\frac{
\upsilon\,B_1\,B_2 
\left ( k_d^2- \frac {e_k} {B_1} \right ) \left ( k_d^2- \frac {e_k} {B_2} \right )
 +  {\mathbf  K}^2 
- \gamma_{d-1} \left( \mathbf  \Gamma \cdot \mathbf  K \right)
\left[ \upsilon\,B_1
\left ( k_d^2- \frac {e_k} {B_1} \right )
 + B_2 \left ( k_d^2- \frac {e_k} {B_2} \right )  \right ]
}
{ \left[ {\mathbf  K}^2 + B_1^2 
\left ( k_d^2- \frac {e_k} {B_1} \right ) ^2 \right ]
\left[ {\mathbf  K}^2 + \upsilon^2 \,B_2^2 \left ( k_d^2- \frac{e_k} {B_2} \right )^2 \right ]
} \nn
&= \frac{ e^2 \,\mu^{x_e} } {2} 
\frac{{ d^{d-1}\mathbf K}} {(2\, \pi)^{d+1}}
\int_{e_k>0}  de_k   
\int_{0}^\infty \frac{d u} { \sqrt u } \,
\frac{1 } 
{ u +  \frac{\beta_d \, e^2 \, \mu^{x_e} }
{\sqrt{\upsilon \,(\upsilon-\kappa)}} \,
 \frac{  |{\mathbf K}|^{ d - 1} } {\sqrt e_k} } \,
\nn & \hspace{ 3.5 cm}  \times
\frac{
\upsilon\,B_1\,B_2 
\left ( k_d^2- \frac {e_k} {B_1} \right ) \left ( k_d^2- \frac {e_k} {B_2} \right )
+ {\mathbf  K}^2 
- \gamma_{d-1} \left( \mathbf  \Gamma \cdot \mathbf  K \right)
\left[ \upsilon\,B_1
\left ( k_d^2- \frac {e_k} {B_1} \right )
 + B_2 \left ( k_d^2- \frac {e_k} {B_2} \right )  \right ]
}
{ \left[ {\mathbf  K}^2 + B_1^2 
\left ( u- \frac {e_k} {B_1} \right ) ^2 \right ]
\left[ {\mathbf  K}^2 + \upsilon^2 \,B_2^2 \left ( u - \frac{e_k} {B_2} \right )^2 \right ]
} \nn
&= \frac{ e^2 \,\mu^{x_e}  } {2} 
 \int \frac{{ d^{d-1}\mathbf K}} {(2\, \pi)^{d+1}}
\int_{e_k>0}  de_k   
\int_{-\frac{e_k} {B_2}}^\infty \frac{d u} 
{ \sqrt {u+\frac{e_k} {B_2}} } \,
\frac{1 } 
{ u + \frac{\beta_d \, e^2 \, \mu^{x_e} }
{\sqrt{\upsilon \,(\upsilon-\kappa)}} \,
 \frac{  |{\mathbf K}|^{ d - 1} } {\sqrt e_k} } \,
\nn & \hspace{ 4 cm}  \times
\frac{
\upsilon\,B_1\,B_2 
\left (u- \frac {e_k} {B_1} +\frac {e_k} {B_2} \right ) \,u
+ {\mathbf  K}^2 
- \gamma_{d-1} \left( \mathbf  \Gamma \cdot \mathbf  K \right)
\left[ \upsilon\,B_1
\left ( u- \frac {e_k} {B_1} +\frac {e_k} {B_2}\right )
 + B_2\, u  \right ]
}
{ \left[ {\mathbf  K}^2 + B_1^2 
\left ( u- \frac {e_k} {B_1}+\frac{e_k} {B_2} \right ) ^2 \right ]
\left[ {\mathbf  K}^2 + \upsilon^2 \,B_2^2 \, u^2 \right ]
} \nn
& \simeq  \frac{ e^2 \,\mu^{x_e}} {2}   \int
\frac{{ d^{d-1}\mathbf K}} {(2\, \pi)^{d+1}}
\int_{e_k>0}  \frac{ de_k} { \sqrt {\frac{e_k} {B_1}} }  
\, \frac{1 } 
{ \frac{e_k} {B_1} +  \frac{\beta_d \, e^2 \, \mu^{x_e} }
{\sqrt{\upsilon \,(\upsilon-\kappa)}} \,
 \frac{  |{\mathbf K}|^{ d - 1} } {\sqrt e_k} } \,
\nn & \hspace{ 3.5 cm}  \times
\int_{-\infty}^\infty {d u} \,
\frac{
\upsilon\,B_1\,B_2 
\left (u- \frac {e_k} {B_1} +\frac {e_k} {B_2} \right ) \,u
+ {\mathbf  K}^2 
- \gamma_{d-1} \left( \mathbf  \Gamma \cdot \mathbf  K \right)
\left[ \upsilon\,B_1
\left ( u- \frac {e_k} {B_1} +\frac {e_k} {B_2}\right )
 + B_2\, u  \right ]
}
{ \left[ {\mathbf  K}^2 + B_1^2 
\left ( u- \frac {e_k} {B_1}+\frac{e_k} {B_2} \right ) ^2 \right ]
\left[ {\mathbf  K}^2 + \upsilon^2 \,B_2^2 \, u^2 \right ]
} \nn & = 0 \,.
\end{align}
This is because the integral over $u$ vanishes identically.

\bibliography{ref}

\begin{thebibliography}{97}%
\makeatletter
\providecommand \@ifxundefined [1]{%
 \@ifx{#1\undefined}
}%
\providecommand \@ifnum [1]{%
 \ifnum #1\expandafter \@firstoftwo
 \else \expandafter \@secondoftwo
 \fi
}%
\providecommand \@ifx [1]{%
 \ifx #1\expandafter \@firstoftwo
 \else \expandafter \@secondoftwo
 \fi
}%
\providecommand \natexlab [1]{#1}%
\providecommand \enquote  [1]{``#1''}%
\providecommand \bibnamefont  [1]{#1}%
\providecommand \bibfnamefont [1]{#1}%
\providecommand \citenamefont [1]{#1}%
\providecommand \href@noop [0]{\@secondoftwo}%
\providecommand \href [0]{\begingroup \@sanitize@url \@href}%
\providecommand \@href[1]{\@@startlink{#1}\@@href}%
\providecommand \@@href[1]{\endgroup#1\@@endlink}%
\providecommand \@sanitize@url [0]{\catcode `\\12\catcode `\$12\catcode
  `\&12\catcode `\#12\catcode `\^12\catcode `\_12\catcode `\%12\relax}%
\providecommand \@@startlink[1]{}%
\providecommand \@@endlink[0]{}%
\providecommand \url  [0]{\begingroup\@sanitize@url \@url }%
\providecommand \@url [1]{\endgroup\@href {#1}{\urlprefix }}%
\providecommand \urlprefix  [0]{URL }%
\providecommand \Eprint [0]{\href }%
\providecommand \doibase [0]{https://doi.org/}%
\providecommand \selectlanguage [0]{\@gobble}%
\providecommand \bibinfo  [0]{\@secondoftwo}%
\providecommand \bibfield  [0]{\@secondoftwo}%
\providecommand \translation [1]{[#1]}%
\providecommand \BibitemOpen [0]{}%
\providecommand \bibitemStop [0]{}%
\providecommand \bibitemNoStop [0]{.\EOS\space}%
\providecommand \EOS [0]{\spacefactor3000\relax}%
\providecommand \BibitemShut  [1]{\csname bibitem#1\endcsname}%
\let\auto@bib@innerbib\@empty
\bibitem [{\citenamefont {Holstein}\ \emph {et~al.}(1973)\citenamefont
  {Holstein}, \citenamefont {Norton},\ and\ \citenamefont {Pincus}}]{holstein}%
  \BibitemOpen
  \bibfield  {author} {\bibinfo {author} {\bibfnamefont {T.}~\bibnamefont
  {Holstein}}, \bibinfo {author} {\bibfnamefont {R.~E.}\ \bibnamefont
  {Norton}},\ and\ \bibinfo {author} {\bibfnamefont {P.}~\bibnamefont
  {Pincus}},\ }\bibfield  {title} {\bibinfo {title} {de {H}aas-van {A}lphen
  effect and the specific heat of an electron gas},\ }\href
  {https://doi.org/10.1103/PhysRevB.8.2649} {\bibfield  {journal} {\bibinfo
  {journal} {Phys. Rev. B}\ }\textbf {\bibinfo {volume} {8}},\ \bibinfo {pages}
  {2649} (\bibinfo {year} {1973})}\BibitemShut {NoStop}%
\bibitem [{\citenamefont {Baskaran}\ and\ \citenamefont
  {Anderson}(1988)}]{baskaran}%
  \BibitemOpen
  \bibfield  {author} {\bibinfo {author} {\bibfnamefont {G.}~\bibnamefont
  {Baskaran}}\ and\ \bibinfo {author} {\bibfnamefont {P.~W.}\ \bibnamefont
  {Anderson}},\ }\bibfield  {title} {\bibinfo {title} {Gauge theory of
  high-temperature superconductors and strongly correlated {F}ermi systems},\
  }\href {https://doi.org/10.1103/PhysRevB.37.580} {\bibfield  {journal}
  {\bibinfo  {journal} {Phys. Rev. B}\ }\textbf {\bibinfo {volume} {37}},\
  \bibinfo {pages} {580} (\bibinfo {year} {1988})}\BibitemShut {NoStop}%
\bibitem [{\citenamefont {Ioffe}\ and\ \citenamefont {Larkin}(1989)}]{larkin}%
  \BibitemOpen
  \bibfield  {author} {\bibinfo {author} {\bibfnamefont {L.~B.}\ \bibnamefont
  {Ioffe}}\ and\ \bibinfo {author} {\bibfnamefont {A.~I.}\ \bibnamefont
  {Larkin}},\ }\bibfield  {title} {\bibinfo {title} {Gapless fermions and gauge
  fields in dielectrics},\ }\href {https://doi.org/10.1103/PhysRevB.39.8988}
  {\bibfield  {journal} {\bibinfo  {journal} {Phys. Rev. B}\ }\textbf {\bibinfo
  {volume} {39}},\ \bibinfo {pages} {8988} (\bibinfo {year}
  {1989})}\BibitemShut {NoStop}%
\bibitem [{\citenamefont {Lee}(1989)}]{PhysRevLett.63.680}%
  \BibitemOpen
  \bibfield  {author} {\bibinfo {author} {\bibfnamefont {P.~A.}\ \bibnamefont
  {Lee}},\ }\bibfield  {title} {\bibinfo {title} {{Gauge field, Aharonov-Bohm
  flux, and high-T$_{c}$ superconductivity}},\ }\href
  {https://doi.org/10.1103/PhysRevLett.63.680} {\bibfield  {journal} {\bibinfo
  {journal} {Phys. Rev. Lett.}\ }\textbf {\bibinfo {volume} {63}},\ \bibinfo
  {pages} {680} (\bibinfo {year} {1989})}\BibitemShut {NoStop}%
\bibitem [{\citenamefont {Lee}\ and\ \citenamefont {Nagaosa}(1992)}]{leenag}%
  \BibitemOpen
  \bibfield  {author} {\bibinfo {author} {\bibfnamefont {P.~A.}\ \bibnamefont
  {Lee}}\ and\ \bibinfo {author} {\bibfnamefont {N.}~\bibnamefont {Nagaosa}},\
  }\bibfield  {title} {\bibinfo {title} {Gauge theory of the normal state of
  high-{T}$_{\mathit{c}}$ superconductors},\ }\href
  {https://doi.org/10.1103/PhysRevB.46.5621} {\bibfield  {journal} {\bibinfo
  {journal} {Phys. Rev. B}\ }\textbf {\bibinfo {volume} {46}},\ \bibinfo
  {pages} {5621} (\bibinfo {year} {1992})}\BibitemShut {NoStop}%
\bibitem [{\citenamefont {Blok}\ and\ \citenamefont {Monien}(1993)}]{blok}%
  \BibitemOpen
  \bibfield  {author} {\bibinfo {author} {\bibfnamefont {B.}~\bibnamefont
  {Blok}}\ and\ \bibinfo {author} {\bibfnamefont {H.}~\bibnamefont {Monien}},\
  }\bibfield  {title} {\bibinfo {title} {{Gauge theories of high-T$_c$
  superconductors}},\ }\href {https://doi.org/10.1103/PhysRevB.47.3454}
  {\bibfield  {journal} {\bibinfo  {journal} {Phys. Rev. B}\ }\textbf {\bibinfo
  {volume} {47}},\ \bibinfo {pages} {3454} (\bibinfo {year}
  {1993})}\BibitemShut {NoStop}%
\bibitem [{\citenamefont {Ubbens}\ and\ \citenamefont {Lee}(1994)}]{ubbens}%
  \BibitemOpen
  \bibfield  {author} {\bibinfo {author} {\bibfnamefont {M.~U.}\ \bibnamefont
  {Ubbens}}\ and\ \bibinfo {author} {\bibfnamefont {P.~A.}\ \bibnamefont
  {Lee}},\ }\bibfield  {title} {\bibinfo {title} {Superconductivity phase
  diagram in the gauge-field description of the t-{J} model},\ }\href
  {https://doi.org/10.1103/PhysRevB.49.6853} {\bibfield  {journal} {\bibinfo
  {journal} {Phys. Rev. B}\ }\textbf {\bibinfo {volume} {49}},\ \bibinfo
  {pages} {6853} (\bibinfo {year} {1994})}\BibitemShut {NoStop}%
\bibitem [{\citenamefont {{Nayak}}\ and\ \citenamefont
  {{Wilczek}}(1994{\natexlab{a}})}]{nayak1}%
  \BibitemOpen
  \bibfield  {author} {\bibinfo {author} {\bibfnamefont {C.}~\bibnamefont
  {{Nayak}}}\ and\ \bibinfo {author} {\bibfnamefont {F.}~\bibnamefont
  {{Wilczek}}},\ }\bibfield  {title} {\bibinfo {title} {{Non-Fermi liquid fixed
  point in 2+1 dimensions}},\ }\href
  {https://doi.org/10.1016/0550-3213(94)90477-4} {\bibfield  {journal}
  {\bibinfo  {journal} {Nuclear Physics B}\ }\textbf {\bibinfo {volume}
  {417}},\ \bibinfo {pages} {359} (\bibinfo {year}
  {1994}{\natexlab{a}})}\BibitemShut {NoStop}%
\bibitem [{\citenamefont {{Chakravarty}}\ \emph {et~al.}(1995)\citenamefont
  {{Chakravarty}}, \citenamefont {{Norton}},\ and\ \citenamefont
  {{Sylju{\aa}sen}}}]{sudip}%
  \BibitemOpen
  \bibfield  {author} {\bibinfo {author} {\bibfnamefont {S.}~\bibnamefont
  {{Chakravarty}}}, \bibinfo {author} {\bibfnamefont {R.~E.}\ \bibnamefont
  {{Norton}}},\ and\ \bibinfo {author} {\bibfnamefont {O.~F.}\ \bibnamefont
  {{Sylju{\aa}sen}}},\ }\bibfield  {title} {\bibinfo {title} {Transverse gauge
  interactions and the vanquished {F}ermi liquid},\ }\href
  {https://doi.org/10.1103/PhysRevLett.74.1423} {\bibfield  {journal} {\bibinfo
   {journal} {Physical Review Letters}\ }\textbf {\bibinfo {volume} {74}},\
  \bibinfo {pages} {1423} (\bibinfo {year} {1995})}\BibitemShut {NoStop}%
\bibitem [{\citenamefont {Metlitski}\ and\ \citenamefont
  {Sachdev}(2010{\natexlab{a}})}]{max-isn}%
  \BibitemOpen
  \bibfield  {author} {\bibinfo {author} {\bibfnamefont {M.~A.}\ \bibnamefont
  {Metlitski}}\ and\ \bibinfo {author} {\bibfnamefont {S.}~\bibnamefont
  {Sachdev}},\ }\bibfield  {title} {\bibinfo {title} {{Quantum phase
  transitions of metals in two spatial dimensions. I. Ising-nematic order}},\
  }\href {https://doi.org/10.1103/PhysRevB.82.075127} {\bibfield  {journal}
  {\bibinfo  {journal} {Phys. Rev. B}\ }\textbf {\bibinfo {volume} {82}},\
  \bibinfo {pages} {075127} (\bibinfo {year} {2010}{\natexlab{a}})}\BibitemShut
  {NoStop}%
\bibitem [{\citenamefont {Oganesyan}\ \emph {et~al.}(2001)\citenamefont
  {Oganesyan}, \citenamefont {Kivelson},\ and\ \citenamefont
  {Fradkin}}]{ogankivfr}%
  \BibitemOpen
  \bibfield  {author} {\bibinfo {author} {\bibfnamefont {V.}~\bibnamefont
  {Oganesyan}}, \bibinfo {author} {\bibfnamefont {S.~A.}\ \bibnamefont
  {Kivelson}},\ and\ \bibinfo {author} {\bibfnamefont {E.}~\bibnamefont
  {Fradkin}},\ }\bibfield  {title} {\bibinfo {title} {Quantum theory of a
  nematic {F}ermi fluid},\ }\href {https://doi.org/10.1103/PhysRevB.64.195109}
  {\bibfield  {journal} {\bibinfo  {journal} {Phys. Rev. B}\ }\textbf {\bibinfo
  {volume} {64}},\ \bibinfo {pages} {195109} (\bibinfo {year}
  {2001})}\BibitemShut {NoStop}%
\bibitem [{\citenamefont {Metzner}\ \emph {et~al.}(2003)\citenamefont
  {Metzner}, \citenamefont {Rohe},\ and\ \citenamefont
  {Andergassen}}]{metzner}%
  \BibitemOpen
  \bibfield  {author} {\bibinfo {author} {\bibfnamefont {W.}~\bibnamefont
  {Metzner}}, \bibinfo {author} {\bibfnamefont {D.}~\bibnamefont {Rohe}},\ and\
  \bibinfo {author} {\bibfnamefont {S.}~\bibnamefont {Andergassen}},\
  }\bibfield  {title} {\bibinfo {title} {Soft {F}ermi surfaces and breakdown of
  {F}ermi-liquid behavior},\ }\href
  {https://doi.org/10.1103/PhysRevLett.91.066402} {\bibfield  {journal}
  {\bibinfo  {journal} {Phys. Rev. Lett.}\ }\textbf {\bibinfo {volume} {91}},\
  \bibinfo {pages} {066402} (\bibinfo {year} {2003})}\BibitemShut {NoStop}%
\bibitem [{\citenamefont {Dell'Anna}\ and\ \citenamefont
  {Metzner}(2006)}]{delanna}%
  \BibitemOpen
  \bibfield  {author} {\bibinfo {author} {\bibfnamefont {L.}~\bibnamefont
  {Dell'Anna}}\ and\ \bibinfo {author} {\bibfnamefont {W.}~\bibnamefont
  {Metzner}},\ }\bibfield  {title} {\bibinfo {title} {Fermi surface
  fluctuations and single electron excitations near {P}omeranchuk instability
  in two dimensions},\ }\href {https://doi.org/10.1103/PhysRevB.73.045127}
  {\bibfield  {journal} {\bibinfo  {journal} {Phys. Rev. B}\ }\textbf {\bibinfo
  {volume} {73}},\ \bibinfo {pages} {045127} (\bibinfo {year}
  {2006})}\BibitemShut {NoStop}%
\bibitem [{\citenamefont {Kee}\ \emph {et~al.}(2003)\citenamefont {Kee},
  \citenamefont {Kim},\ and\ \citenamefont {Chung}}]{kee}%
  \BibitemOpen
  \bibfield  {author} {\bibinfo {author} {\bibfnamefont {H.-Y.}\ \bibnamefont
  {Kee}}, \bibinfo {author} {\bibfnamefont {E.~H.}\ \bibnamefont {Kim}},\ and\
  \bibinfo {author} {\bibfnamefont {C.-H.}\ \bibnamefont {Chung}},\ }\bibfield
  {title} {\bibinfo {title} {Signatures of an electronic nematic phase at the
  isotropic-nematic phase transition},\ }\href
  {https://doi.org/10.1103/PhysRevB.68.245109} {\bibfield  {journal} {\bibinfo
  {journal} {Phys. Rev. B}\ }\textbf {\bibinfo {volume} {68}},\ \bibinfo
  {pages} {245109} (\bibinfo {year} {2003})}\BibitemShut {NoStop}%
\bibitem [{\citenamefont {{Lawler}}\ \emph {et~al.}(2006)\citenamefont
  {{Lawler}}, \citenamefont {{Barci}}, \citenamefont {{Fern{\'a}ndez}},
  \citenamefont {{Fradkin}},\ and\ \citenamefont {{Oxman}}}]{lawler1}%
  \BibitemOpen
  \bibfield  {author} {\bibinfo {author} {\bibfnamefont {M.~J.}\ \bibnamefont
  {{Lawler}}}, \bibinfo {author} {\bibfnamefont {D.~G.}\ \bibnamefont
  {{Barci}}}, \bibinfo {author} {\bibfnamefont {V.}~\bibnamefont
  {{Fern{\'a}ndez}}}, \bibinfo {author} {\bibfnamefont {E.}~\bibnamefont
  {{Fradkin}}},\ and\ \bibinfo {author} {\bibfnamefont {L.}~\bibnamefont
  {{Oxman}}},\ }\bibfield  {title} {\bibinfo {title} {{Nonperturbative behavior
  of the quantum phase transition to a nematic {F}ermi fluid}},\ }\href
  {https://doi.org/10.1103/PhysRevB.73.085101} {\bibfield  {journal} {\bibinfo
  {journal} {\prb}\ }\textbf {\bibinfo {volume} {73}},\ \bibinfo {eid} {085101}
  (\bibinfo {year} {2006})}\BibitemShut {NoStop}%
\bibitem [{\citenamefont {Rech}\ \emph {et~al.}(2006)\citenamefont {Rech},
  \citenamefont {P\'epin},\ and\ \citenamefont {Chubukov}}]{rech}%
  \BibitemOpen
  \bibfield  {author} {\bibinfo {author} {\bibfnamefont {J.}~\bibnamefont
  {Rech}}, \bibinfo {author} {\bibfnamefont {C.}~\bibnamefont {P\'epin}},\ and\
  \bibinfo {author} {\bibfnamefont {A.~V.}\ \bibnamefont {Chubukov}},\
  }\bibfield  {title} {\bibinfo {title} {{Quantum critical behavior in
  itinerant electron systems: Eliashberg theory and instability of a
  ferromagnetic quantum critical point}},\ }\href
  {https://doi.org/10.1103/PhysRevB.74.195126} {\bibfield  {journal} {\bibinfo
  {journal} {Phys. Rev. B}\ }\textbf {\bibinfo {volume} {74}},\ \bibinfo
  {pages} {195126} (\bibinfo {year} {2006})}\BibitemShut {NoStop}%
\bibitem [{\citenamefont {{W{\"o}lfle}}\ and\ \citenamefont
  {{Rosch}}(2007)}]{wolfle}%
  \BibitemOpen
  \bibfield  {author} {\bibinfo {author} {\bibfnamefont {P.}~\bibnamefont
  {{W{\"o}lfle}}}\ and\ \bibinfo {author} {\bibfnamefont {A.}~\bibnamefont
  {{Rosch}}},\ }\bibfield  {title} {\bibinfo {title} {Fermi liquid near a
  quantum critical point},\ }\href {https://doi.org/10.1007/s10909-007-9308-y}
  {\bibfield  {journal} {\bibinfo  {journal} {Journal of Low Temperature
  Physics}\ }\textbf {\bibinfo {volume} {147}},\ \bibinfo {pages} {165}
  (\bibinfo {year} {2007})}\BibitemShut {NoStop}%
\bibitem [{\citenamefont {Maslov}\ and\ \citenamefont
  {Chubukov}(2010)}]{maslov}%
  \BibitemOpen
  \bibfield  {author} {\bibinfo {author} {\bibfnamefont {D.~L.}\ \bibnamefont
  {Maslov}}\ and\ \bibinfo {author} {\bibfnamefont {A.~V.}\ \bibnamefont
  {Chubukov}},\ }\bibfield  {title} {\bibinfo {title} {Fermi liquid near
  {P}omeranchuk quantum criticality},\ }\href
  {https://doi.org/10.1103/PhysRevB.81.045110} {\bibfield  {journal} {\bibinfo
  {journal} {Phys. Rev. B}\ }\textbf {\bibinfo {volume} {81}},\ \bibinfo
  {pages} {045110} (\bibinfo {year} {2010})}\BibitemShut {NoStop}%
\bibitem [{\citenamefont {{Quintanilla}}\ and\ \citenamefont
  {{Schofield}}(2006)}]{quintanilla}%
  \BibitemOpen
  \bibfield  {author} {\bibinfo {author} {\bibfnamefont {J.}~\bibnamefont
  {{Quintanilla}}}\ and\ \bibinfo {author} {\bibfnamefont {A.~J.}\ \bibnamefont
  {{Schofield}}},\ }\bibfield  {title} {\bibinfo {title} {{Pomeranchuk and
  topological {F}ermi surface instabilities from central interactions}},\
  }\href {https://doi.org/10.1103/PhysRevB.74.115126} {\bibfield  {journal}
  {\bibinfo  {journal} {\prb}\ }\textbf {\bibinfo {volume} {74}},\ \bibinfo
  {eid} {115126} (\bibinfo {year} {2006})}\BibitemShut {NoStop}%
\bibitem [{\citenamefont {{Yamase}}\ and\ \citenamefont
  {{Kohno}}(2000)}]{yamase1}%
  \BibitemOpen
  \bibfield  {author} {\bibinfo {author} {\bibfnamefont {H.}~\bibnamefont
  {{Yamase}}}\ and\ \bibinfo {author} {\bibfnamefont {H.}~\bibnamefont
  {{Kohno}}},\ }\bibfield  {title} {\bibinfo {title} {Instability toward
  formation of quasi-one-dimensional {F}ermi surface in two-dimensional t-{J}
  model},\ }\href {https://doi.org/10.1143/JPSJ.69.2151} {\bibfield  {journal}
  {\bibinfo  {journal} {Journal of the Physical Society of Japan}\ }\textbf
  {\bibinfo {volume} {69}},\ \bibinfo {pages} {2151} (\bibinfo {year}
  {2000})}\BibitemShut {NoStop}%
\bibitem [{\citenamefont {Yamase}\ \emph {et~al.}(2005)\citenamefont {Yamase},
  \citenamefont {Oganesyan},\ and\ \citenamefont {Metzner}}]{yamase2}%
  \BibitemOpen
  \bibfield  {author} {\bibinfo {author} {\bibfnamefont {H.}~\bibnamefont
  {Yamase}}, \bibinfo {author} {\bibfnamefont {V.}~\bibnamefont {Oganesyan}},\
  and\ \bibinfo {author} {\bibfnamefont {W.}~\bibnamefont {Metzner}},\
  }\bibfield  {title} {\bibinfo {title} {Mean-field theory for
  symmetry-breaking {F}ermi surface deformations on a square lattice},\ }\href
  {https://doi.org/10.1103/PhysRevB.72.035114} {\bibfield  {journal} {\bibinfo
  {journal} {Phys. Rev. B}\ }\textbf {\bibinfo {volume} {72}},\ \bibinfo
  {pages} {035114} (\bibinfo {year} {2005})}\BibitemShut {NoStop}%
\bibitem [{\citenamefont {Halboth}\ and\ \citenamefont
  {Metzner}(2000)}]{halboth}%
  \BibitemOpen
  \bibfield  {author} {\bibinfo {author} {\bibfnamefont {C.~J.}\ \bibnamefont
  {Halboth}}\ and\ \bibinfo {author} {\bibfnamefont {W.}~\bibnamefont
  {Metzner}},\ }\bibfield  {title} {\bibinfo {title} {d-wave superconductivity
  and {P}omeranchuk instability in the two-dimensional {H}ubbard model},\
  }\href {https://doi.org/10.1103/PhysRevLett.85.5162} {\bibfield  {journal}
  {\bibinfo  {journal} {Phys. Rev. Lett.}\ }\textbf {\bibinfo {volume} {85}},\
  \bibinfo {pages} {5162} (\bibinfo {year} {2000})}\BibitemShut {NoStop}%
\bibitem [{\citenamefont {Jakubczyk}\ \emph {et~al.}(2008)\citenamefont
  {Jakubczyk}, \citenamefont {Strack}, \citenamefont {Katanin},\ and\
  \citenamefont {Metzner}}]{jakub}%
  \BibitemOpen
  \bibfield  {author} {\bibinfo {author} {\bibfnamefont {P.}~\bibnamefont
  {Jakubczyk}}, \bibinfo {author} {\bibfnamefont {P.}~\bibnamefont {Strack}},
  \bibinfo {author} {\bibfnamefont {A.~A.}\ \bibnamefont {Katanin}},\ and\
  \bibinfo {author} {\bibfnamefont {W.}~\bibnamefont {Metzner}},\ }\bibfield
  {title} {\bibinfo {title} {Renormalization group for phases with broken
  discrete symmetry near quantum critical points},\ }\href
  {https://doi.org/10.1103/PhysRevB.77.195120} {\bibfield  {journal} {\bibinfo
  {journal} {Phys. Rev. B}\ }\textbf {\bibinfo {volume} {77}},\ \bibinfo
  {pages} {195120} (\bibinfo {year} {2008})}\BibitemShut {NoStop}%
\bibitem [{\citenamefont {Zacharias}\ \emph {et~al.}(2009)\citenamefont
  {Zacharias}, \citenamefont {W\"olfle},\ and\ \citenamefont
  {Garst}}]{zacharias}%
  \BibitemOpen
  \bibfield  {author} {\bibinfo {author} {\bibfnamefont {M.}~\bibnamefont
  {Zacharias}}, \bibinfo {author} {\bibfnamefont {P.}~\bibnamefont
  {W\"olfle}},\ and\ \bibinfo {author} {\bibfnamefont {M.}~\bibnamefont
  {Garst}},\ }\bibfield  {title} {\bibinfo {title} {Multiscale quantum
  criticality: {P}omeranchuk instability in isotropic metals},\ }\href
  {https://doi.org/10.1103/PhysRevB.80.165116} {\bibfield  {journal} {\bibinfo
  {journal} {Phys. Rev. B}\ }\textbf {\bibinfo {volume} {80}},\ \bibinfo
  {pages} {165116} (\bibinfo {year} {2009})}\BibitemShut {NoStop}%
\bibitem [{\citenamefont {{Kim}}\ \emph {et~al.}(2008)\citenamefont {{Kim}},
  \citenamefont {{Lawler}}, \citenamefont {{Oreto}}, \citenamefont {{Sachdev}},
  \citenamefont {{Fradkin}},\ and\ \citenamefont {{Kivelson}}}]{eaKim}%
  \BibitemOpen
  \bibfield  {author} {\bibinfo {author} {\bibfnamefont {E.-A.}\ \bibnamefont
  {{Kim}}}, \bibinfo {author} {\bibfnamefont {M.~J.}\ \bibnamefont {{Lawler}}},
  \bibinfo {author} {\bibfnamefont {P.}~\bibnamefont {{Oreto}}}, \bibinfo
  {author} {\bibfnamefont {S.}~\bibnamefont {{Sachdev}}}, \bibinfo {author}
  {\bibfnamefont {E.}~\bibnamefont {{Fradkin}}},\ and\ \bibinfo {author}
  {\bibfnamefont {S.~A.}\ \bibnamefont {{Kivelson}}},\ }\bibfield  {title}
  {\bibinfo {title} {{Theory of the nodal nematic quantum phase transition in
  superconductors}},\ }\href {https://doi.org/10.1103/PhysRevB.77.184514}
  {\bibfield  {journal} {\bibinfo  {journal} {\prb}\ }\textbf {\bibinfo
  {volume} {77}},\ \bibinfo {eid} {184514} (\bibinfo {year}
  {2008})}\BibitemShut {NoStop}%
\bibitem [{\citenamefont {{Huh}}\ and\ \citenamefont {{Sachdev}}(2008)}]{huh}%
  \BibitemOpen
  \bibfield  {author} {\bibinfo {author} {\bibfnamefont {Y.}~\bibnamefont
  {{Huh}}}\ and\ \bibinfo {author} {\bibfnamefont {S.}~\bibnamefont
  {{Sachdev}}},\ }\bibfield  {title} {\bibinfo {title} {{Renormalization group
  theory of nematic ordering in d-wave superconductors}},\ }\href
  {https://doi.org/10.1103/PhysRevB.78.064512} {\bibfield  {journal} {\bibinfo
  {journal} {\prb}\ }\textbf {\bibinfo {volume} {78}},\ \bibinfo {eid} {064512}
  (\bibinfo {year} {2008})}\BibitemShut {NoStop}%
\bibitem [{\citenamefont {Dalidovich}\ and\ \citenamefont {Lee}(2013)}]{denis}%
  \BibitemOpen
  \bibfield  {author} {\bibinfo {author} {\bibfnamefont {D.}~\bibnamefont
  {Dalidovich}}\ and\ \bibinfo {author} {\bibfnamefont {S.-S.}\ \bibnamefont
  {Lee}},\ }\bibfield  {title} {\bibinfo {title} {Perturbative non-{F}ermi
  liquids from dimensional regularization},\ }\href
  {https://doi.org/10.1103/PhysRevB.88.245106} {\bibfield  {journal} {\bibinfo
  {journal} {Phys. Rev. B}\ }\textbf {\bibinfo {volume} {88}},\ \bibinfo
  {pages} {245106} (\bibinfo {year} {2013})}\BibitemShut {NoStop}%
\bibitem [{\citenamefont {Mandal}\ and\ \citenamefont {Lee}(2015)}]{ips-lee}%
  \BibitemOpen
  \bibfield  {author} {\bibinfo {author} {\bibfnamefont {I.}~\bibnamefont
  {Mandal}}\ and\ \bibinfo {author} {\bibfnamefont {S.-S.}\ \bibnamefont
  {Lee}},\ }\bibfield  {title} {\bibinfo {title} {Ultraviolet/infrared mixing
  in non-{F}ermi liquids},\ }\href {https://doi.org/10.1103/PhysRevB.92.035141}
  {\bibfield  {journal} {\bibinfo  {journal} {Phys. Rev. B}\ }\textbf {\bibinfo
  {volume} {92}},\ \bibinfo {pages} {035141} (\bibinfo {year}
  {2015})}\BibitemShut {NoStop}%
\bibitem [{\citenamefont {Mandal}(2016{\natexlab{a}})}]{ips-uv-ir2}%
  \BibitemOpen
  \bibfield  {author} {\bibinfo {author} {\bibfnamefont {I.}~\bibnamefont
  {Mandal}},\ }\bibfield  {title} {\bibinfo {title} {{UV/IR mixing in non-Fermi
  liquids: Higher-loop corrections in different energy ranges}},\ }\href
  {https://doi.org/10.1140/epjb/e2016-70509-4} {\bibfield  {journal} {\bibinfo
  {journal} {Eur. Phys. J. B}\ }\textbf {\bibinfo {volume} {89}},\ \bibinfo
  {pages} {278} (\bibinfo {year} {2016}{\natexlab{a}})}\BibitemShut {NoStop}%
\bibitem [{\citenamefont {Eberlein}\ \emph {et~al.}(2016)\citenamefont
  {Eberlein}, \citenamefont {Mandal},\ and\ \citenamefont
  {Sachdev}}]{ips-subir}%
  \BibitemOpen
  \bibfield  {author} {\bibinfo {author} {\bibfnamefont {A.}~\bibnamefont
  {Eberlein}}, \bibinfo {author} {\bibfnamefont {I.}~\bibnamefont {Mandal}},\
  and\ \bibinfo {author} {\bibfnamefont {S.}~\bibnamefont {Sachdev}},\
  }\bibfield  {title} {\bibinfo {title} {Hyperscaling violation at the
  {I}sing-nematic quantum critical point in two-dimensional metals},\ }\href
  {https://doi.org/10.1103/PhysRevB.94.045133} {\bibfield  {journal} {\bibinfo
  {journal} {Phys. Rev. B}\ }\textbf {\bibinfo {volume} {94}},\ \bibinfo
  {pages} {045133} (\bibinfo {year} {2016})}\BibitemShut {NoStop}%
\bibitem [{\citenamefont {Mandal}(2016{\natexlab{b}})}]{ips-sc}%
  \BibitemOpen
  \bibfield  {author} {\bibinfo {author} {\bibfnamefont {I.}~\bibnamefont
  {Mandal}},\ }\bibfield  {title} {\bibinfo {title} {Superconducting
  instability in non-{F}ermi liquids},\ }\href
  {https://doi.org/10.1103/PhysRevB.94.115138} {\bibfield  {journal} {\bibinfo
  {journal} {Phys. Rev. B}\ }\textbf {\bibinfo {volume} {94}},\ \bibinfo
  {pages} {115138} (\bibinfo {year} {2016}{\natexlab{b}})}\BibitemShut
  {NoStop}%
\bibitem [{\citenamefont {Metlitski}\ and\ \citenamefont
  {Sachdev}(2010{\natexlab{b}})}]{max-sdw}%
  \BibitemOpen
  \bibfield  {author} {\bibinfo {author} {\bibfnamefont {M.~A.}\ \bibnamefont
  {Metlitski}}\ and\ \bibinfo {author} {\bibfnamefont {S.}~\bibnamefont
  {Sachdev}},\ }\bibfield  {title} {\bibinfo {title} {{Quantum phase
  transitions of metals in two spatial dimensions. II. Spin density wave
  order}},\ }\href {https://doi.org/10.1103/PhysRevB.82.075128} {\bibfield
  {journal} {\bibinfo  {journal} {Phys. Rev. B}\ }\textbf {\bibinfo {volume}
  {82}},\ \bibinfo {pages} {075128} (\bibinfo {year}
  {2010}{\natexlab{b}})}\BibitemShut {NoStop}%
\bibitem [{\citenamefont {{Abanov}}\ and\ \citenamefont
  {{Chubukov}}(2004)}]{chubukov1}%
  \BibitemOpen
  \bibfield  {author} {\bibinfo {author} {\bibfnamefont {A.}~\bibnamefont
  {{Abanov}}}\ and\ \bibinfo {author} {\bibfnamefont {A.}~\bibnamefont
  {{Chubukov}}},\ }\bibfield  {title} {\bibinfo {title} {Anomalous scaling at
  the quantum critical point in itinerant antiferromagnets},\ }\href
  {https://doi.org/10.1103/PhysRevLett.93.255702} {\bibfield  {journal}
  {\bibinfo  {journal} {Physical Review Letters}\ }\textbf {\bibinfo {volume}
  {93}},\ \bibinfo {eid} {255702} (\bibinfo {year} {2004})}\BibitemShut
  {NoStop}%
\bibitem [{\citenamefont {{Abanov}}\ and\ \citenamefont
  {{Chubukov}}(2000)}]{Chubukov}%
  \BibitemOpen
  \bibfield  {author} {\bibinfo {author} {\bibfnamefont {A.}~\bibnamefont
  {{Abanov}}}\ and\ \bibinfo {author} {\bibfnamefont {A.~V.}\ \bibnamefont
  {{Chubukov}}},\ }\bibfield  {title} {\bibinfo {title} {Spin-fermion model
  near the quantum critical point: {O}ne-loop renormalization group results},\
  }\href {https://doi.org/10.1103/PhysRevLett.84.5608} {\bibfield  {journal}
  {\bibinfo  {journal} {Physical Review Letters}\ }\textbf {\bibinfo {volume}
  {84}},\ \bibinfo {pages} {5608} (\bibinfo {year} {2000})}\BibitemShut
  {NoStop}%
\bibitem [{\citenamefont {Sur}\ and\ \citenamefont {Lee}(2015)}]{shouvik2}%
  \BibitemOpen
  \bibfield  {author} {\bibinfo {author} {\bibfnamefont {S.}~\bibnamefont
  {Sur}}\ and\ \bibinfo {author} {\bibfnamefont {S.-S.}\ \bibnamefont {Lee}},\
  }\bibfield  {title} {\bibinfo {title} {Quasilocal strange metal},\ }\href
  {https://doi.org/10.1103/PhysRevB.91.125136} {\bibfield  {journal} {\bibinfo
  {journal} {Phys. Rev. B}\ }\textbf {\bibinfo {volume} {91}},\ \bibinfo
  {pages} {125136} (\bibinfo {year} {2015})}\BibitemShut {NoStop}%
\bibitem [{\citenamefont {Mandal}(2017)}]{ips-c2}%
  \BibitemOpen
  \bibfield  {author} {\bibinfo {author} {\bibfnamefont {I.}~\bibnamefont
  {Mandal}},\ }\bibfield  {title} {\bibinfo {title} {Scaling behaviour and
  superconducting instability in anisotropic non-{F}ermi liquids},\ }\href
  {https://doi.org/https://doi.org/10.1016/j.aop.2016.11.009} {\bibfield
  {journal} {\bibinfo  {journal} {Annals of Physics}\ }\textbf {\bibinfo
  {volume} {376}},\ \bibinfo {pages} {89 } (\bibinfo {year}
  {2017})}\BibitemShut {NoStop}%
\bibitem [{\citenamefont {Schlief}\ \emph {et~al.}(2017)\citenamefont
  {Schlief}, \citenamefont {Lunts},\ and\ \citenamefont {Lee}}]{andres1}%
  \BibitemOpen
  \bibfield  {author} {\bibinfo {author} {\bibfnamefont {A.}~\bibnamefont
  {Schlief}}, \bibinfo {author} {\bibfnamefont {P.}~\bibnamefont {Lunts}},\
  and\ \bibinfo {author} {\bibfnamefont {S.-S.}\ \bibnamefont {Lee}},\
  }\bibfield  {title} {\bibinfo {title} {Exact critical exponents for the
  antiferromagnetic quantum critical metal in two dimensions},\ }\href
  {https://doi.org/10.1103/PhysRevX.7.021010} {\bibfield  {journal} {\bibinfo
  {journal} {Phys. Rev. X}\ }\textbf {\bibinfo {volume} {7}},\ \bibinfo {pages}
  {021010} (\bibinfo {year} {2017})}\BibitemShut {NoStop}%
\bibitem [{\citenamefont {Lunts}\ \emph {et~al.}(2017)\citenamefont {Lunts},
  \citenamefont {Schlief},\ and\ \citenamefont {Lee}}]{andres2}%
  \BibitemOpen
  \bibfield  {author} {\bibinfo {author} {\bibfnamefont {P.}~\bibnamefont
  {Lunts}}, \bibinfo {author} {\bibfnamefont {A.}~\bibnamefont {Schlief}},\
  and\ \bibinfo {author} {\bibfnamefont {S.-S.}\ \bibnamefont {Lee}},\
  }\bibfield  {title} {\bibinfo {title} {Emergence of a control parameter for
  the antiferromagnetic quantum critical metal},\ }\href
  {https://doi.org/10.1103/PhysRevB.95.245109} {\bibfield  {journal} {\bibinfo
  {journal} {Phys. Rev. B}\ }\textbf {\bibinfo {volume} {95}},\ \bibinfo
  {pages} {245109} (\bibinfo {year} {2017})}\BibitemShut {NoStop}%
\bibitem [{\citenamefont {Reizer}(1989)}]{reizer}%
  \BibitemOpen
  \bibfield  {author} {\bibinfo {author} {\bibfnamefont {M.~Y.}\ \bibnamefont
  {Reizer}},\ }\bibfield  {title} {\bibinfo {title} {Relativistic effects in
  the electron density of states, specific heat, and the electron spectrum of
  normal metals},\ }\href {https://doi.org/10.1103/PhysRevB.40.11571}
  {\bibfield  {journal} {\bibinfo  {journal} {Phys. Rev. B}\ }\textbf {\bibinfo
  {volume} {40}},\ \bibinfo {pages} {11571} (\bibinfo {year}
  {1989})}\BibitemShut {NoStop}%
\bibitem [{\citenamefont {Halperin}\ \emph {et~al.}(1993)\citenamefont
  {Halperin}, \citenamefont {Lee},\ and\ \citenamefont {Read}}]{HALPERIN}%
  \BibitemOpen
  \bibfield  {author} {\bibinfo {author} {\bibfnamefont {B.~I.}\ \bibnamefont
  {Halperin}}, \bibinfo {author} {\bibfnamefont {P.~A.}\ \bibnamefont {Lee}},\
  and\ \bibinfo {author} {\bibfnamefont {N.}~\bibnamefont {Read}},\ }\bibfield
  {title} {\bibinfo {title} {Theory of the half-filled {L}andau level},\ }\href
  {https://doi.org/10.1103/PhysRevB.47.7312} {\bibfield  {journal} {\bibinfo
  {journal} {Phys. Rev. B}\ }\textbf {\bibinfo {volume} {47}},\ \bibinfo
  {pages} {7312} (\bibinfo {year} {1993})}\BibitemShut {NoStop}%
\bibitem [{\citenamefont {Polchinski}(1994)}]{polchinski}%
  \BibitemOpen
  \bibfield  {author} {\bibinfo {author} {\bibfnamefont {J.}~\bibnamefont
  {Polchinski}},\ }\bibfield  {title} {\bibinfo {title} {Low-energy dynamics of
  the spinon-gauge system},\ }\href
  {https://doi.org/https://doi.org/10.1016/0550-3213(94)90449-9} {\bibfield
  {journal} {\bibinfo  {journal} {Nuclear Physics B}\ }\textbf {\bibinfo
  {volume} {422}},\ \bibinfo {pages} {617} (\bibinfo {year}
  {1994})}\BibitemShut {NoStop}%
\bibitem [{\citenamefont {Altshuler}\ \emph {et~al.}(1994)\citenamefont
  {Altshuler}, \citenamefont {Ioffe},\ and\ \citenamefont
  {Millis}}]{ALTSHULER}%
  \BibitemOpen
  \bibfield  {author} {\bibinfo {author} {\bibfnamefont {B.~L.}\ \bibnamefont
  {Altshuler}}, \bibinfo {author} {\bibfnamefont {L.~B.}\ \bibnamefont
  {Ioffe}},\ and\ \bibinfo {author} {\bibfnamefont {A.~J.}\ \bibnamefont
  {Millis}},\ }\bibfield  {title} {\bibinfo {title} {Low-energy properties of
  fermions with singular interactions},\ }\href
  {https://doi.org/10.1103/PhysRevB.50.14048} {\bibfield  {journal} {\bibinfo
  {journal} {Phys. Rev. B}\ }\textbf {\bibinfo {volume} {50}},\ \bibinfo
  {pages} {14048} (\bibinfo {year} {1994})}\BibitemShut {NoStop}%
\bibitem [{\citenamefont {{Nayak}}\ and\ \citenamefont
  {{Wilczek}}(1994{\natexlab{b}})}]{nayak}%
  \BibitemOpen
  \bibfield  {author} {\bibinfo {author} {\bibfnamefont {C.}~\bibnamefont
  {{Nayak}}}\ and\ \bibinfo {author} {\bibfnamefont {F.}~\bibnamefont
  {{Wilczek}}},\ }\bibfield  {title} {\bibinfo {title} {Renormalization group
  approach to low temperature properties of a non-{F}ermi liquid metal},\
  }\href {https://doi.org/10.1016/0550-3213(94)90158-9} {\bibfield  {journal}
  {\bibinfo  {journal} {Nuclear Physics B}\ }\textbf {\bibinfo {volume}
  {430}},\ \bibinfo {pages} {534} (\bibinfo {year}
  {1994}{\natexlab{b}})}\BibitemShut {NoStop}%
\bibitem [{\citenamefont {Lee}(2009)}]{SSLee}%
  \BibitemOpen
  \bibfield  {author} {\bibinfo {author} {\bibfnamefont {S.-S.}\ \bibnamefont
  {Lee}},\ }\bibfield  {title} {\bibinfo {title} {Low-energy effective theory
  of {F}ermi surface coupled with {U}(1) gauge field in $2+1$ dimensions},\
  }\href {https://doi.org/10.1103/PhysRevB.80.165102} {\bibfield  {journal}
  {\bibinfo  {journal} {Phys. Rev. B}\ }\textbf {\bibinfo {volume} {80}},\
  \bibinfo {pages} {165102} (\bibinfo {year} {2009})}\BibitemShut {NoStop}%
\bibitem [{\citenamefont {Mross}\ \emph {et~al.}(2010)\citenamefont {Mross},
  \citenamefont {McGreevy}, \citenamefont {Liu},\ and\ \citenamefont
  {Senthil}}]{mross}%
  \BibitemOpen
  \bibfield  {author} {\bibinfo {author} {\bibfnamefont {D.~F.}\ \bibnamefont
  {Mross}}, \bibinfo {author} {\bibfnamefont {J.}~\bibnamefont {McGreevy}},
  \bibinfo {author} {\bibfnamefont {H.}~\bibnamefont {Liu}},\ and\ \bibinfo
  {author} {\bibfnamefont {T.}~\bibnamefont {Senthil}},\ }\bibfield  {title}
  {\bibinfo {title} {Controlled expansion for certain non-{F}ermi-liquid
  metals},\ }\href {https://doi.org/10.1103/PhysRevB.82.045121} {\bibfield
  {journal} {\bibinfo  {journal} {Phys. Rev. B}\ }\textbf {\bibinfo {volume}
  {82}},\ \bibinfo {pages} {045121} (\bibinfo {year} {2010})}\BibitemShut
  {NoStop}%
\bibitem [{\citenamefont {{Jiang}}\ \emph {et~al.}(2013)\citenamefont
  {{Jiang}}, \citenamefont {{Block}}, \citenamefont {{Mishmash}}, \citenamefont
  {{Garrison}}, \citenamefont {{Sheng}}, \citenamefont {{Motrunich}},\ and\
  \citenamefont {{Fisher}}}]{jiang}%
  \BibitemOpen
  \bibfield  {author} {\bibinfo {author} {\bibfnamefont {H.-C.}\ \bibnamefont
  {{Jiang}}}, \bibinfo {author} {\bibfnamefont {M.~S.}\ \bibnamefont
  {{Block}}}, \bibinfo {author} {\bibfnamefont {R.~V.}\ \bibnamefont
  {{Mishmash}}}, \bibinfo {author} {\bibfnamefont {J.~R.}\ \bibnamefont
  {{Garrison}}}, \bibinfo {author} {\bibfnamefont {D.~N.}\ \bibnamefont
  {{Sheng}}}, \bibinfo {author} {\bibfnamefont {O.~I.}\ \bibnamefont
  {{Motrunich}}},\ and\ \bibinfo {author} {\bibfnamefont {M.~P.~A.}\
  \bibnamefont {{Fisher}}},\ }\bibfield  {title} {\bibinfo {title}
  {{Non-{F}ermi-liquid d-wave metal phase of strongly interacting electrons}},\
  }\href {https://doi.org/10.1038/nature11732} {\bibfield  {journal} {\bibinfo
  {journal} {\nat}\ }\textbf {\bibinfo {volume} {493}},\ \bibinfo {pages} {39}
  (\bibinfo {year} {2013})}\BibitemShut {NoStop}%
\bibitem [{\citenamefont {Chung}\ \emph {et~al.}(2013)\citenamefont {Chung},
  \citenamefont {Mandal}, \citenamefont {Raghu},\ and\ \citenamefont
  {Chakravarty}}]{ips2}%
  \BibitemOpen
  \bibfield  {author} {\bibinfo {author} {\bibfnamefont {S.~B.}\ \bibnamefont
  {Chung}}, \bibinfo {author} {\bibfnamefont {I.}~\bibnamefont {Mandal}},
  \bibinfo {author} {\bibfnamefont {S.}~\bibnamefont {Raghu}},\ and\ \bibinfo
  {author} {\bibfnamefont {S.}~\bibnamefont {Chakravarty}},\ }\bibfield
  {title} {\bibinfo {title} {Higher angular momentum pairing from transverse
  gauge interactions},\ }\href {https://doi.org/10.1103/PhysRevB.88.045127}
  {\bibfield  {journal} {\bibinfo  {journal} {Phys. Rev. B}\ }\textbf {\bibinfo
  {volume} {88}},\ \bibinfo {pages} {045127} (\bibinfo {year}
  {2013})}\BibitemShut {NoStop}%
\bibitem [{\citenamefont {Wang}\ \emph {et~al.}(2014)\citenamefont {Wang},
  \citenamefont {Mandal}, \citenamefont {Chung},\ and\ \citenamefont
  {Chakravarty}}]{ips3}%
  \BibitemOpen
  \bibfield  {author} {\bibinfo {author} {\bibfnamefont {Z.}~\bibnamefont
  {Wang}}, \bibinfo {author} {\bibfnamefont {I.}~\bibnamefont {Mandal}},
  \bibinfo {author} {\bibfnamefont {S.~B.}\ \bibnamefont {Chung}},\ and\
  \bibinfo {author} {\bibfnamefont {S.}~\bibnamefont {Chakravarty}},\
  }\bibfield  {title} {\bibinfo {title} {Pairing in half-filled {L}andau
  level},\ }\href {https://doi.org/http://dx.doi.org/10.1016/j.aop.2014.09.021}
  {\bibfield  {journal} {\bibinfo  {journal} {Annals of Physics}\ }\textbf
  {\bibinfo {volume} {351}},\ \bibinfo {pages} {727 } (\bibinfo {year}
  {2014})}\BibitemShut {NoStop}%
\bibitem [{\citenamefont {Sur}\ and\ \citenamefont {Lee}(2014)}]{Shouvik1}%
  \BibitemOpen
  \bibfield  {author} {\bibinfo {author} {\bibfnamefont {S.}~\bibnamefont
  {Sur}}\ and\ \bibinfo {author} {\bibfnamefont {S.-S.}\ \bibnamefont {Lee}},\
  }\bibfield  {title} {\bibinfo {title} {Chiral non-{F}ermi liquids},\ }\href
  {https://doi.org/10.1103/PhysRevB.90.045121} {\bibfield  {journal} {\bibinfo
  {journal} {Phys. Rev. B}\ }\textbf {\bibinfo {volume} {90}},\ \bibinfo
  {pages} {045121} (\bibinfo {year} {2014})}\BibitemShut {NoStop}%
\bibitem [{\citenamefont {Lee}(2018)}]{Lee_2018}%
  \BibitemOpen
  \bibfield  {author} {\bibinfo {author} {\bibfnamefont {S.-S.}\ \bibnamefont
  {Lee}},\ }\bibfield  {title} {\bibinfo {title} {Recent developments in
  non-{F}ermi liquid theory},\ }\href
  {https://doi.org/10.1146/annurev-conmatphys-031016-025531} {\bibfield
  {journal} {\bibinfo  {journal} {Annual Review of Condensed Matter Physics}\
  }\textbf {\bibinfo {volume} {9}},\ \bibinfo {pages} {227–244} (\bibinfo
  {year} {2018})}\BibitemShut {NoStop}%
\bibitem [{\citenamefont {Pimenov}\ \emph {et~al.}(2018)\citenamefont
  {Pimenov}, \citenamefont {Mandal}, \citenamefont {Piazza},\ and\
  \citenamefont {Punk}}]{ips-fflo}%
  \BibitemOpen
  \bibfield  {author} {\bibinfo {author} {\bibfnamefont {D.}~\bibnamefont
  {Pimenov}}, \bibinfo {author} {\bibfnamefont {I.}~\bibnamefont {Mandal}},
  \bibinfo {author} {\bibfnamefont {F.}~\bibnamefont {Piazza}},\ and\ \bibinfo
  {author} {\bibfnamefont {M.}~\bibnamefont {Punk}},\ }\bibfield  {title}
  {\bibinfo {title} {{Non-Fermi liquid at the FFLO quantum critical point}},\
  }\href {https://doi.org/10.1103/PhysRevB.98.024510} {\bibfield  {journal}
  {\bibinfo  {journal} {Phys. Rev. B}\ }\textbf {\bibinfo {volume} {98}},\
  \bibinfo {pages} {024510} (\bibinfo {year} {2018})}\BibitemShut {NoStop}%
\bibitem [{\citenamefont {Mandal}(2020)}]{ips-u1}%
  \BibitemOpen
  \bibfield  {author} {\bibinfo {author} {\bibfnamefont {I.}~\bibnamefont
  {Mandal}},\ }\bibfield  {title} {\bibinfo {title} {Critical {F}ermi surfaces
  in generic dimensions arising from transverse gauge field interactions},\
  }\href {https://doi.org/10.1103/PhysRevResearch.2.043277} {\bibfield
  {journal} {\bibinfo  {journal} {Phys. Rev. Research}\ }\textbf {\bibinfo
  {volume} {2}},\ \bibinfo {pages} {043277} (\bibinfo {year}
  {2020})}\BibitemShut {NoStop}%
\bibitem [{\citenamefont {Mandal}\ and\ \citenamefont
  {Fernandes}(2023)}]{ips-rafael}%
  \BibitemOpen
  \bibfield  {author} {\bibinfo {author} {\bibfnamefont {I.}~\bibnamefont
  {Mandal}}\ and\ \bibinfo {author} {\bibfnamefont {R.~M.}\ \bibnamefont
  {Fernandes}},\ }\bibfield  {title} {\bibinfo {title} {Valley-polarized
  nematic order in twisted moir\'e systems: In-plane orbital magnetism and
  crossover from non-{F}ermi liquid to {F}ermi liquid},\ }\href
  {https://doi.org/10.1103/PhysRevB.107.125142} {\bibfield  {journal} {\bibinfo
   {journal} {Phys. Rev. B}\ }\textbf {\bibinfo {volume} {107}},\ \bibinfo
  {pages} {125142} (\bibinfo {year} {2023})}\BibitemShut {NoStop}%
\bibitem [{\citenamefont {Abrikosov}(1974)}]{abrikosov}%
  \BibitemOpen
  \bibfield  {author} {\bibinfo {author} {\bibfnamefont {A.~A.}\ \bibnamefont
  {Abrikosov}},\ }\bibfield  {title} {\bibinfo {title} {Calculation of critical
  indices for zero-gap semiconductors},\ }\href@noop {} {\bibfield  {journal}
  {\bibinfo  {journal} {Journal of Experimental and Theoretical Physics}\
  }\textbf {\bibinfo {volume} {39}},\ \bibinfo {pages} {709} (\bibinfo {year}
  {1974})}\BibitemShut {NoStop}%
\bibitem [{\citenamefont {Moon}\ \emph {et~al.}(2013)\citenamefont {Moon},
  \citenamefont {Xu}, \citenamefont {Kim},\ and\ \citenamefont
  {Balents}}]{moon-xu}%
  \BibitemOpen
  \bibfield  {author} {\bibinfo {author} {\bibfnamefont {E.-G.}\ \bibnamefont
  {Moon}}, \bibinfo {author} {\bibfnamefont {C.}~\bibnamefont {Xu}}, \bibinfo
  {author} {\bibfnamefont {Y.~B.}\ \bibnamefont {Kim}},\ and\ \bibinfo {author}
  {\bibfnamefont {L.}~\bibnamefont {Balents}},\ }\bibfield  {title} {\bibinfo
  {title} {Non-{F}ermi-liquid and topological states with strong spin-orbit
  coupling},\ }\href {https://doi.org/10.1103/PhysRevLett.111.206401}
  {\bibfield  {journal} {\bibinfo  {journal} {Phys. Rev. Lett.}\ }\textbf
  {\bibinfo {volume} {111}},\ \bibinfo {pages} {206401} (\bibinfo {year}
  {2013})}\BibitemShut {NoStop}%
\bibitem [{\citenamefont {Nandkishore}\ and\ \citenamefont
  {Parameswaran}(2017)}]{rahul-sid}%
  \BibitemOpen
  \bibfield  {author} {\bibinfo {author} {\bibfnamefont {R.~M.}\ \bibnamefont
  {Nandkishore}}\ and\ \bibinfo {author} {\bibfnamefont {S.~A.}\ \bibnamefont
  {Parameswaran}},\ }\bibfield  {title} {\bibinfo {title} {Disorder-driven
  destruction of a non-{F}ermi liquid semimetal studied by renormalization
  group analysis},\ }\href {https://doi.org/10.1103/PhysRevB.95.205106}
  {\bibfield  {journal} {\bibinfo  {journal} {Phys. Rev. B}\ }\textbf {\bibinfo
  {volume} {95}},\ \bibinfo {pages} {205106} (\bibinfo {year}
  {2017})}\BibitemShut {NoStop}%
\bibitem [{\citenamefont {{Mandal}}\ and\ \citenamefont
  {{Nandkishore}}(2018)}]{ips-rahul}%
  \BibitemOpen
  \bibfield  {author} {\bibinfo {author} {\bibfnamefont {I.}~\bibnamefont
  {{Mandal}}}\ and\ \bibinfo {author} {\bibfnamefont {R.~M.}\ \bibnamefont
  {{Nandkishore}}},\ }\bibfield  {title} {\bibinfo {title} {{Interplay of
  Coulomb interactions and disorder in three-dimensional quadratic band
  crossings without time-reversal symmetry and with unequal masses for
  conduction and valence bands}},\ }\href
  {https://doi.org/10.1103/PhysRevB.97.125121} {\bibfield  {journal} {\bibinfo
  {journal} {\prb}\ }\textbf {\bibinfo {volume} {97}},\ \bibinfo {eid} {125121}
  (\bibinfo {year} {2018})}\BibitemShut {NoStop}%
\bibitem [{\citenamefont {Mandal}(2018)}]{ips-qbt-sc}%
  \BibitemOpen
  \bibfield  {author} {\bibinfo {author} {\bibfnamefont {I.}~\bibnamefont
  {Mandal}},\ }\bibfield  {title} {\bibinfo {title} {Fate of superconductivity
  in three-dimensional disordered {L}uttinger semimetals},\ }\href
  {https://doi.org/https://doi.org/10.1016/j.aop.2018.03.004} {\bibfield
  {journal} {\bibinfo  {journal} {Annals of Physics}\ }\textbf {\bibinfo
  {volume} {392}},\ \bibinfo {pages} {179 } (\bibinfo {year}
  {2018})}\BibitemShut {NoStop}%
\bibitem [{\citenamefont {Mandal}\ and\ \citenamefont
  {Freire}(2021)}]{ips-hermann}%
  \BibitemOpen
  \bibfield  {author} {\bibinfo {author} {\bibfnamefont {I.}~\bibnamefont
  {Mandal}}\ and\ \bibinfo {author} {\bibfnamefont {H.}~\bibnamefont
  {Freire}},\ }\bibfield  {title} {\bibinfo {title} {Transport in the
  non-{F}ermi liquid phase of isotropic {L}uttinger semimetals},\ }\href
  {https://doi.org/10.1103/PhysRevB.103.195116} {\bibfield  {journal} {\bibinfo
   {journal} {Phys. Rev. B}\ }\textbf {\bibinfo {volume} {103}},\ \bibinfo
  {pages} {195116} (\bibinfo {year} {2021})}\BibitemShut {NoStop}%
\bibitem [{\citenamefont {Freire}\ and\ \citenamefont
  {Mandal}(2021)}]{ips-hermann2}%
  \BibitemOpen
  \bibfield  {author} {\bibinfo {author} {\bibfnamefont {H.}~\bibnamefont
  {Freire}}\ and\ \bibinfo {author} {\bibfnamefont {I.}~\bibnamefont
  {Mandal}},\ }\bibfield  {title} {\bibinfo {title} {Thermoelectric and thermal
  properties of the weakly disordered non-{F}ermi liquid phase of {L}uttinger
  semimetals},\ }\href
  {https://doi.org/https://doi.org/10.1016/j.physleta.2021.127470} {\bibfield
  {journal} {\bibinfo  {journal} {Physics Letters A}\ }\textbf {\bibinfo
  {volume} {407}},\ \bibinfo {pages} {127470} (\bibinfo {year}
  {2021})}\BibitemShut {NoStop}%
\bibitem [{\citenamefont {Mandal}\ and\ \citenamefont
  {Freire}(2022)}]{ips-hermann3}%
  \BibitemOpen
  \bibfield  {author} {\bibinfo {author} {\bibfnamefont {I.}~\bibnamefont
  {Mandal}}\ and\ \bibinfo {author} {\bibfnamefont {H.}~\bibnamefont
  {Freire}},\ }\bibfield  {title} {\bibinfo {title} {Raman response and shear
  viscosity in the non-{F}ermi liquid phase of {L}uttinger semimetals},\ }\href
  {https://doi.org/10.1088/1361-648x/ac6785} {\bibfield  {journal} {\bibinfo
  {journal} {Journal of Physics: Condensed Matter}\ }\textbf {\bibinfo {volume}
  {34}},\ \bibinfo {pages} {275604} (\bibinfo {year} {2022})}\BibitemShut
  {NoStop}%
\bibitem [{\citenamefont {Roy}\ \emph {et~al.}(2018)\citenamefont {Roy},
  \citenamefont {Kennett}, \citenamefont {Yang},\ and\ \citenamefont
  {Juri\ifmmode \check{c}\else \v{c}\fi{}i\ifmmode~\acute{c}\else
  \'{c}\fi{}}}]{juricic}%
  \BibitemOpen
  \bibfield  {author} {\bibinfo {author} {\bibfnamefont {B.}~\bibnamefont
  {Roy}}, \bibinfo {author} {\bibfnamefont {M.~P.}\ \bibnamefont {Kennett}},
  \bibinfo {author} {\bibfnamefont {K.}~\bibnamefont {Yang}},\ and\ \bibinfo
  {author} {\bibfnamefont {V.}~\bibnamefont {Juri\ifmmode \check{c}\else
  \v{c}\fi{}i\ifmmode~\acute{c}\else \'{c}\fi{}}},\ }\bibfield  {title}
  {\bibinfo {title} {From birefringent electrons to a marginal or non-{F}ermi
  liquid of relativistic spin-$1/2$ fermions: {A}n emergent
  superuniversality},\ }\href {https://doi.org/10.1103/PhysRevLett.121.157602}
  {\bibfield  {journal} {\bibinfo  {journal} {Phys. Rev. Lett.}\ }\textbf
  {\bibinfo {volume} {121}},\ \bibinfo {pages} {157602} (\bibinfo {year}
  {2018})}\BibitemShut {NoStop}%
\bibitem [{\citenamefont {Mandal}(2021)}]{ips-birefringent}%
  \BibitemOpen
  \bibfield  {author} {\bibinfo {author} {\bibfnamefont {I.}~\bibnamefont
  {Mandal}},\ }\bibfield  {title} {\bibinfo {title} {Robust marginal {F}ermi
  liquid in birefringent semimetals},\ }\href
  {https://doi.org/https://doi.org/10.1016/j.physleta.2021.127707} {\bibfield
  {journal} {\bibinfo  {journal} {Physics Letters A}\ }\textbf {\bibinfo
  {volume} {418}},\ \bibinfo {pages} {127707} (\bibinfo {year}
  {2021})}\BibitemShut {NoStop}%
\bibitem [{\citenamefont {Mandal}\ and\ \citenamefont
  {Freire}(2024)}]{ips-hermann-review}%
  \BibitemOpen
  \bibfield  {author} {\bibinfo {author} {\bibfnamefont {I.}~\bibnamefont
  {Mandal}}\ and\ \bibinfo {author} {\bibfnamefont {H.}~\bibnamefont
  {Freire}},\ }\bibfield  {title} {\bibinfo {title} {{Transport properties in
  non-Fermi liquid phases of nodal-point semimetals}},\ }\href
  {https://doi.org/10.1088/1361-648X/ad665e} {\bibfield  {journal} {\bibinfo
  {journal} {Journal of Physics: Condensed Matter}\ }\textbf {\bibinfo {volume}
  {36}},\ \bibinfo {pages} {443002} (\bibinfo {year} {2024})}\BibitemShut
  {NoStop}%
\bibitem [{\citenamefont {Mandal}(2024)}]{ips-sc_err}%
  \BibitemOpen
  \bibfield  {author} {\bibinfo {author} {\bibfnamefont {I.}~\bibnamefont
  {Mandal}},\ }\bibfield  {title} {\bibinfo {title} {{Erratum: Superconducting
  instability in non-Fermi liquids [Phys. Rev. B 94, 115138 (2016)]}},\ }\href
  {https://doi.org/10.1103/PhysRevB.109.079902} {\bibfield  {journal} {\bibinfo
   {journal} {Phys. Rev. B}\ }\textbf {\bibinfo {volume} {109}},\ \bibinfo
  {pages} {079902} (\bibinfo {year} {2024})}\BibitemShut {NoStop}%
\bibitem [{\citenamefont {{Mandal}}(2024)}]{ips-2kf}%
  \BibitemOpen
  \bibfield  {author} {\bibinfo {author} {\bibfnamefont {I.}~\bibnamefont
  {{Mandal}}},\ }\bibfield  {title} {\bibinfo {title} {{Stable non-Fermi liquid
  fixed point at the onset of incommensurate $2k_F$ charge density wave
  order}},\ }\href {https://doi.org/10.1016/j.nuclphysb.2024.116586} {\bibfield
   {journal} {\bibinfo  {journal} {Nucl. Phys. B}\ }\textbf {\bibinfo {volume}
  {1005}},\ \bibinfo {pages} {116586} (\bibinfo {year} {2024})}\BibitemShut
  {NoStop}%
\bibitem [{\citenamefont {Holder}\ and\ \citenamefont
  {Metzner}(2014)}]{metzner1}%
  \BibitemOpen
  \bibfield  {author} {\bibinfo {author} {\bibfnamefont {T.}~\bibnamefont
  {Holder}}\ and\ \bibinfo {author} {\bibfnamefont {W.}~\bibnamefont
  {Metzner}},\ }\bibfield  {title} {\bibinfo {title} {{Non-Fermi-liquid
  behavior at the onset of incommensurate $2{k}_{F}$ charge- or spin-density
  wave order in two dimensions}},\ }\href
  {https://doi.org/10.1103/PhysRevB.90.161106} {\bibfield  {journal} {\bibinfo
  {journal} {Phys. Rev. B}\ }\textbf {\bibinfo {volume} {90}},\ \bibinfo
  {pages} {161106} (\bibinfo {year} {2014})}\BibitemShut {NoStop}%
\bibitem [{\citenamefont {S\'ykora}\ \emph {et~al.}(2018)\citenamefont
  {S\'ykora}, \citenamefont {Holder},\ and\ \citenamefont
  {Metzner}}]{metzner2}%
  \BibitemOpen
  \bibfield  {author} {\bibinfo {author} {\bibfnamefont {J.}~\bibnamefont
  {S\'ykora}}, \bibinfo {author} {\bibfnamefont {T.}~\bibnamefont {Holder}},\
  and\ \bibinfo {author} {\bibfnamefont {W.}~\bibnamefont {Metzner}},\
  }\bibfield  {title} {\bibinfo {title} {Fluctuation effects at the onset of
  the $2{k}_{F}$ density wave order with one pair of hot spots in
  two-dimensional metals},\ }\href {https://doi.org/10.1103/PhysRevB.97.155159}
  {\bibfield  {journal} {\bibinfo  {journal} {Phys. Rev. B}\ }\textbf {\bibinfo
  {volume} {97}},\ \bibinfo {pages} {155159} (\bibinfo {year}
  {2018})}\BibitemShut {NoStop}%
\bibitem [{\citenamefont {Schlawin}\ and\ \citenamefont
  {Jaksch}(2019)}]{jaksch}%
  \BibitemOpen
  \bibfield  {author} {\bibinfo {author} {\bibfnamefont {F.}~\bibnamefont
  {Schlawin}}\ and\ \bibinfo {author} {\bibfnamefont {D.}~\bibnamefont
  {Jaksch}},\ }\bibfield  {title} {\bibinfo {title} {Cavity-mediated
  unconventional pairing in ultracold fermionic atoms},\ }\href
  {https://doi.org/10.1103/PhysRevLett.123.133601} {\bibfield  {journal}
  {\bibinfo  {journal} {Phys. Rev. Lett.}\ }\textbf {\bibinfo {volume} {123}},\
  \bibinfo {pages} {133601} (\bibinfo {year} {2019})}\BibitemShut {NoStop}%
\bibitem [{\citenamefont {Sheikhan}\ and\ \citenamefont
  {Kollath}(2019)}]{kollath}%
  \BibitemOpen
  \bibfield  {author} {\bibinfo {author} {\bibfnamefont {A.}~\bibnamefont
  {Sheikhan}}\ and\ \bibinfo {author} {\bibfnamefont {C.}~\bibnamefont
  {Kollath}},\ }\bibfield  {title} {\bibinfo {title} {Cavity-induced
  superconducting and $4{k}_{F}$ charge-density-wave states},\ }\href
  {https://doi.org/10.1103/PhysRevA.99.053611} {\bibfield  {journal} {\bibinfo
  {journal} {Phys. Rev. A}\ }\textbf {\bibinfo {volume} {99}},\ \bibinfo
  {pages} {053611} (\bibinfo {year} {2019})}\BibitemShut {NoStop}%
\bibitem [{\citenamefont {Li}\ and\ \citenamefont {Eckstein}(2020)}]{eckstein}%
  \BibitemOpen
  \bibfield  {author} {\bibinfo {author} {\bibfnamefont {J.}~\bibnamefont
  {Li}}\ and\ \bibinfo {author} {\bibfnamefont {M.}~\bibnamefont {Eckstein}},\
  }\bibfield  {title} {\bibinfo {title} {Manipulating intertwined orders in
  solids with quantum light},\ }\href
  {https://doi.org/10.1103/PhysRevLett.125.217402} {\bibfield  {journal}
  {\bibinfo  {journal} {Phys. Rev. Lett.}\ }\textbf {\bibinfo {volume} {125}},\
  \bibinfo {pages} {217402} (\bibinfo {year} {2020})}\BibitemShut {NoStop}%
\bibitem [{\citenamefont {Mivehvar}\ \emph {et~al.}(2019)\citenamefont
  {Mivehvar}, \citenamefont {Ritsch},\ and\ \citenamefont {Piazza}}]{farokh}%
  \BibitemOpen
  \bibfield  {author} {\bibinfo {author} {\bibfnamefont {F.}~\bibnamefont
  {Mivehvar}}, \bibinfo {author} {\bibfnamefont {H.}~\bibnamefont {Ritsch}},\
  and\ \bibinfo {author} {\bibfnamefont {F.}~\bibnamefont {Piazza}},\
  }\bibfield  {title} {\bibinfo {title} {Cavity-quantum-electrodynamical
  toolbox for quantum magnetism},\ }\href
  {https://doi.org/10.1103/PhysRevLett.122.113603} {\bibfield  {journal}
  {\bibinfo  {journal} {Phys. Rev. Lett.}\ }\textbf {\bibinfo {volume} {122}},\
  \bibinfo {pages} {113603} (\bibinfo {year} {2019})}\BibitemShut {NoStop}%
\bibitem [{\citenamefont {Ashida}\ \emph {et~al.}(2020)\citenamefont {Ashida},
  \citenamefont {\ifmmode \dot{I}\else \.{I}\fi{}mamo\ifmmode~\breve{g}\else
  \u{g}\fi{}lu}, \citenamefont {Faist}, \citenamefont {Jaksch}, \citenamefont
  {Cavalleri},\ and\ \citenamefont {Demler}}]{demler}%
  \BibitemOpen
  \bibfield  {author} {\bibinfo {author} {\bibfnamefont {Y.}~\bibnamefont
  {Ashida}}, \bibinfo {author} {\bibfnamefont {A.~m.~c.}\ \bibnamefont
  {\ifmmode \dot{I}\else \.{I}\fi{}mamo\ifmmode~\breve{g}\else \u{g}\fi{}lu}},
  \bibinfo {author} {\bibfnamefont {J.}~\bibnamefont {Faist}}, \bibinfo
  {author} {\bibfnamefont {D.}~\bibnamefont {Jaksch}}, \bibinfo {author}
  {\bibfnamefont {A.}~\bibnamefont {Cavalleri}},\ and\ \bibinfo {author}
  {\bibfnamefont {E.}~\bibnamefont {Demler}},\ }\bibfield  {title} {\bibinfo
  {title} {Quantum electrodynamic control of matter: {C}avity-enhanced
  ferroelectric phase transition},\ }\href
  {https://doi.org/10.1103/PhysRevX.10.041027} {\bibfield  {journal} {\bibinfo
  {journal} {Phys. Rev. X}\ }\textbf {\bibinfo {volume} {10}},\ \bibinfo
  {pages} {041027} (\bibinfo {year} {2020})}\BibitemShut {NoStop}%
\bibitem [{\citenamefont {{Chiocchetta}}\ \emph {et~al.}(2021)\citenamefont
  {{Chiocchetta}}, \citenamefont {{Kiese}}, \citenamefont {{Zelle}},
  \citenamefont {{Piazza}},\ and\ \citenamefont {{Diehl}}}]{diehl}%
  \BibitemOpen
  \bibfield  {author} {\bibinfo {author} {\bibfnamefont {A.}~\bibnamefont
  {{Chiocchetta}}}, \bibinfo {author} {\bibfnamefont {D.}~\bibnamefont
  {{Kiese}}}, \bibinfo {author} {\bibfnamefont {C.~P.}\ \bibnamefont
  {{Zelle}}}, \bibinfo {author} {\bibfnamefont {F.}~\bibnamefont {{Piazza}}},\
  and\ \bibinfo {author} {\bibfnamefont {S.}~\bibnamefont {{Diehl}}},\
  }\bibfield  {title} {\bibinfo {title} {{Cavity-induced quantum spin
  liquids}},\ }\href {https://doi.org/10.1038/s41467-021-26076-3} {\bibfield
  {journal} {\bibinfo  {journal} {Nature Communications}\ }\textbf {\bibinfo
  {volume} {12}},\ \bibinfo {eid} {5901} (\bibinfo {year} {2021})}\BibitemShut
  {NoStop}%
\bibitem [{\citenamefont {Chakraborty}\ and\ \citenamefont
  {Piazza}(2021)}]{ahana}%
  \BibitemOpen
  \bibfield  {author} {\bibinfo {author} {\bibfnamefont {A.}~\bibnamefont
  {Chakraborty}}\ and\ \bibinfo {author} {\bibfnamefont {F.}~\bibnamefont
  {Piazza}},\ }\bibfield  {title} {\bibinfo {title} {Long-range photon
  fluctuations enhance photon-mediated electron pairing and
  superconductivity},\ }\href {https://doi.org/10.1103/PhysRevLett.127.177002}
  {\bibfield  {journal} {\bibinfo  {journal} {Phys. Rev. Lett.}\ }\textbf
  {\bibinfo {volume} {127}},\ \bibinfo {pages} {177002} (\bibinfo {year}
  {2021})}\BibitemShut {NoStop}%
\bibitem [{\citenamefont {{Roux}}\ \emph {et~al.}(2020)\citenamefont {{Roux}},
  \citenamefont {{Konishi}}, \citenamefont {{Helson}},\ and\ \citenamefont
  {{Brantut}}}]{roux}%
  \BibitemOpen
  \bibfield  {author} {\bibinfo {author} {\bibfnamefont {K.}~\bibnamefont
  {{Roux}}}, \bibinfo {author} {\bibfnamefont {H.}~\bibnamefont {{Konishi}}},
  \bibinfo {author} {\bibfnamefont {V.}~\bibnamefont {{Helson}}},\ and\
  \bibinfo {author} {\bibfnamefont {J.-P.}\ \bibnamefont {{Brantut}}},\
  }\bibfield  {title} {\bibinfo {title} {{Strongly correlated fermions strongly
  coupled to light}},\ }\href {https://doi.org/10.1038/s41467-020-16767-8}
  {\bibfield  {journal} {\bibinfo  {journal} {Nature Communications}\ }\textbf
  {\bibinfo {volume} {11}},\ \bibinfo {eid} {2974} (\bibinfo {year}
  {2020})}\BibitemShut {NoStop}%
\bibitem [{\citenamefont {{Mivehvar}}\ \emph {et~al.}(2021)\citenamefont
  {{Mivehvar}}, \citenamefont {{Piazza}}, \citenamefont {{Donner}},\ and\
  \citenamefont {{Ritsch}}}]{piazza_qed}%
  \BibitemOpen
  \bibfield  {author} {\bibinfo {author} {\bibfnamefont {F.}~\bibnamefont
  {{Mivehvar}}}, \bibinfo {author} {\bibfnamefont {F.}~\bibnamefont
  {{Piazza}}}, \bibinfo {author} {\bibfnamefont {T.}~\bibnamefont {{Donner}}},\
  and\ \bibinfo {author} {\bibfnamefont {H.}~\bibnamefont {{Ritsch}}},\
  }\bibfield  {title} {\bibinfo {title} {{Cavity QED with quantum gases: new
  paradigms in many-body physics}},\ }\href
  {https://doi.org/10.1080/00018732.2021.1969727} {\bibfield  {journal}
  {\bibinfo  {journal} {Advances in Physics}\ }\textbf {\bibinfo {volume}
  {70}},\ \bibinfo {pages} {1} (\bibinfo {year} {2021})}\BibitemShut {NoStop}%
\bibitem [{\citenamefont {Zhang}\ \emph {et~al.}(2021)\citenamefont {Zhang},
  \citenamefont {Chen}, \citenamefont {Wu}, \citenamefont {Wang}, \citenamefont
  {Fan}, \citenamefont {Deng},\ and\ \citenamefont {Wu}}]{zhang}%
  \BibitemOpen
  \bibfield  {author} {\bibinfo {author} {\bibfnamefont {X.}~\bibnamefont
  {Zhang}}, \bibinfo {author} {\bibfnamefont {Y.}~\bibnamefont {Chen}},
  \bibinfo {author} {\bibfnamefont {Z.}~\bibnamefont {Wu}}, \bibinfo {author}
  {\bibfnamefont {J.}~\bibnamefont {Wang}}, \bibinfo {author} {\bibfnamefont
  {J.}~\bibnamefont {Fan}}, \bibinfo {author} {\bibfnamefont {S.}~\bibnamefont
  {Deng}},\ and\ \bibinfo {author} {\bibfnamefont {H.}~\bibnamefont {Wu}},\
  }\bibfield  {title} {\bibinfo {title} {{Observation of a superradiant quantum
  phase transition in an intracavity degenerate Fermi gas}},\ }\href
  {https://doi.org/10.1126/science.abd4385} {\bibfield  {journal} {\bibinfo
  {journal} {Science}\ }\textbf {\bibinfo {volume} {373}},\ \bibinfo {pages}
  {1359} (\bibinfo {year} {2021})}\BibitemShut {NoStop}%
\bibitem [{\citenamefont {Piazza}\ and\ \citenamefont
  {Strack}(2014)}]{piazza_superrad}%
  \BibitemOpen
  \bibfield  {author} {\bibinfo {author} {\bibfnamefont {F.}~\bibnamefont
  {Piazza}}\ and\ \bibinfo {author} {\bibfnamefont {P.}~\bibnamefont
  {Strack}},\ }\bibfield  {title} {\bibinfo {title} {Umklapp superradiance with
  a collisionless quantum degenerate {F}ermi gas},\ }\href
  {https://doi.org/10.1103/PhysRevLett.112.143003} {\bibfield  {journal}
  {\bibinfo  {journal} {Phys. Rev. Lett.}\ }\textbf {\bibinfo {volume} {112}},\
  \bibinfo {pages} {143003} (\bibinfo {year} {2014})}\BibitemShut {NoStop}%
\bibitem [{\citenamefont {Keeling}\ \emph {et~al.}(2014)\citenamefont
  {Keeling}, \citenamefont {Bhaseen},\ and\ \citenamefont {Simons}}]{bhaseen}%
  \BibitemOpen
  \bibfield  {author} {\bibinfo {author} {\bibfnamefont {J.}~\bibnamefont
  {Keeling}}, \bibinfo {author} {\bibfnamefont {M.~J.}\ \bibnamefont
  {Bhaseen}},\ and\ \bibinfo {author} {\bibfnamefont {B.~D.}\ \bibnamefont
  {Simons}},\ }\bibfield  {title} {\bibinfo {title} {Fermionic superradiance in
  a transversely pumped optical cavity},\ }\href
  {https://doi.org/10.1103/PhysRevLett.112.143002} {\bibfield  {journal}
  {\bibinfo  {journal} {Phys. Rev. Lett.}\ }\textbf {\bibinfo {volume} {112}},\
  \bibinfo {pages} {143002} (\bibinfo {year} {2014})}\BibitemShut {NoStop}%
\bibitem [{\citenamefont {Chen}\ \emph {et~al.}(2014)\citenamefont {Chen},
  \citenamefont {Yu},\ and\ \citenamefont {Zhai}}]{zhai}%
  \BibitemOpen
  \bibfield  {author} {\bibinfo {author} {\bibfnamefont {Y.}~\bibnamefont
  {Chen}}, \bibinfo {author} {\bibfnamefont {Z.}~\bibnamefont {Yu}},\ and\
  \bibinfo {author} {\bibfnamefont {H.}~\bibnamefont {Zhai}},\ }\bibfield
  {title} {\bibinfo {title} {Superradiance of degenerate fermi gases in a
  cavity},\ }\href {https://doi.org/10.1103/PhysRevLett.112.143004} {\bibfield
  {journal} {\bibinfo  {journal} {Phys. Rev. Lett.}\ }\textbf {\bibinfo
  {volume} {112}},\ \bibinfo {pages} {143004} (\bibinfo {year}
  {2014})}\BibitemShut {NoStop}%
\bibitem [{\citenamefont {Nataf}\ \emph {et~al.}(2019)\citenamefont {Nataf},
  \citenamefont {Champel}, \citenamefont {Blatter},\ and\ \citenamefont
  {Basko}}]{basko}%
  \BibitemOpen
  \bibfield  {author} {\bibinfo {author} {\bibfnamefont {P.}~\bibnamefont
  {Nataf}}, \bibinfo {author} {\bibfnamefont {T.}~\bibnamefont {Champel}},
  \bibinfo {author} {\bibfnamefont {G.}~\bibnamefont {Blatter}},\ and\ \bibinfo
  {author} {\bibfnamefont {D.~M.}\ \bibnamefont {Basko}},\ }\bibfield  {title}
  {\bibinfo {title} {Rashba cavity {QED}: {A} route towards the superradiant
  quantum phase transition},\ }\href
  {https://doi.org/10.1103/PhysRevLett.123.207402} {\bibfield  {journal}
  {\bibinfo  {journal} {Phys. Rev. Lett.}\ }\textbf {\bibinfo {volume} {123}},\
  \bibinfo {pages} {207402} (\bibinfo {year} {2019})}\BibitemShut {NoStop}%
\bibitem [{\citenamefont {Andolina}\ \emph {et~al.}(2019)\citenamefont
  {Andolina}, \citenamefont {Pellegrino}, \citenamefont {Giovannetti},
  \citenamefont {MacDonald},\ and\ \citenamefont {Polini}}]{polini}%
  \BibitemOpen
  \bibfield  {author} {\bibinfo {author} {\bibfnamefont {G.~M.}\ \bibnamefont
  {Andolina}}, \bibinfo {author} {\bibfnamefont {F.~M.~D.}\ \bibnamefont
  {Pellegrino}}, \bibinfo {author} {\bibfnamefont {V.}~\bibnamefont
  {Giovannetti}}, \bibinfo {author} {\bibfnamefont {A.~H.}\ \bibnamefont
  {MacDonald}},\ and\ \bibinfo {author} {\bibfnamefont {M.}~\bibnamefont
  {Polini}},\ }\bibfield  {title} {\bibinfo {title} {{Cavity quantum
  electrodynamics of strongly correlated electron systems: A no-go theorem for
  photon condensation}},\ }\href {https://doi.org/10.1103/PhysRevB.100.121109}
  {\bibfield  {journal} {\bibinfo  {journal} {Phys. Rev. B}\ }\textbf {\bibinfo
  {volume} {100}},\ \bibinfo {pages} {121109} (\bibinfo {year}
  {2019})}\BibitemShut {NoStop}%
\bibitem [{\citenamefont {Guerci}\ \emph {et~al.}(2020)\citenamefont {Guerci},
  \citenamefont {Simon},\ and\ \citenamefont {Mora}}]{pascal}%
  \BibitemOpen
  \bibfield  {author} {\bibinfo {author} {\bibfnamefont {D.}~\bibnamefont
  {Guerci}}, \bibinfo {author} {\bibfnamefont {P.}~\bibnamefont {Simon}},\ and\
  \bibinfo {author} {\bibfnamefont {C.}~\bibnamefont {Mora}},\ }\bibfield
  {title} {\bibinfo {title} {Superradiant phase transition in electronic
  systems and emergent topological phases},\ }\href
  {https://doi.org/10.1103/PhysRevLett.125.257604} {\bibfield  {journal}
  {\bibinfo  {journal} {Phys. Rev. Lett.}\ }\textbf {\bibinfo {volume} {125}},\
  \bibinfo {pages} {257604} (\bibinfo {year} {2020})}\BibitemShut {NoStop}%
\bibitem [{\citenamefont {Rao}\ and\ \citenamefont {Piazza}(2023)}]{peng}%
  \BibitemOpen
  \bibfield  {author} {\bibinfo {author} {\bibfnamefont {P.}~\bibnamefont
  {Rao}}\ and\ \bibinfo {author} {\bibfnamefont {F.}~\bibnamefont {Piazza}},\
  }\bibfield  {title} {\bibinfo {title} {Non-{F}ermi-liquid behavior from
  cavity electromagnetic vacuum fluctuations at the superradiant transition},\
  }\href {https://doi.org/10.1103/PhysRevLett.130.083603} {\bibfield  {journal}
  {\bibinfo  {journal} {Phys. Rev. Lett.}\ }\textbf {\bibinfo {volume} {130}},\
  \bibinfo {pages} {083603} (\bibinfo {year} {2023})}\BibitemShut {NoStop}%
\bibitem [{\citenamefont {Schlawin}\ \emph {et~al.}(2022)\citenamefont
  {Schlawin}, \citenamefont {Kennes},\ and\ \citenamefont {Sentef}}]{sentef}%
  \BibitemOpen
  \bibfield  {author} {\bibinfo {author} {\bibfnamefont {F.}~\bibnamefont
  {Schlawin}}, \bibinfo {author} {\bibfnamefont {D.~M.}\ \bibnamefont
  {Kennes}},\ and\ \bibinfo {author} {\bibfnamefont {M.~A.}\ \bibnamefont
  {Sentef}},\ }\bibfield  {title} {\bibinfo {title} {{Cavity quantum
  materials}},\ }\href {https://doi.org/10.1063/5.0083825} {\bibfield
  {journal} {\bibinfo  {journal} {Applied Physics Reviews}\ }\textbf {\bibinfo
  {volume} {9}},\ \bibinfo {pages} {011312} (\bibinfo {year}
  {2022})}\BibitemShut {NoStop}%
\bibitem [{\citenamefont {Skribanowitz}\ \emph {et~al.}(1973)\citenamefont
  {Skribanowitz}, \citenamefont {Herman}, \citenamefont {MacGillivray},\ and\
  \citenamefont {Feld}}]{ref22cav}%
  \BibitemOpen
  \bibfield  {author} {\bibinfo {author} {\bibfnamefont {N.}~\bibnamefont
  {Skribanowitz}}, \bibinfo {author} {\bibfnamefont {I.~P.}\ \bibnamefont
  {Herman}}, \bibinfo {author} {\bibfnamefont {J.~C.}\ \bibnamefont
  {MacGillivray}},\ and\ \bibinfo {author} {\bibfnamefont {M.~S.}\ \bibnamefont
  {Feld}},\ }\bibfield  {title} {\bibinfo {title} {{Observation of Dicke
  Superradiance in Optically Pumped HF Gas}},\ }\href
  {https://doi.org/10.1103/PhysRevLett.30.309} {\bibfield  {journal} {\bibinfo
  {journal} {Phys. Rev. Lett.}\ }\textbf {\bibinfo {volume} {30}},\ \bibinfo
  {pages} {309} (\bibinfo {year} {1973})}\BibitemShut {NoStop}%
\bibitem [{\citenamefont {Scheibner}\ \emph {et~al.}(2007)\citenamefont
  {Scheibner}, \citenamefont {Schmidt}, \citenamefont {Worschech},
  \citenamefont {Forchel}, \citenamefont {Bacher}, \citenamefont {Passow},\
  and\ \citenamefont {Hommel}}]{ref23cav}%
  \BibitemOpen
  \bibfield  {author} {\bibinfo {author} {\bibfnamefont {M.}~\bibnamefont
  {Scheibner}}, \bibinfo {author} {\bibfnamefont {T.}~\bibnamefont {Schmidt}},
  \bibinfo {author} {\bibfnamefont {L.}~\bibnamefont {Worschech}}, \bibinfo
  {author} {\bibfnamefont {A.}~\bibnamefont {Forchel}}, \bibinfo {author}
  {\bibfnamefont {G.}~\bibnamefont {Bacher}}, \bibinfo {author} {\bibfnamefont
  {T.}~\bibnamefont {Passow}},\ and\ \bibinfo {author} {\bibfnamefont
  {D.}~\bibnamefont {Hommel}},\ }\bibfield  {title} {\bibinfo {title}
  {Superradiance of quantum dots},\ }\href {https://doi.org/10.1038/nphys494}
  {\bibfield  {journal} {\bibinfo  {journal} {Nat. Phys.}\ }\textbf {\bibinfo
  {volume} {3}},\ \bibinfo {pages} {106} (\bibinfo {year} {2007})}\BibitemShut
  {NoStop}%
\bibitem [{\citenamefont {{Timothy Noe}}\ \emph {et~al.}(2012)\citenamefont
  {{Timothy Noe}}, \citenamefont {{Kim}}, \citenamefont {{Lee}}, \citenamefont
  {{Wang}}, \citenamefont {{W{\'o}jcik}}, \citenamefont {{McGill}},
  \citenamefont {{Reitze}}, \citenamefont {{Belyanin}},\ and\ \citenamefont
  {{Kono}}}]{ref24cav}%
  \BibitemOpen
  \bibfield  {author} {\bibinfo {author} {\bibfnamefont {I.}~\bibnamefont
  {{Timothy Noe}}, \bibfnamefont {G.}}, \bibinfo {author} {\bibfnamefont
  {J.-H.}\ \bibnamefont {{Kim}}}, \bibinfo {author} {\bibfnamefont
  {J.}~\bibnamefont {{Lee}}}, \bibinfo {author} {\bibfnamefont
  {Y.}~\bibnamefont {{Wang}}}, \bibinfo {author} {\bibfnamefont {A.~K.}\
  \bibnamefont {{W{\'o}jcik}}}, \bibinfo {author} {\bibfnamefont {S.~A.}\
  \bibnamefont {{McGill}}}, \bibinfo {author} {\bibfnamefont {D.~H.}\
  \bibnamefont {{Reitze}}}, \bibinfo {author} {\bibfnamefont {A.~A.}\
  \bibnamefont {{Belyanin}}},\ and\ \bibinfo {author} {\bibfnamefont
  {J.}~\bibnamefont {{Kono}}},\ }\bibfield  {title} {\bibinfo {title} {{Giant
  superfluorescent bursts from a semiconductor magneto-plasma}},\ }\href
  {https://doi.org/10.1038/nphys2207} {\bibfield  {journal} {\bibinfo
  {journal} {Nat. Phys.}\ }\textbf {\bibinfo {volume} {8}},\ \bibinfo {pages}
  {219} (\bibinfo {year} {2012})}\BibitemShut {NoStop}%
\bibitem [{\citenamefont {{Baumann}}\ \emph {et~al.}(2010)\citenamefont
  {{Baumann}}, \citenamefont {{Guerlin}}, \citenamefont {{Brennecke}},\ and\
  \citenamefont {{Esslinger}}}]{ref25cav}%
  \BibitemOpen
  \bibfield  {author} {\bibinfo {author} {\bibfnamefont {K.}~\bibnamefont
  {{Baumann}}}, \bibinfo {author} {\bibfnamefont {C.}~\bibnamefont
  {{Guerlin}}}, \bibinfo {author} {\bibfnamefont {F.}~\bibnamefont
  {{Brennecke}}},\ and\ \bibinfo {author} {\bibfnamefont {T.}~\bibnamefont
  {{Esslinger}}},\ }\bibfield  {title} {\bibinfo {title} {{Dicke quantum phase
  transition with a superfluid gas in an optical cavity}},\ }\href
  {https://doi.org/10.1038/nature09009} {\bibfield  {journal} {\bibinfo
  {journal} {Nature}\ }\textbf {\bibinfo {volume} {464}},\ \bibinfo {pages}
  {1301} (\bibinfo {year} {2010})}\BibitemShut {NoStop}%
\bibitem [{\citenamefont {Ando}\ \emph {et~al.}(1998)\citenamefont {Ando},
  \citenamefont {Nakanishi},\ and\ \citenamefont {Saito}}]{ando1998}%
  \BibitemOpen
  \bibfield  {author} {\bibinfo {author} {\bibfnamefont {T.}~\bibnamefont
  {Ando}}, \bibinfo {author} {\bibfnamefont {T.}~\bibnamefont {Nakanishi}},\
  and\ \bibinfo {author} {\bibfnamefont {R.}~\bibnamefont {Saito}},\ }\bibfield
   {title} {\bibinfo {title} {Berry's phase and absence of back scattering in
  carbon nanotubes},\ }\href {https://doi.org/10.1143/JPSJ.67.2857} {\bibfield
  {journal} {\bibinfo  {journal} {Journal of the Physical Society of Japan}\
  }\textbf {\bibinfo {volume} {67}},\ \bibinfo {pages} {2857} (\bibinfo {year}
  {1998})}\BibitemShut {NoStop}%
\bibitem [{\citenamefont {{Dresselhaus}}\ and\ \citenamefont
  {{Dresselhaus}}(2002)}]{dresselhaus}%
  \BibitemOpen
  \bibfield  {author} {\bibinfo {author} {\bibfnamefont {M.~S.}\ \bibnamefont
  {{Dresselhaus}}}\ and\ \bibinfo {author} {\bibfnamefont {G.}~\bibnamefont
  {{Dresselhaus}}},\ }\bibfield  {title} {\bibinfo {title} {{Intercalation
  compounds of graphite}},\ }\href {https://doi.org/10.1080/00018730110113644}
  {\bibfield  {journal} {\bibinfo  {journal} {Advances in Physics}\ }\textbf
  {\bibinfo {volume} {51}},\ \bibinfo {pages} {1} (\bibinfo {year}
  {2002})}\BibitemShut {NoStop}%
\bibitem [{\citenamefont {Castro~Neto}\ \emph {et~al.}(2009)\citenamefont
  {Castro~Neto}, \citenamefont {Guinea}, \citenamefont {Peres}, \citenamefont
  {Novoselov},\ and\ \citenamefont {Geim}}]{neto}%
  \BibitemOpen
  \bibfield  {author} {\bibinfo {author} {\bibfnamefont {A.~H.}\ \bibnamefont
  {Castro~Neto}}, \bibinfo {author} {\bibfnamefont {F.}~\bibnamefont {Guinea}},
  \bibinfo {author} {\bibfnamefont {N.~M.~R.}\ \bibnamefont {Peres}}, \bibinfo
  {author} {\bibfnamefont {K.~S.}\ \bibnamefont {Novoselov}},\ and\ \bibinfo
  {author} {\bibfnamefont {A.~K.}\ \bibnamefont {Geim}},\ }\bibfield  {title}
  {\bibinfo {title} {The electronic properties of graphene},\ }\href
  {https://doi.org/10.1103/RevModPhys.81.109} {\bibfield  {journal} {\bibinfo
  {journal} {Rev. Mod. Phys.}\ }\textbf {\bibinfo {volume} {81}},\ \bibinfo
  {pages} {109} (\bibinfo {year} {2009})}\BibitemShut {NoStop}%
\bibitem [{\citenamefont {Sharma}\ \emph {et~al.}(2021)\citenamefont {Sharma},
  \citenamefont {Principi},\ and\ \citenamefont {Maslov}}]{maslov-dfl}%
  \BibitemOpen
  \bibfield  {author} {\bibinfo {author} {\bibfnamefont {P.}~\bibnamefont
  {Sharma}}, \bibinfo {author} {\bibfnamefont {A.}~\bibnamefont {Principi}},\
  and\ \bibinfo {author} {\bibfnamefont {D.~L.}\ \bibnamefont {Maslov}},\
  }\bibfield  {title} {\bibinfo {title} {{Optical conductivity of a Dirac-Fermi
  liquid}},\ }\href {https://doi.org/10.1103/PhysRevB.104.045142} {\bibfield
  {journal} {\bibinfo  {journal} {Phys. Rev. B}\ }\textbf {\bibinfo {volume}
  {104}},\ \bibinfo {pages} {045142} (\bibinfo {year} {2021})}\BibitemShut
  {NoStop}%
\bibitem [{\citenamefont {{Dong}}\ \emph {et~al.}(2024)\citenamefont {{Dong}},
  \citenamefont {{Lee}},\ and\ \citenamefont {{Levitov}}}]{levitov}%
  \BibitemOpen
  \bibfield  {author} {\bibinfo {author} {\bibfnamefont {Z.}~\bibnamefont
  {{Dong}}}, \bibinfo {author} {\bibfnamefont {P.~A.}\ \bibnamefont {{Lee}}},\
  and\ \bibinfo {author} {\bibfnamefont {L.}~\bibnamefont {{Levitov}}},\
  }\bibfield  {title} {\bibinfo {title} {{Charge and spin density wave orders
  in field-biased Bernal bilayer graphene}},\ }\href@noop {} {\bibfield
  {journal} {\bibinfo  {journal} {arXiv e-prints}\ } (\bibinfo {year}
  {2024})},\ \Eprint {https://arxiv.org/abs/2404.18073} {arXiv:2404.18073
  [cond-mat.str-el]} \BibitemShut {NoStop}%
\bibitem [{\citenamefont {{'t Hooft}}(1973)}]{thooft}%
  \BibitemOpen
  \bibfield  {author} {\bibinfo {author} {\bibfnamefont {G.}~\bibnamefont {{'t
  Hooft}}},\ }\bibfield  {title} {\bibinfo {title} {Dimensional regularization
  and the renormalization group},\ }\href
  {https://doi.org/https://doi.org/10.1016/0550-3213(73)90376-3} {\bibfield
  {journal} {\bibinfo  {journal} {Nuclear Physics B}\ }\textbf {\bibinfo
  {volume} {61}},\ \bibinfo {pages} {455} (\bibinfo {year} {1973})}\BibitemShut
  {NoStop}%
\bibitem [{\citenamefont {Weinberg}(1973)}]{weinberg}%
  \BibitemOpen
  \bibfield  {author} {\bibinfo {author} {\bibfnamefont {S.}~\bibnamefont
  {Weinberg}},\ }\bibfield  {title} {\bibinfo {title} {New approach to the
  renormalization group},\ }\href {https://doi.org/10.1103/PhysRevD.8.3497}
  {\bibfield  {journal} {\bibinfo  {journal} {Phys. Rev. D}\ }\textbf {\bibinfo
  {volume} {8}},\ \bibinfo {pages} {3497} (\bibinfo {year} {1973})}\BibitemShut
  {NoStop}%
\end{thebibliography}%

\end{document}